\documentclass[review,12pt,letterpaper]{elsarticle}

\usepackage[utf8]{inputenc}
\usepackage[T1]{fontenc}
\usepackage{graphicx}
\usepackage{mathrsfs}
\usepackage[intlimits,centertags]{amsmath}
\usepackage{amssymb,amsfonts}
\usepackage[pdftex]{hyperref}
\usepackage[x11names]{xcolor}
\usepackage{aas_macros}

\hypersetup{pdftitle={Origin of Small-Scale Anisotropies in Galactic Cosmic Rays},
pdfsubject={Origin of Small-Scale Anisotropies in Galactic Cosmic Rays},
pdfauthor={Markus Ahlers and Philipp Mertsch},
pdfstartview={FitH},
colorlinks=true,
bookmarksopen=false,
bookmarksnumbered=false,
bookmarksopenlevel=0,
linkcolor=Blue1!60!black,
citecolor=Green1!50!black,
urlcolor=Blue1!70!black
}

\journal{Progress in Particle and Nuclear Physics}

\begin{document}

\begin{frontmatter}

\title{\vspace*{1cm}Origin of Small--Scale Anisotropies in Galactic Cosmic Rays}

\author{Markus Ahlers}
\address{WIPAC \& Department of Physics, University of Wisconsin--Madison, Madison, WI 53706, USA}

\author{Philipp Mertsch}
\address{Niels Bohr International Academy, Niels Bohr Institute, Blegdamsvej 17, 2100 Copenhagen, Denmark}

\begin{abstract}

The arrival directions of Galactic cosmic rays (CRs) are highly isotropic. This is expected from the presence of turbulent magnetic fields in our Galactic environment that repeatedly scatter charged CRs during propagation. However, various CR observatories have identified weak anisotropies of various angular sizes and with relative intensities of up to a level of 1 part in 1,000. Whereas large--scale anisotropies are generally predicted by standard diffusion models, the appearance of small--scale anisotropies down to an angular size of $10^\circ$ is surprising. In this review, we summarise the current experimental situation for both the large--scale and small--scale anisotropies. We address some of the issues in comparing different experimental results and remaining questions in interpreting the observed large--scale anisotropies. We then review the standard diffusive picture and its difficulty in producing the small--scale anisotropies. Having set the stage, we review the various ideas and models put forward for explaining the small--scale anisotropies.
\end{abstract}


\end{frontmatter}

%--------------------------------------------------------------------------------------------------------------------------------
%--------------------------------------------------------------------------------------------------------------------------------
%--------------------------------------------------------------------------------------------------------------------------------

\section{Introduction}
\label{sec:Introduction}

The Earth's atmosphere is constantly bombarded by a flux of charged particles, called cosmic rays (CRs). There is a consensus that at energies between a few hundreds of MeV and a few PeV ($10^{15} \, \text{eV}$), CRs are of Galactic origin and most likely connected to the deaths of massive stars~\cite{Baade1934,TheOriginofCosmicRays1964}: supernova remnants (SNRs), pulsars, or pulsar wind nebulae. These Galactic sources are mostly distributed in the Galactic disk. Therefore, if CRs were propagating rectilinearly, these sources would be visible in the distribution of arrival directions, very much like the sources of electromagnetic radiation. However, the observed distribution of CR arrival directions is highly isotropic, to better than 1 part in 1,000 or even 10,000 depending on energy. This implies a mechanism that efficiently randomizes the arrival directions over Galactic distance scales.

In the presence of a turbulent magnetic field, a CR nucleus with charge $Z$ can scatter resonantly with turbulence modes with a wavelength of the order of the gyroradius $r_g\simeq 1.1 (\mathcal{R}/{\rm PV})/(B/{\rm \mu G}) \, \text{pc}$ ($1\,{\rm pc}\simeq 3 \times 10^{18} \, \text{cm}$)~\cite{1966ApJ...146..480J,1966PhFl....9.2377K,1967PhFl...10.2620H,1970ApJ...162.1049H}, where $\mathcal{R} \equiv p c/(Z e)$ is the CR's rigidity. (In a static magnetic field, the trajectory of a CR depends only on this ratio of its momentum $p$ and charge $Z$.) CRs are thus performing a random walk and are losing any correlation with their initial directions over a few scattering times. After long time scales this results in a diffusion process and it is this diffusion that quickly erases the information on the distribution of sources. Since CRs with larger rigidity can escape the Galactic environment more quickly, the local CR spectrum is softer than the initial CR emission spectrum from diffusive shock acceleration~\cite{1977SPhD...22..327K,1977ICRC...11..132A,Bell1978a,Bell1978b,Blandford1978} which in its simplest incarnation is $\propto \mathcal{R}^{-2}$. This is consistent with the relatively soft observed spectrum $\propto \mathcal{R}^{-2.7}$.

Another hint for diffusion being the most important mechanism of CR transport comes from the observation of so-called CR secondary species (e.g.\ Lithium, Beryllium, Boron, sub-Iron elements). The relative contribution of these CRs is larger than the observed solar abundance, which is believed to be representative for the abundances at CR sources. Consequently, all the observed CR secondaries must be produced during the propagation of CR primaries (e.g.\ protons, Helium, Nitrogen, Oxygen, Carbon, Iron) by spallation on interstellar gas. The integrated matter density (``grammage''), that needs to be traversed in the interstellar medium ($91 \, \%$ p, $9 \, \%$ Helium by number~\cite{2001RvMP...73.1031F}) to produce the secondary fluxes, is inferred to be of the order of a few $\text{g} \, \text{cm}^{-2}$. With the typical distance scale for Galactic sources of the order of a few kiloparsec and a number density of $n_{\text{gas}} \simeq 1 \, \text{cm}^{-3}$ in the Galactic disk, the grammage for rectilinear propagation is falling three orders of magnitude short. This requires that the observed flux of CRs must have traversed the Galactic disk many times after emission, which is also implied by diffusion.

However, it can easily be seen that diffusion does not imply that the arrival directions of CRs are completely isotropic. For instance, the relative motion of the observer with respect to a frame in which the CR distribution was completely isotropic would induce a weak dipole anisotropy in the direction of the motion, the Compton--Getting effect~\cite{CG1935,1968Ap&SS...2..431G}. Moreover, an asymmetric distribution of sources introduces a local density gradient which implies, by Fick's law, the presence of a net flux. In the case of isotropic diffusion, this will be visible in the CR arrival direction as a dipole anisotropy pointing into the upstream direction. This has been advertised~\cite{1995ICRC....3...56P,Buesching:2008hr,DiBernardo:2010is,Borriello:2010qh,Linden:2013mqa} as a means of finding the direction of the bulk of sources or even young nearby sources which can be dominating the local CR gradient. However, none of these predictions have so far been unambiguously identified in the CR data. In particular, simple models of isotropic CR diffusion predict dipole anisotropies of TeV-PeV CRs that are much larger than the observed values~\cite{2005JPhG...31R..95H,Erlykin:2006ri,Ptuskin2006,Blasi:2011fm,Evoli:2012ha,Pohl:2012xs,Kumar:2014dma,Sveshnikova:2013ui}. This discrepancy has been dubbed the CR ``anisotropy problem''~\cite{2005JPhG...31R..95H}.

Together with the overall CR spectrum and the relative abundances of different species, anisotropies constitute one of the classical observables of CR physics. First hints of a large--scale anisotropy were already observed in the early 1930's, but the systematic and statistical uncertainties of these observations were quite large~\cite{Wollan1939}. A systematic study of the small effect became possible in the 1950's due to data collected by large underground muon detectors and extended air shower arrays, see~\cite{DiSciascio:2014jwa}. We refer to the comprehensive review by Di Sciascio \& Iuppa~~\cite{DiSciascio:2014jwa} for the history of CR anisotropy studies. Only rather recently, however, have experiments achieved the necessary level of statistics to be able to find anisotropies of the order of $10^{-3}$ or even $10^{-4}$~\citep{Amenomori:2005dy,Amenomori:2006bx,Guillian:2005wp,Abdo:2008kr,Abdo:2008aw,ARGO-YBJ:2013gya,Aglietta:2009mu,Abbasi:2011ai,Aartsen:2013lla,Abeysekara:2014sna}. To the surprise of many, besides the expected large--scale anisotropy mentioned above, there is structure in the maps of arrival directions on much smaller scales, at least down to $10^{\circ}$. 

Anisotropies on scales smaller than the dipole are not predicted by the simple diffusion picture and this has been the topic of much attention and theoretical modeling efforts. The suggested explanations consider effects of the heliosphere, non--diffusive propagation, modifications of pitch--angle diffusion, stochasticity effects and also more exotic explanations. The interpretations of these small--scale anisotropies will be the main focus of this review. Further details on Galactic CRs can be found in the standard textbook by Berezinsky, Bulanov, Dogiel \& Ptuskin~\cite{Ginzburg:1990sk}, and more recent reviews by Strong, Moskalenko \& Ptuskin~\cite{Strong:2007nh}, or Grenier, Black \& Strong~\cite{Grenier:2015egx}. For details on wave--particle interactions we refer to the detailed textbook of Schlickheiser~\cite{Schlickeiser:2002pg}.

We start in Sec.~\ref{sec:Observation} with a summary of observations of TeV-PeV anisotropies on various angular scales. Here, we will focus on recent data observed by high-statistics experiments in the past two decades and refer to the review~\cite{DiSciascio:2014jwa} for an extended list of observations. Apart from sky maps, we present the observational results of the classical harmonic analysis, which at lowest order parametrises the anisotropy as a dipole. The angular power spectrum, which is another way of presenting the information on the small--scale anisotropies, will be discussed. Throughout, we will point out the difficulties and limitations of anisotropy reconstruction with ground--based observatories, that are crucial for a correct interpretation of results. We will also review some of the various techniques used by observatories, which do not always yield the same results and have to be treated with some care. To provide a background for interpretations of the small--scale anisotropies, we  recall in Sec.~\ref{sec:StandardPicture} the derivation of the diffuse dipole in the presence of a CR gradient as well as the dipole due to the Compton--Getting effect. We will confront models for the large--scale anisotropies with the observations and offer a few remarks on open issues. Explanations put forward for the small--scale anisotropies will be covered in Sec.~\ref{sec:InterpretationSmall} and we will put emphasis on the attractive suggestions that the specific realization of the local turbulent magnetic field can be the source of the small--scale anisotropies. We summarize and provide an outlook in Sec.~\ref{sec:SummaryOutlook}.

%--------------------------------------------------------------------------------------------------------------------------------
%--------------------------------------------------------------------------------------------------------------------------------
%--------------------------------------------------------------------------------------------------------------------------------
\section{Observation}
\label{sec:Observation}

The anisotropy of CR arrival directions is typically defined as the {\it relative intensity} of CRs as a function of arrival direction. If $\phi^{\rm iso}$ denotes the isotropic average of the CR flux $\phi({\bf n})$ from the arrival direction parametrized by a unit vector ${\bf n}$,
then the relative intensity $I$ can be defined as
\begin{equation}\label{eq:relint}
I({\bf n}) \equiv \frac{\phi({\bf n})}{\phi^{\rm iso}} \equiv 1 + \delta I({\bf n}) \,.
\end{equation}
There is some confusion in the literature, whether $I$ or its residual $\delta I$ should be called relative intensity. In this review we will denote $\delta I$ as the {\it anisotropy}, but note that some of the figures extracted from various experimental publications might use a different convention. 

\begin{figure}[t]\centering
\includegraphics[width=0.9\linewidth]{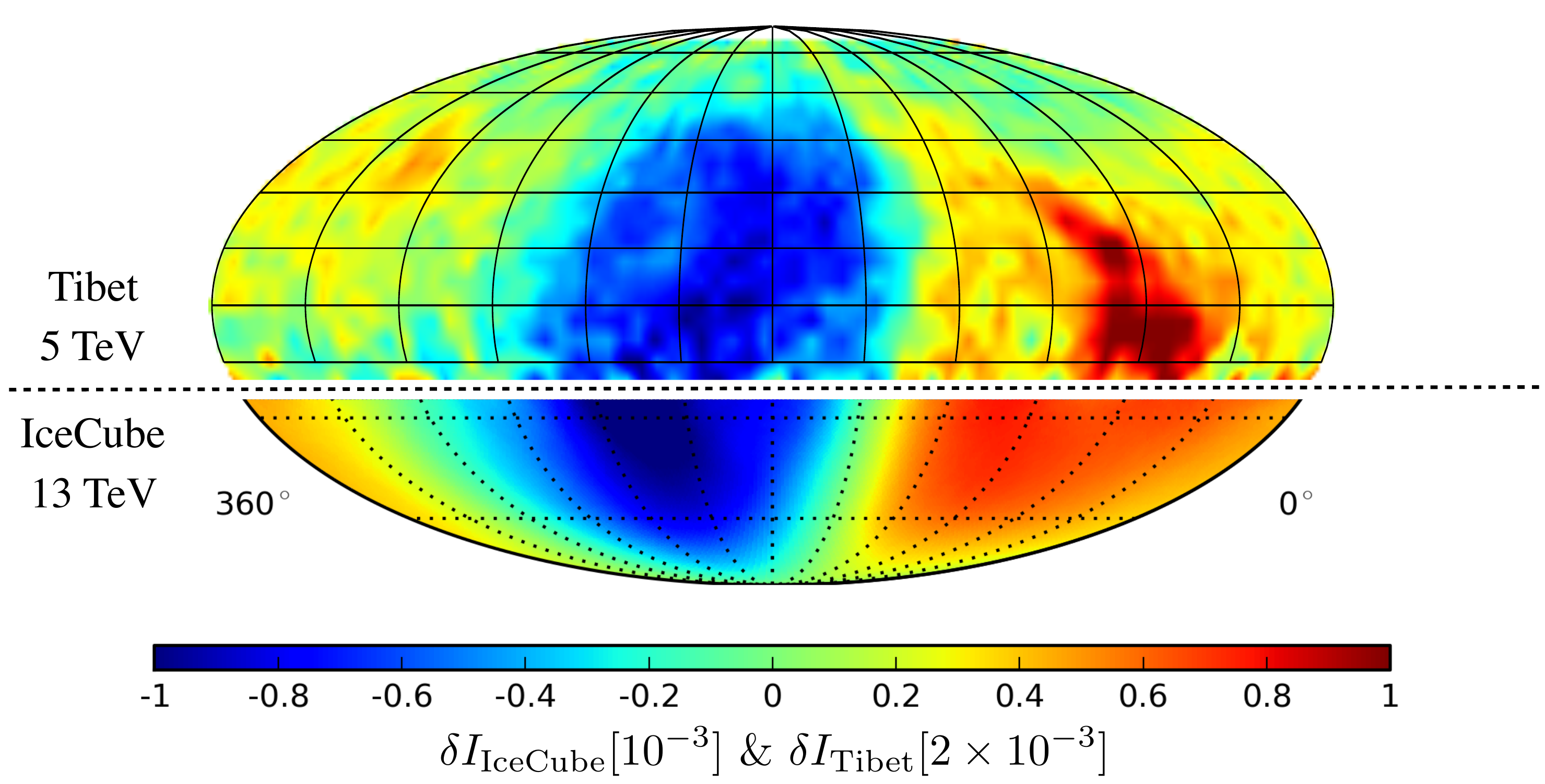}
\caption[]{Combined CR anisotropy of Tibet-AS$\gamma$~\cite{Amenomori:2010yr,Saito2011} and IceCube~\cite{Aartsen:2016ivj} in the equatorial coordinate system. The Tibet result (units of $2\times10^{-3}$) is reconstructed from $4.9\times10^{10}$ events observed over nine years with a median energy of 5~TeV. The declination range $-10^\circ\lesssim\delta\lesssim70^\circ$ reflects the location of the Tibet array at $30^\circ$ North and an effective zenith range $\theta<40^\circ$. The IceCube result (units of $10^{-3}$) is inferred from $3.2\times10^{11}$ CR events observed over a period of six years and has a larger median energy of 13~TeV. The observable zenith range limits the FOV to declinations $\delta\lesssim-25^\circ$. Note that the anisotropy map of IceCube is smoothed with a top-hat kernel with radius of $10^\circ$, which explains the absence of smaller features visible in the Tibet map.}\label{fig:combinedmaps}
\end{figure} 

Without loss of generality, we will discuss the CR anisotropy in the equatorial coordinate system parametrized by right ascension $\alpha\in[0,360^\circ]$ and declination $\delta\in[-90^\circ,90^\circ]$. The Earth's equator projected onto the sky is located at $\delta=0^\circ$ and the meridian $\alpha=0^\circ$ intersects the equator at the location of March equinox, i.e.\ the location of the Sun as it passes through the equatorial plane from South ($\delta<0^\circ$) to North ($\delta>0^\circ$). A unit vector in the equatorial coordinate system can be defined via 
\begin{equation}\label{eq:n}
{\bf n}(\alpha,\delta)=(\cos\alpha\cos\delta,\sin\alpha\cos\delta,\sin{\delta})\,,
\end{equation}
where the direction of geographic North defines $\delta=90^\circ$. As we will explain in more detail at the end of this section, the equatorial system is the {\it natural} coordinate system of ground--based astronomy.

\begin{figure}[t]\centering
\includegraphics[width=0.7\linewidth,viewport=10 0 830 430]{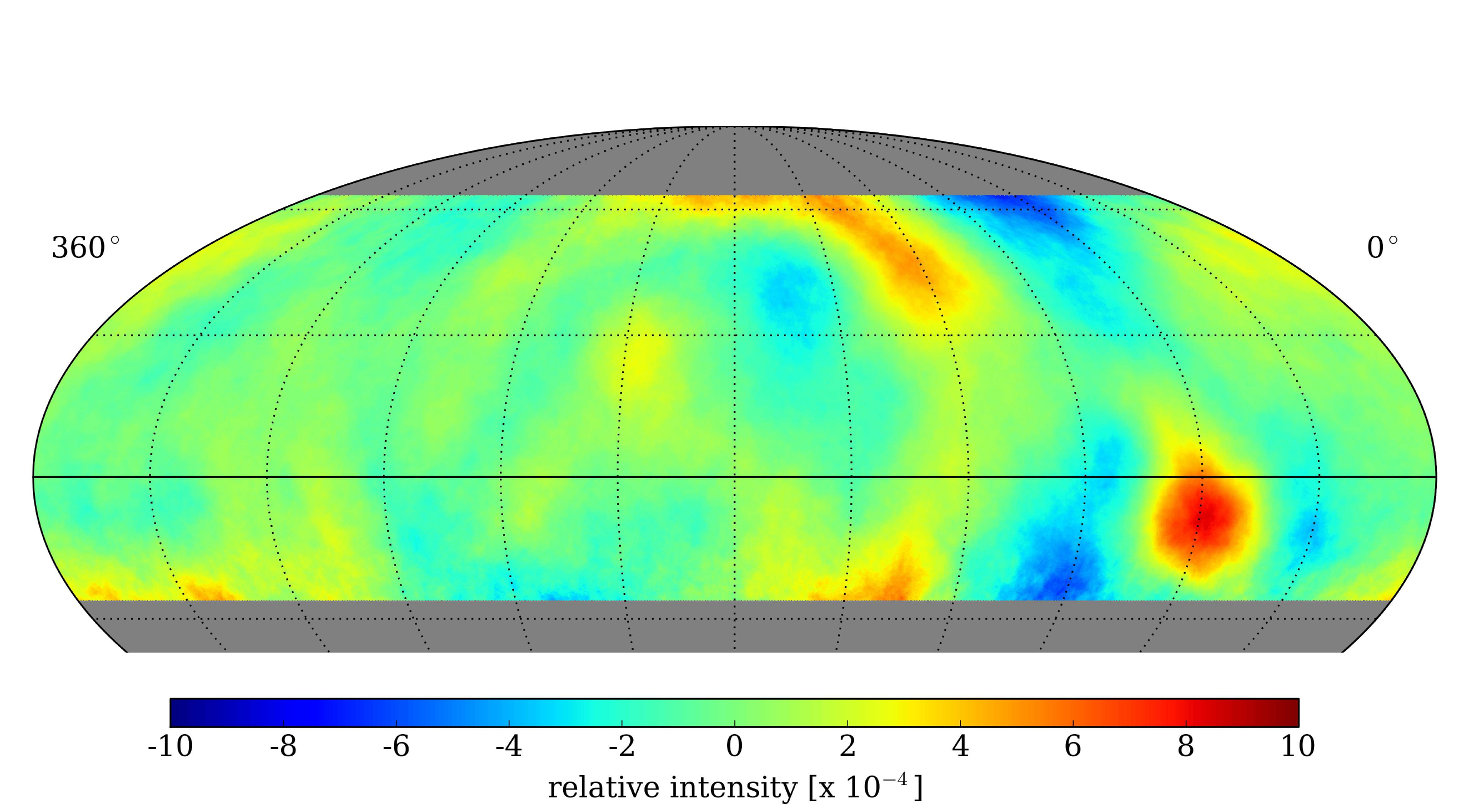}\\
\includegraphics[width=0.7\linewidth,viewport=10 0 830 430,clip=true]{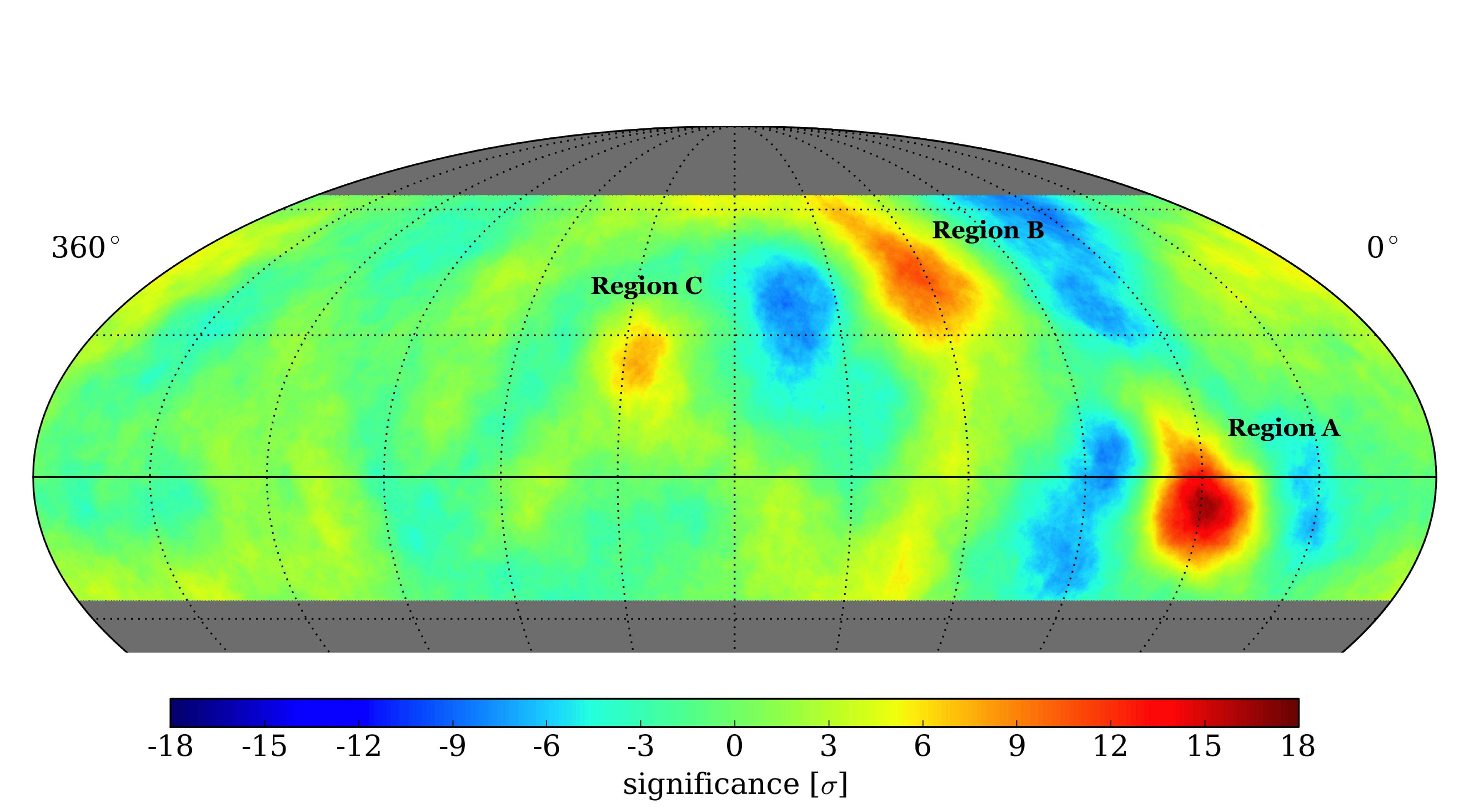}
\caption[]{Equatorial map of small--scale anisotropies of CRs (top) and their significance (bottom) observed by HAWC~\cite{Abeysekara:2014sna} after removal of dipole, quadrupole, and octupole features (see Section~\ref{subsec:HarmonicAnalysis}) and after smoothing with a top hat kernel with radius of $10^\circ$. The observation is based on $4.9\times10^{10}$ CR events observed over a period of 113 days with median energy of 2~TeV. Because of HAWC's location at a latitude of $\Phi\simeq19^\circ$ and the zenith cut of $45^\circ$ the time--integrated FOV is limited to $-26^\circ\lesssim\delta\lesssim64^\circ$.}\label{fig:HAWCmap}
\end{figure}

To set the stage and without going into the details of the experimental techniques employed, we consider the a combined CR anisotropy map of Tibet-AS$\gamma$~\cite{Abeysekara:2014sna} and IceCube~\cite{Aartsen:2016ivj} shown in Fig.~\ref{fig:combinedmaps}. In the case of Tibet, the anisotropy is inferred from $4.9\times10^{10}$ CR events observed during nine years with median energy of 5~TeV. IceCube's result is based on the observation of $3.2\times10^{11}$ events during six years with a median energy of 14~TeV. In general, for a maximal zenith range $\theta<\theta_{\rm max}$ of the instantaneous field of view (FOV) the observable declination range is given by $\delta_1<\delta<\delta_2$ with $\delta_{1/2} = \Phi \mp\theta_{\rm max}$, where $\Phi$ is the geographic latitude of the observatory (measured from the equator towards North). For the Tibet array at geographical latitude of $30^\circ$~N and the zenith cut of $\theta_{\rm max}=45^\circ$ this corresponds to $-10^\circ\lesssim\delta\lesssim70^\circ$. The IceCube observatory at the Southpole uses a zenith cut of $65^\circ$ and hence observed the southern hemisphere at declination $\delta<-25^\circ$.

The maps clearly shows anisotropies up to a level of $10^{-3}$ on various angular scales. To highlight small--scale anisotropies one can remove the fit of large scale harmonics (dipole, quadrupole, octupole, etc.) from the anisotropy map. A recent analysis of these small--scale anisotropies is shown in Fig.~\ref{fig:HAWCmap}. The anisotropy is inferred from $4.9\times10^{10}$ CR events with median energy of 2~TeV observed by HAWC~\cite{Abeysekara:2014sna}. The  declination range $-26^\circ\lesssim\delta\lesssim64^\circ$ is due to HAWC's position at $19^\circ$ North combined with the effective zenith range of $\theta<45^\circ$. To suppress shot noise, the anisotropy map is smoothed via averaging over circular regions with a $10^\circ$ half-width.

\begin{figure}[p]\centering
\includegraphics[width=0.49\textwidth]{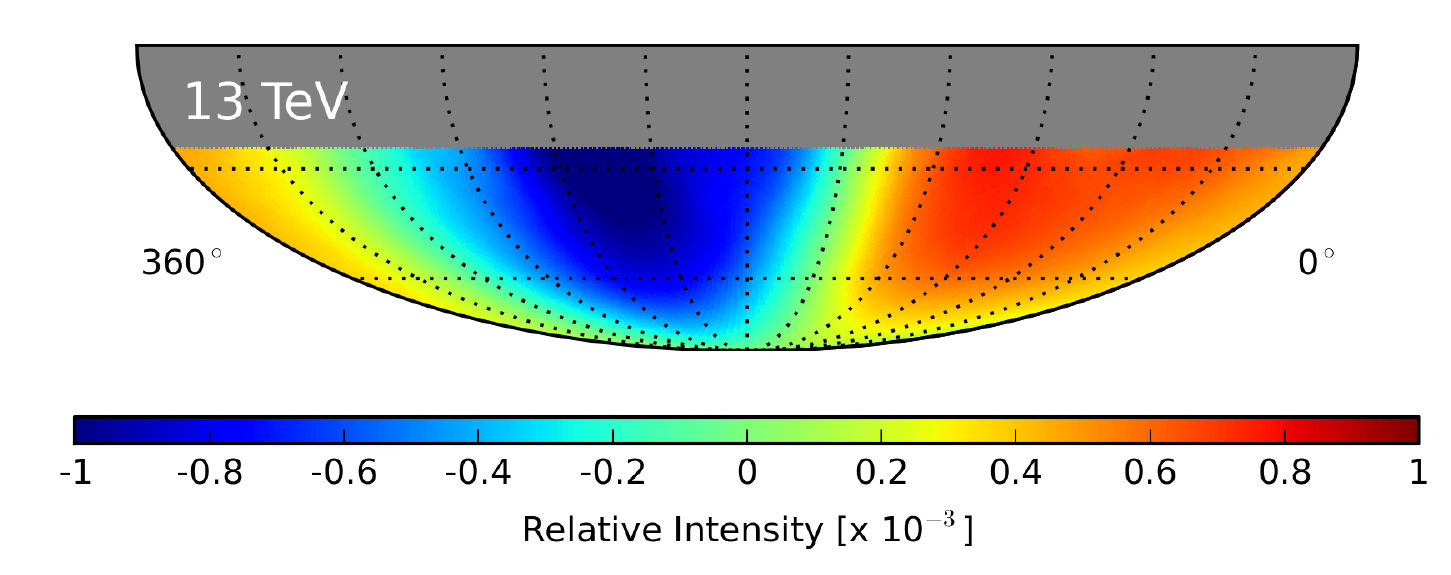}
\includegraphics[width=0.49\textwidth]{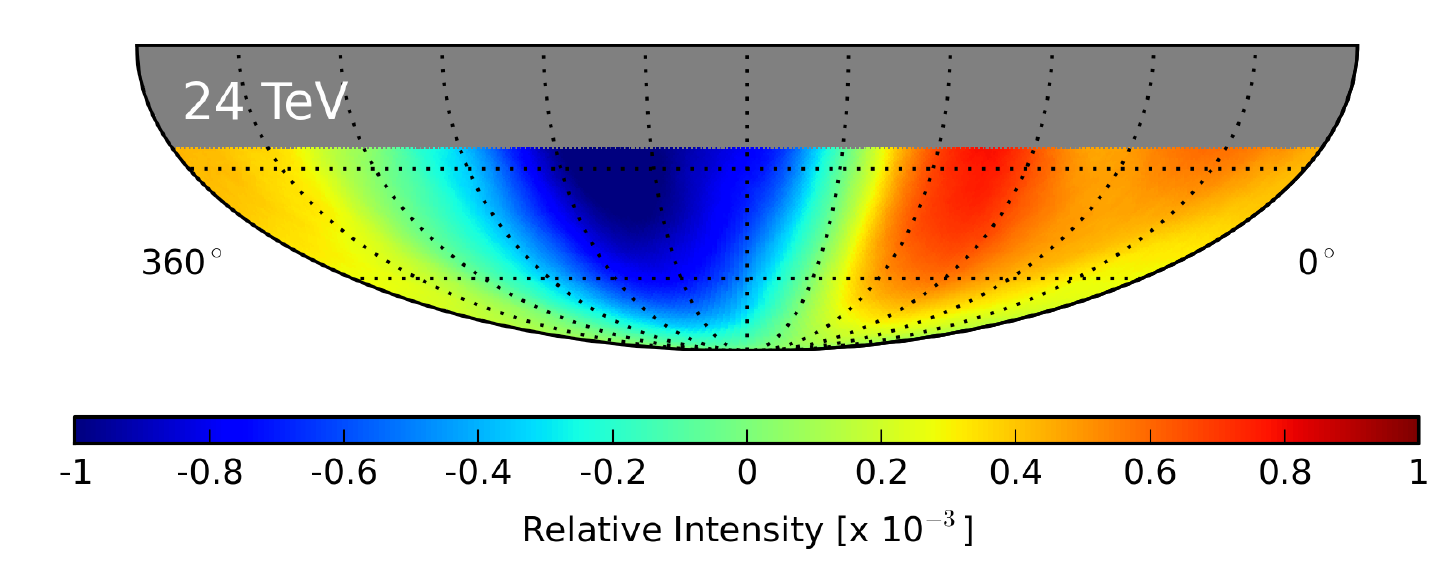}
\includegraphics[width=0.49\textwidth]{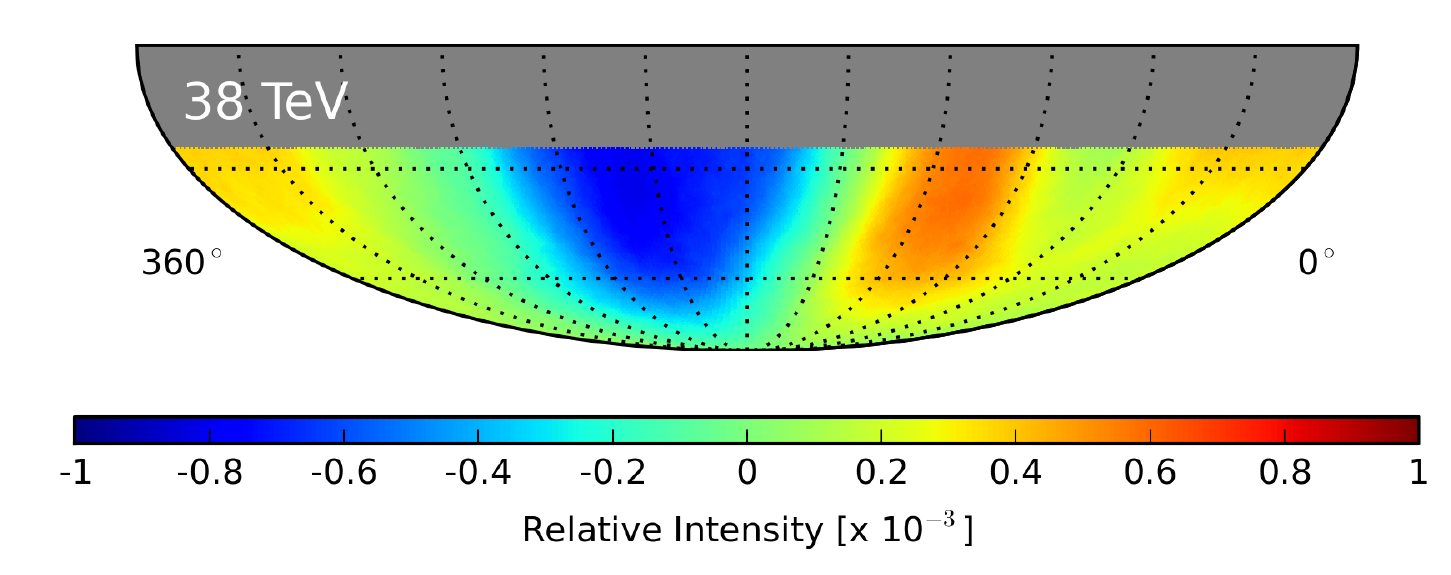}
\includegraphics[width=0.49\textwidth]{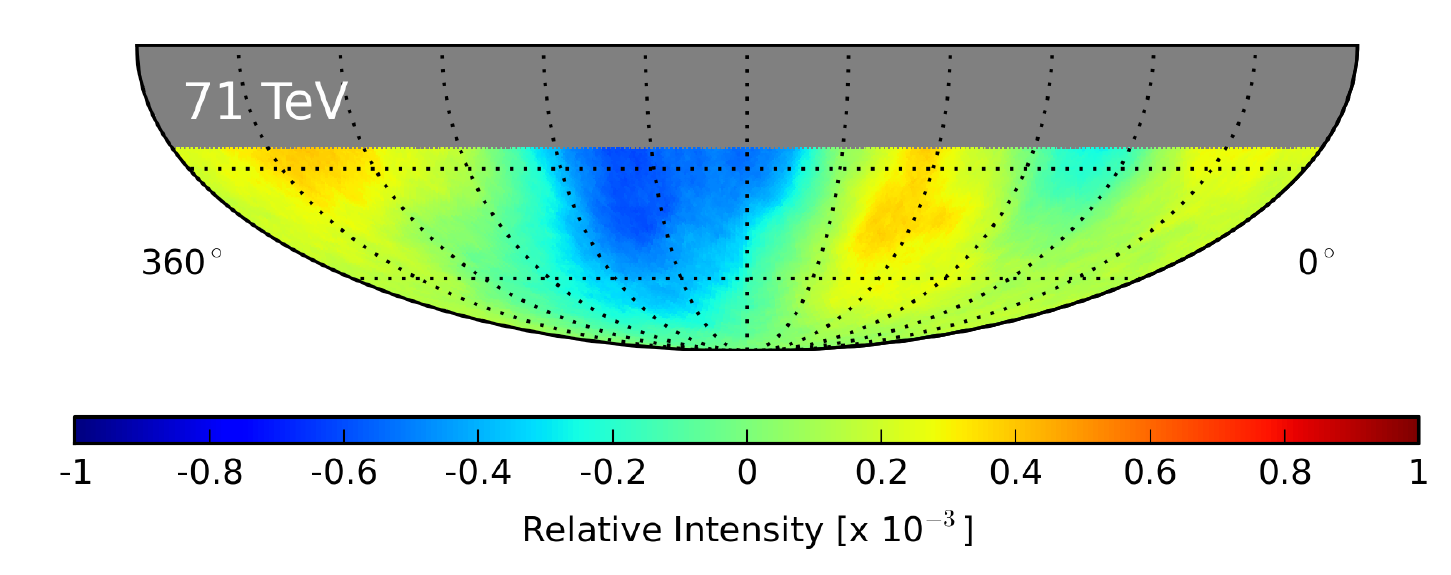}
\includegraphics[width=0.49\textwidth]{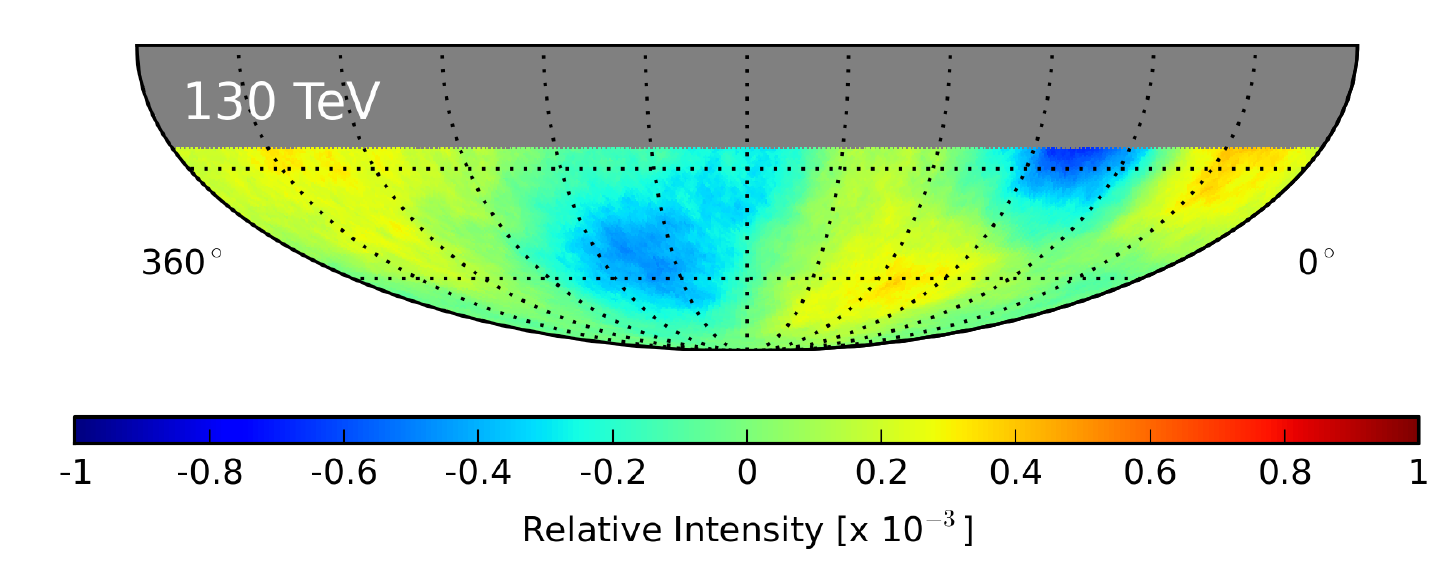}
\includegraphics[width=0.49\textwidth]{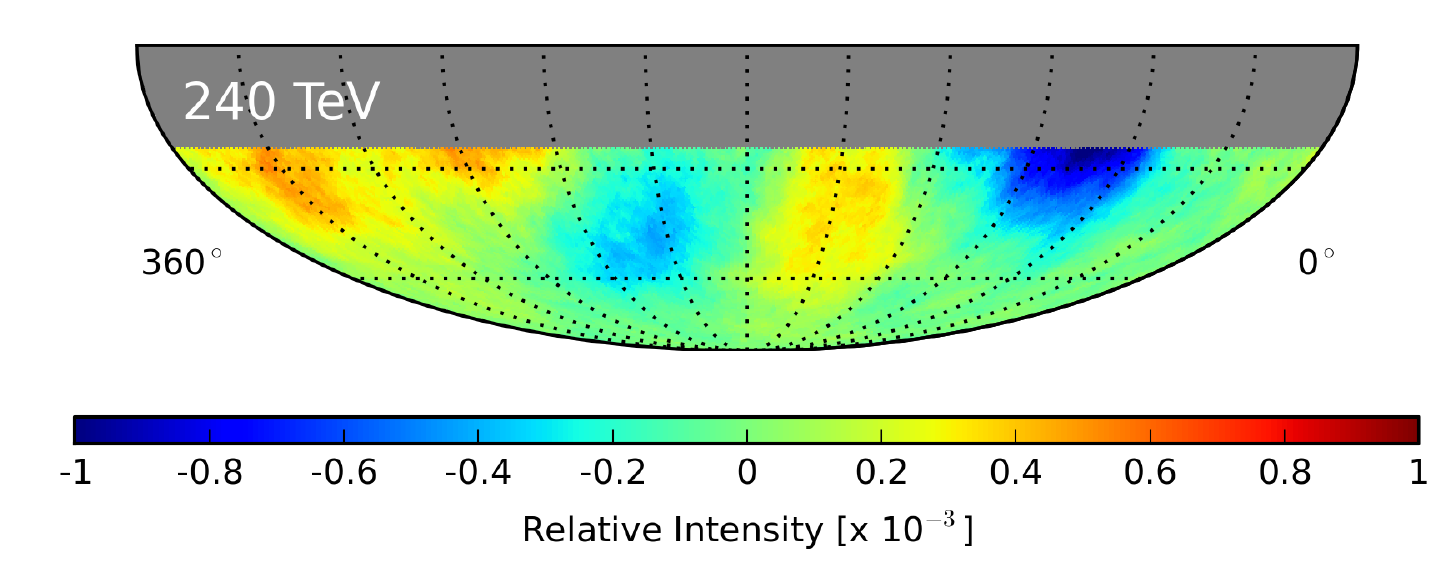}
\includegraphics[width=0.49\textwidth]{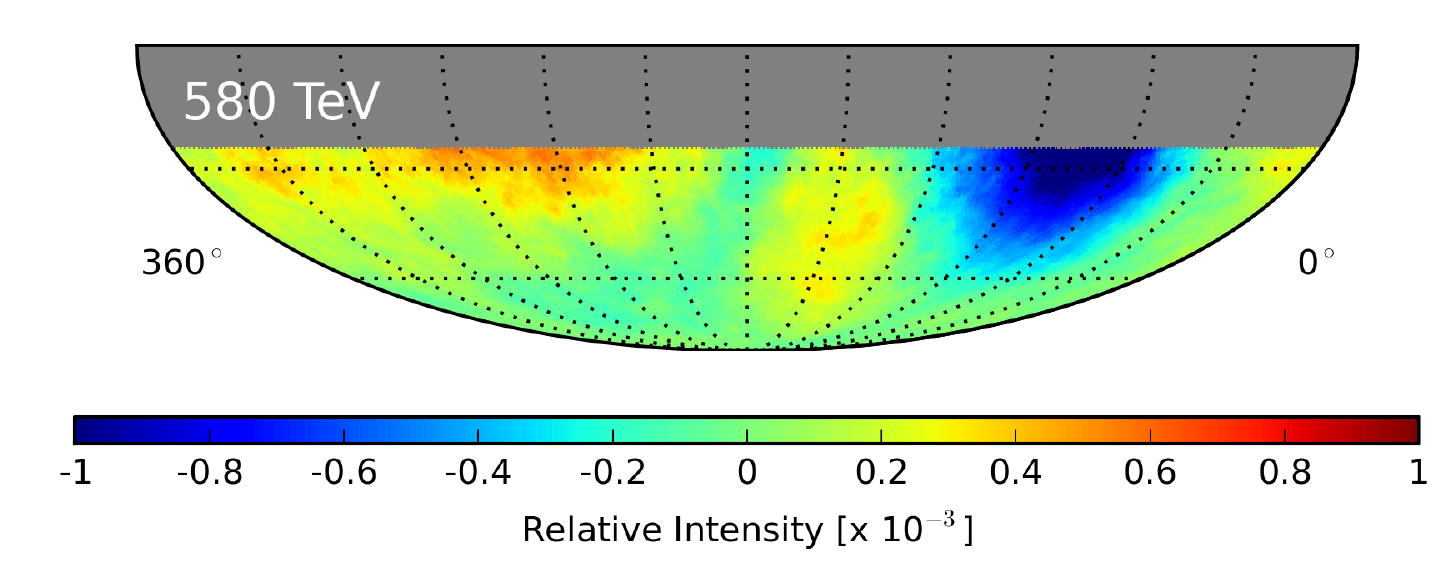}
\includegraphics[width=0.49\textwidth]{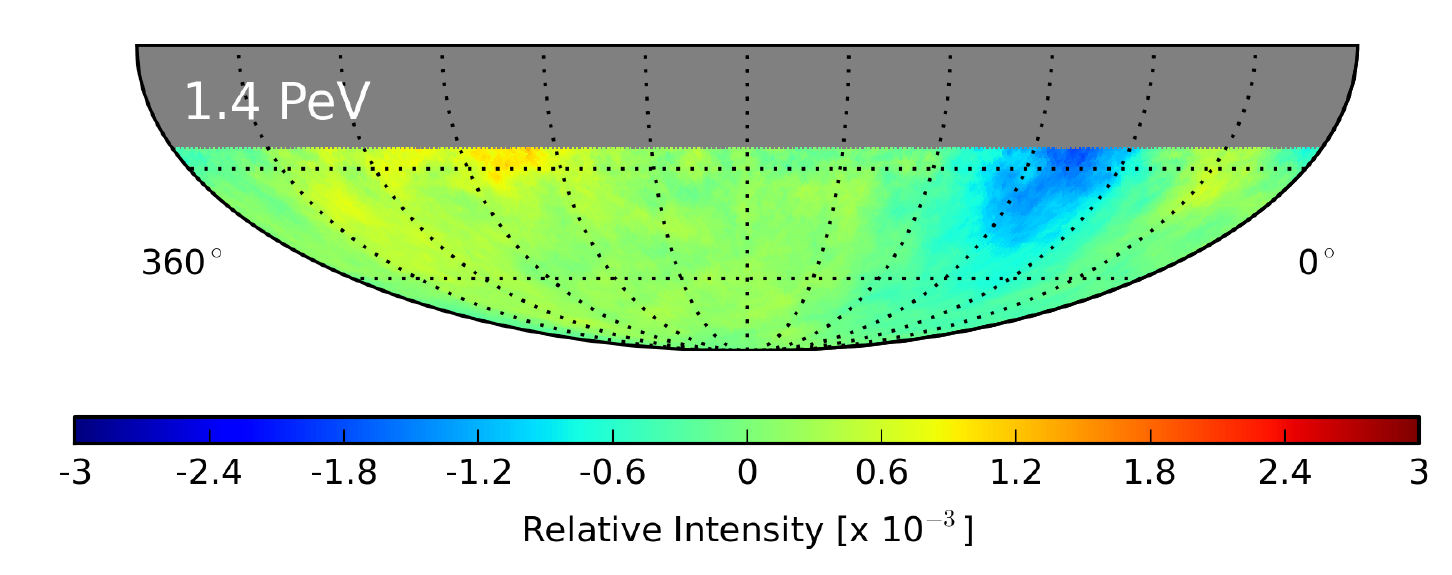}
\includegraphics[width=0.49\textwidth]{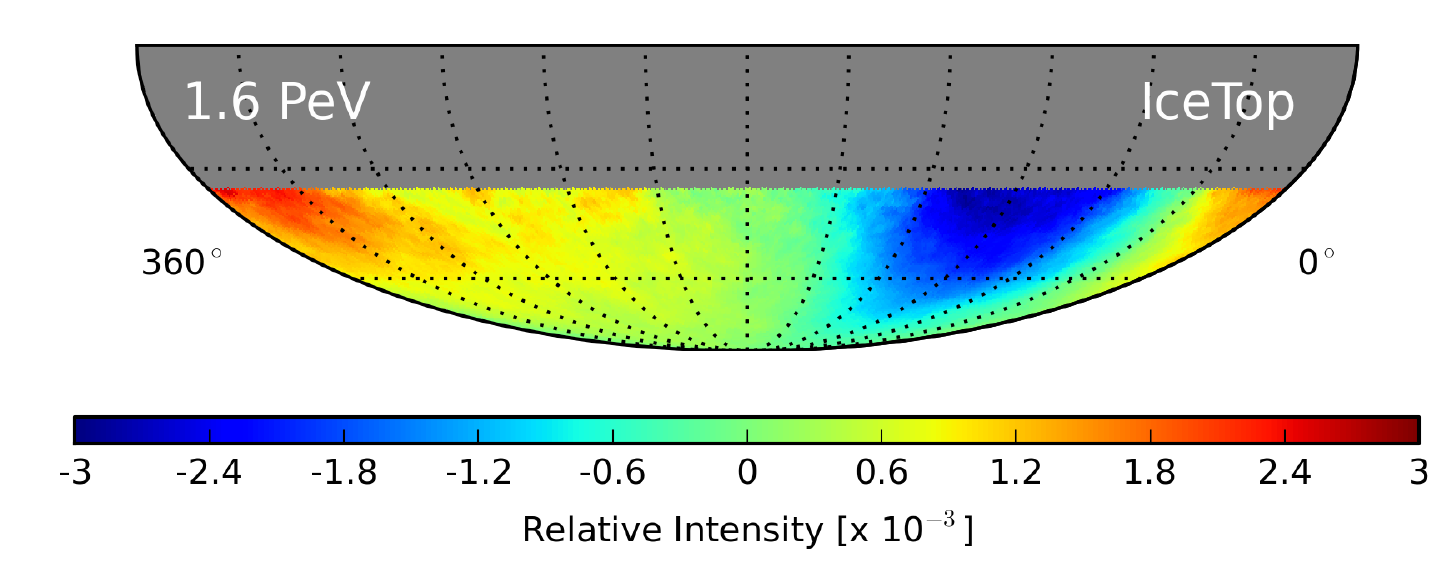}
\includegraphics[width=0.49\textwidth]{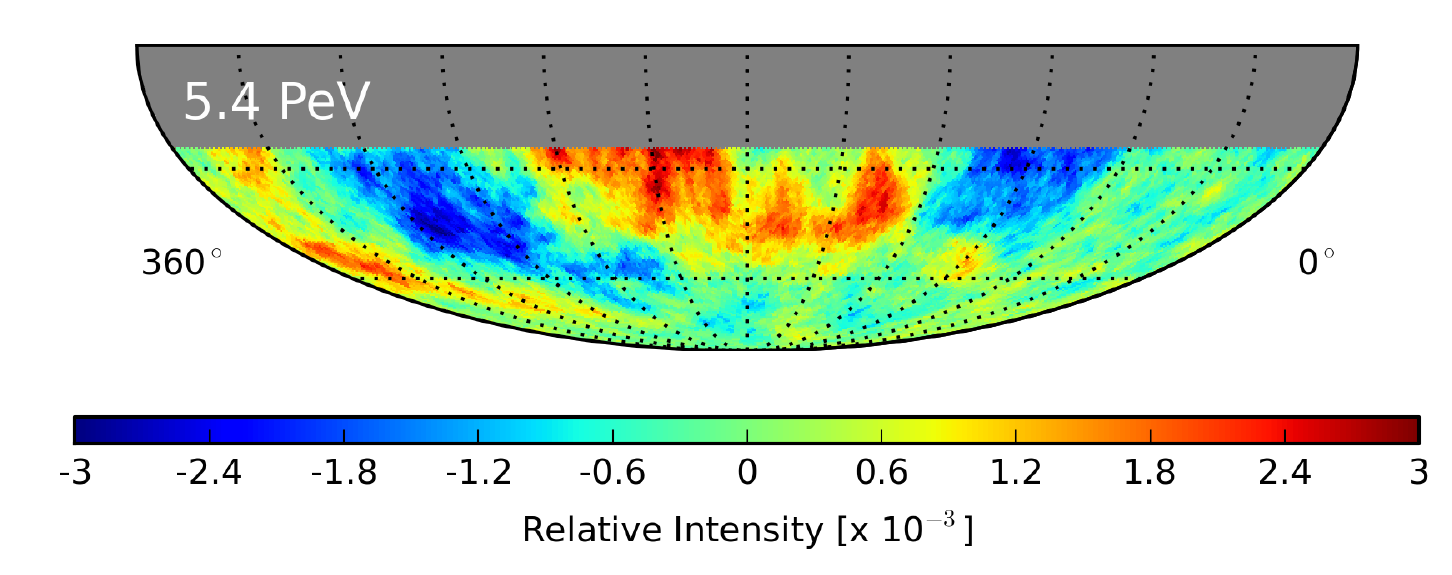}
\caption[]{Anisotropy maps of CRs observed at different energies by IceCube and IceTop~\cite{Aartsen:2016ivj}. The maps are inferred from $3.2\times10^{11}$ CR events observed over a period of six years. The median energy of the data bins range from $13$~TeV to $5.4$~PeV as indicated in the panels. The observable zenith range limits the FOV to declinations $\delta\lesssim25^\circ$ for IceCube and $\delta\lesssim35^\circ$ for IceTop (lower left panel).}\label{fig:ICmaps}
\end{figure}

The smoothed and truncated anisotropy map shows various medium--scale features. To first order, the relative intensity can be estimated from the data by
\begin{equation}
I \simeq \frac{N_{\rm on}}{\alpha N_{\rm off}}\,,
\end{equation}
where $N_{\rm on}$ is the number of events in the excess region ({\it on--source}) and $N_{\rm off}$ the number of events in the reference region for the background estimate ({\it off--source}) with relative exposure $\mathcal{E}_{\rm on}/\mathcal{E}_{\rm off}=\alpha$. The lower panel of Fig.~\ref{fig:HAWCmap} shows the (signed) significance of these excesses and deficits with negative significance indicating deficit regions. For small anisotropies ${\delta I} \equiv I-1\ll 1$ the signed significance $S$ in units of Gaussian standard deviation can be then be approximated as~\cite{Li:1983fv}
\begin{equation}\label{eq:Isens}
S \simeq \sqrt{\frac{N_{\rm on}}{1+\alpha}}\,{\delta I}\,.
\end{equation} 
For instance, in the HAWC anisotropy map shown in the top panel of Fig.~\ref{fig:HAWCmap} with a $10^\circ$ degree half-width smoothing applied, the expected number of on-source events is $N_{\rm on} \simeq 5.6\times10^8$ and hence the sensitivity ($1\sigma$ C.L.) for $\alpha\ll1$ is $\delta I_{1\sigma}  \simeq 4\times10^{-5}$. 

The particularly strong excess regions labeled ``A'', ``B'', and ``C'' indicated in the lower sky map of Fig.~\ref{fig:HAWCmap} have also been observed previously by Tibet~\cite{Amenomori:2006bx}, Milagro~\cite{Abdo:2008kr}, and ARGO-YBJ~\cite{ARGO-YBJ:2013gya}.  Note, that Eq.~(\ref{eq:Isens}) does not account for trials factors (look--elsewhere effect), which somewhat reduces the a posteriori significance of the excess regions. The HAWC collaboration studied the energy dependence of region A and found an increase of the anisotropy in the $1$--$10$~TeV region, with a spectrum harder than the isotropic flux by one power in energy and a possible cutoff at $10$~TeV, in agreement with what had been seen by Milagro earlier~\cite{Abdo:2008kr}.

\begin{figure}[t]\centering
\includegraphics[width=0.95\linewidth]{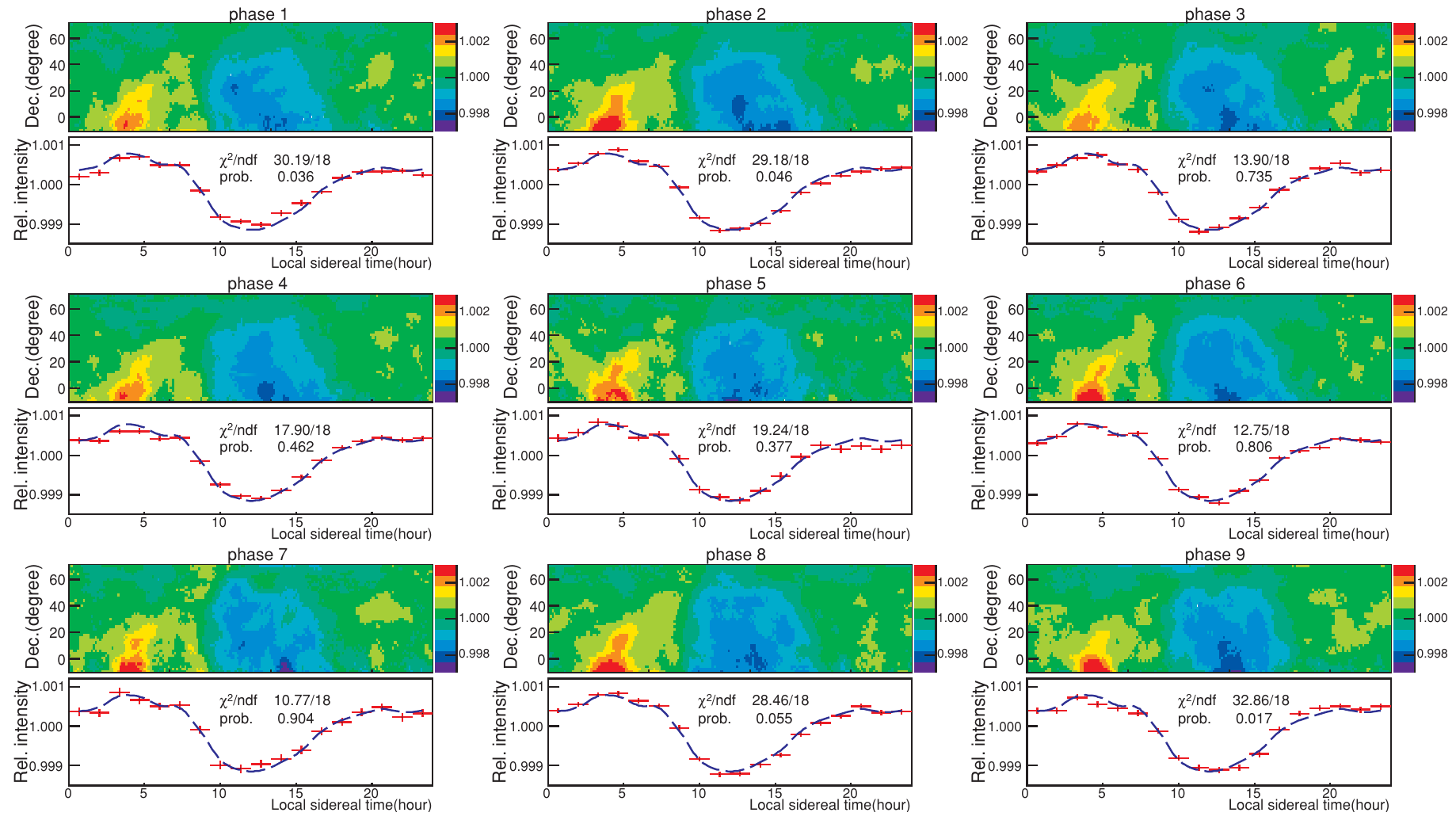}
\caption[]{Time dependence of the relative intensity of Tibet-III~\cite{Amenomori:2010yr}. The two-dimensional map in each panel shows of the relative intensity for nine individual years, in contrast to Fig.~1 showing the combined observation. The lower plots in each panel show the declination-averaged relative intensity for the individual period (red data) compared to the combined result (blue dashed line). No significant time variability is observable.}\label{fig:TibetTime}
\end{figure}

A strong energy dependence of the anisotropy has been observed by various experiments. Figure~\ref{fig:ICmaps} shows the results of a recent analysis of IceCube and IceTop~\cite{Aartsen:2016ivj}. The data is divided into ten bins with median energy ranging from $13$~TeV to $5.4$~PeV as indicated in the panels. The anisotropy map at 13~TeV has already been shown in Fig.~\ref{fig:combinedmaps}. Due to its location at the geographic South Pole the FOV is limited to declinations $\delta\lesssim25^\circ$ for IceCube and $\delta\lesssim35^\circ$ for IceTop. One can notice that the $10$~TeV anisotropy is dominated by a large scale feature that fades away towards $100$~TeV and reappears with reversed sign around $1$~PeV. These maps also show evidence for medium- and small--scale features.

The time--dependence of the large--scale anisotropy is somewhat ambiguous. The Milagro collaboration reported an increasing intensity of the sidereal dipole amplitude over a period of seven years~\cite{Abdo:2008aw}.  However, this result could not be confirmed by Tibet~\cite{Amenomori:2010yr}, covering the same celestial region, nor IceCube~\cite{Aartsen:2013lla} observing the Southern Hemisphere. As an example, Fig.~\ref{fig:TibetTime} shows the results form Tibet-AS$\gamma$~\cite{Amenomori:2010yr} for nine individual years. The combined data was already shown in Fig.~\ref{fig:combinedmaps} as a Mollweide projection. In each panel, the relative intensity is shown in the top over bins of local sidereal time (LST) and declination and at the bottom via its declination-average over LST. (Here, the LST $t$ refers to the moment when the corresponding bin aligns with the meridian and is, therefore, equivalent to right ascension, $\alpha = (t/{\rm h}_{\rm sid})15^\circ$.) No statistically signifiant deviation from the combined result can be observed.

It is also possible to study the CR anisotropy in the {\it solar} reference frame, defined by mean solar time $t$ and declination $\delta$. The mean solar noon (12~{\rm hr}) corresponds to the LST of $0$~{\rm hr}, i.e., the mean Sun lies on the local meridian. Due to Earth's orbital motion around the Sun the length of one {\it sidereal} day, i.e., the proper rotation period of the Earth, is about four minutes shorter than the mean {\it solar} day. As a consequence, the Sun progresses by about one degree in right ascension during one sidereal day. The right ascension angle of the (mean) Sun describes a full circle over the course of a year starting at $\alpha=0^\circ$ at vernal equinox. Therefore, any anisotropy in the sidereal frame will average out in the solar frame over a full year of observation, and {\it vice versa}. However, the orbital motion of the Earth causes a solar Compton--Getting effect in the form of a dipole in the equatorial plane with a maximum at solar time $t\simeq 6$~{\rm hr}. We will discuss this effect in more detail in Section~\ref{subsec:GCeffect}. Here, we only want to point out that the reconstruction of the solar anisotropy follows the same steps and has the same limitations as the reconstruction of the sidereal anisotropy described in the following

%--------------------------------------------------------------------------------------------------------------------------------
%--------------------------------------------------------------------------------------------------------------------------------

\subsection{Reconstruction}
\label{subsec:Reconstruction}

\begin{figure}[t]
\centering
\includegraphics[width=0.6\linewidth]{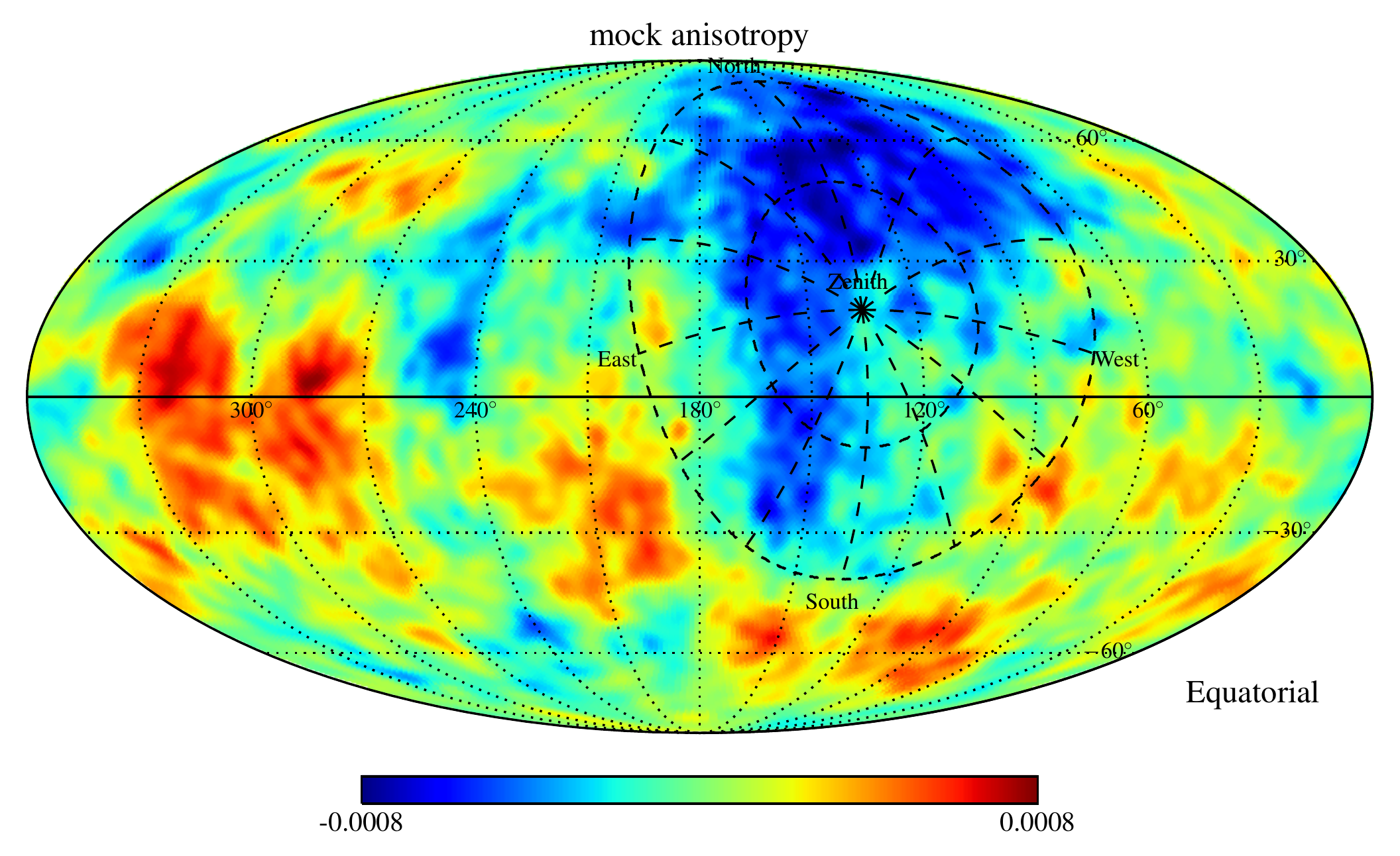}\\\includegraphics[width=0.6\linewidth]{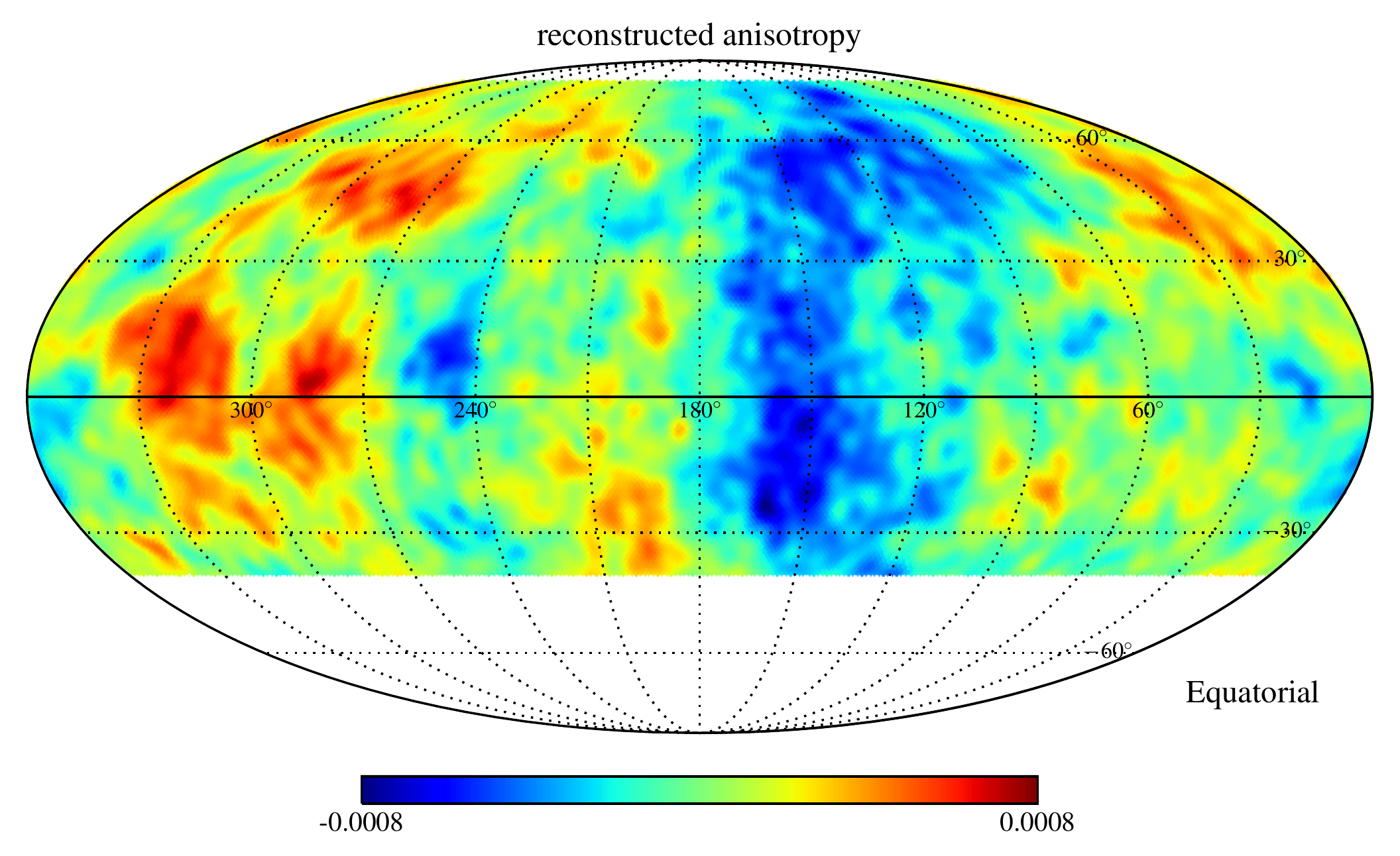}
\caption[]{Top panel: Mock realization of the CR anisotropy in equatorial coordinates. The inset circle indicate the instantaneous FOV of the HAWC observatory (at latitude $19^\circ$ North) at a local sidereal time of 09:00 and a zenith angle cut at $60^\circ$. Bottom panel:  Observable anisotropy accounting for the limited reconstruction capabilities ($a_{\ell0}\to0$, see main text). The time--integrated FOV covered over 24h corresponds to the declination range $-41^\circ<\delta<79^\circ$.}\label{fig:mock}
\end{figure}

Before we analyze the large- and small--scale anisotropy observed in experiments, it is important to realize the subtleties of anisotropy reconstructions in ground--based detectors. The basic observational quantity are the number of CR events observed in the local coordinate system. Typically, the local arrival direction is parametrized by the azimuth angle $\varphi$ (from north increasing to the east) and zenith angle $\theta$. This defines a unit vector 
\begin{equation}\label{eq:nprime}
{\bf n}'(\varphi, \theta)=(\cos\varphi\sin\theta,-\sin\varphi\sin\theta,\cos{\theta})\,,
\end{equation}
in the local (primed) coordinate system (cf.\ top panel of Fig.~\ref{fig:mock}). Over the period of one sidereal day, a fixed position on the local sphere will trace out a circle with a constant declination on the celestial sky and return to its initial position. For instance, an observer at geographic latitude $\Phi$ and longitude $\Lambda$ (measured east from Greenwich) sees the Zenith, ${\bf n}'(0,0)$, at fixed declination $\delta=\Phi$ and right ascension $\alpha = \omega t + \Lambda$ (mod $2\pi$), where $t$ is the local sidereal time and $\omega$ is the angular frequency of the Earth's rotation. It can be expressed as
\begin{equation}
\omega = \omega_{\rm solar} + \omega_{\rm orbit}\,,
\end{equation}
with the apparent solar frequency $\omega_{\rm solar} = 2\pi/24{\rm h}$ and the Earth's orbital frequency $\omega_{\rm orbit} = 2\pi/1{\rm yr}$. 

From the point of view of the ground--based observer the same location in the equatorial sky is given as 
${\bf n}'={\bf R}(t)\cdot{\bf n}$, 
where ${\bf n}$ is the unit vector (\ref{eq:n}) in the equatorial coordinate system and $R$ a time-dependent rotation matrix
\begin{equation}\label{eq:Rmatrix}
{\bf R}(t) =
\begin{pmatrix}
  -\cos \omega t\sin \Phi&-\sin \omega t\sin\Phi&\cos\Phi \\
  \sin \omega t&-\cos \omega t&0\\
  \cos \omega t\cos\Phi&\sin \omega t\cos\Phi&\sin\Phi
\end{pmatrix}\,.
\end{equation}
The inverse transformation is ${\bf n}={\bf R}^T(t)\!\cdot\!{\bf n}'$. For illustration, the mock anisotropy  map in the top panel of Fig.~\ref{fig:mock} indicates the instantaneous FOV of the HAWC detector~\citep{Abeysekara:2014sna} (at latitude $\Phi \simeq19^\circ$) at a local sidereal time of 09:00 (meridian lies along $\alpha=135^\circ$). The dashed circles indicate the zenith angles of $30^\circ$ and $60^\circ$.

The challenge of ground--based observatories is now, to extract $I(\alpha,\delta)$ from the observed number of events in the local reference system with a precision of better than $10^{-3}$. However, the observed event distribution integrated over many sidereal days depends not only on the CR anisotropy but also on the non--uniform and time-dependent detector exposure $\mathcal{E}(t,\varphi,\theta)$. The expected number of events at sidereal time $t$ during a short time interval $\Delta t$ and within a spherical element $\Delta\Omega$ is then given as
\begin{equation}\label{eq:expected}
\mu(t,{\bf n}') = \Delta t \Delta\Omega \mathcal{E}(t,{\bf n}')\phi^{\rm iso}I({\bf R}^T(t)\!\cdot\!{\bf n}')\,.
\end{equation}
Note, that the relative intensity $I$ transformed into the local coordinate system is periodic over one sidereal day. Therefore, the expected number of events $\mu$ and the exposure $\mathcal{E}$ in Eq.~(\ref{eq:expected}) are understood to be accumulated over many consecutive sidereal days. For the moment, we will assume that the accumulated number of days are uniformly distributed over an integer number of years. This ensures that the solar dipole induced by the Earth's motion around the Sun averages out in Eq.~(\ref{eq:expected}).

Now, in order to reconstruct $\delta I$ to a level of $10^{-3}$, we have to know the detector exposure $\mathcal{E}$ in Eq.~(\ref{eq:expected}) to better than $10^{-3}$, accounting for variations in detector up-time, seasonal effects and the relative acceptance of the detector. It is not feasible to know or simulate the detector acceptance at this level of accuracy. However, we can use the fact that the vast majority of CRs arrive isotropically and use this for detector calibration. This method relies on the assumption that the accumulated detector exposure $\mathcal{E}$ can be expressed as a product of its angular--integrated exposure $E$ and relative acceptance $\mathcal{A}$, %
\begin{equation}\label{eq:E}
  \mathcal{E}(t,{\bf n}') \simeq E(t)\mathcal{A}({\bf n}')\,,
\end{equation}
normalized to $\int{\rm d}\Omega \mathcal{A}({\bf n}')=1$. We can now integrate Eq.~(\ref{eq:expected}) over time and use the fact that $I\simeq 1$ to estimate the relative acceptance $\mathcal{A}({\bf n}')$ by comparing to the total number of observed events that arrived from the direction ${\bf n}'$. From this we can derive a first approximation of the relative intensity $\delta I^{(1)}$.

This reconstruction method is known as {\it direct-integration}~\cite{Atkins:2003ep} and {\it time-scrambling}~\citep{Alexandreas1993} and has been used in cosmic-ray anisotropy studies of Super-Kamiokande~\citep{Guillian:2005wp}, Tibet-AS$\gamma$~\citep{Amenomori:2006bx}, Milagro~\citep{Abdo:2008kr}, IceCube~\citep{Abbasi:2011ai,Abbasi:2011zka,Aartsen:2013lla}, IceTop~\citep{Aartsen:2012ma}, HAWC~\citep{Abeysekara:2014sna}, and ARGO-YBJ~\citep{ARGO-YBJ:2013gya}. For the IceCube observatory, where the instantaneous FOV is identical to the time--integrated FOV, this method provides the best estimate of the CR anisotropy. However, all other observatories can improve the anisotropy iteratively, by using the first estimate of $\delta I^{(1)}$ to re-evaluate the relative acceptance by the relative intensity $I\simeq 1+\delta I^{(1)}$ in the next step and optimize the anisotropy to $\delta I^{(2)}$, and so forth. This provides an iterative reconstruction scheme~\cite{Cui2003ICRC,Ahlers:2016njl}, that have been applied to data of Tibet-AS$\gamma$~\citep{Amenomori:2005pn,Amenomori:2010yr,Amenomori:2012uda} and ARGO-YBJ~\citep{Bartoli:2015ysa}.

Unfortunately, there is an important limitation of these data-driven reconstruction methods. Events recorded in a fixed position ${\bf n}'$ in the local coordinate system can only probe the variation of the CR flux in a fixed declination band $\delta$. The expectation values (\ref{eq:expected}) with the ansatz (\ref{eq:E}) are invariant under the rescaling
\begin{eqnarray}
\label{eq:scaleI}
  I &\to & \widetilde{I} \equiv I/a(\delta)/b\,,\\\label{eq:scaleN}
  E&\to &\widetilde{E}\equiv Ebc
\,,\\\label{eq:scaleA}
  \mathcal{A} &\to &\widetilde{\mathcal{A}}\equiv{\mathcal{A}a(\delta)}/{c}\,,\end{eqnarray}
where $a(\delta)$ is an arbitrary function of declination and the normalization factors $b$ and $c$ are defined such that $\int{\rm d}\Omega\widetilde{\mathcal{A}}({\bf n}')=1$ and $\int{\rm d}\Omega \widetilde{I}({\bf n})=1$ for the new values. In other words, the data-driven reconstruction method of ground--based observatories is insensitive to relative intensity variations across declination bands~\citep{Amenomori:2005pn,Iuppa:2013pg,Ahlers:2016njl}. 

This degeneracy requires to choose a convention how to report relative intensities that align with the Earth's rotation axis. A natural choice is that the reconstructed anisotropy is normalized to 
\begin{equation}\label{eq:gauge}
\int{\rm d}\alpha\delta I({\bf n})=0
\end{equation} for all declinations $\delta$, which is consistent with the requirement $\int{\rm d}\Omega \delta I(\alpha,\delta)=0$. As we will see in the following, this condition can also be formulated in terms of an expansion of the relative intensity into spherical harmonics~\cite{Iuppa:2013pg}.

An example, how this limited reconstruction capability affects the anisotropy, is shown in Fig.~\ref{fig:mock}. The top panel shows the true (mock) anisotropy behind the instantaneous FOV of HAWC at a sidereal time of 09:00 and Zenith angle cut $\theta_{\rm max}=60^\circ$. The reconstructed anisotropy in the time--integrated FOV ($-41^\circ<\delta<79^\circ$) is shown in the bottom panel. Clearly, the reconstructed anisotropy is modified and can reduce or enhance small--scale features. This effect is particularly drastic for the dipole reconstruction as we will discuss in the following.

%--------------------------------------------------------------------------------------------------------------------------------
%--------------------------------------------------------------------------------------------------------------------------------
%--------------------------------------------------------------------------------------------------------------------------------

\subsection{Harmonic Analysis}
\label{subsec:HarmonicAnalysis}

In the previous section we discussed the anisotropy of CR arrival directions in terms of relative intensity maps. These observations indicate that the anisotropy is present on various angular scales.  There are large--scale features that broadly divide the relative intensity maps into {\it excess} and {\it deficit} regions (see Fig.~\ref{fig:combinedmaps}) and on top of this there are also small--scale features, like region A (see Fig.~\ref{fig:HAWCmap}), with an angular size of less than $10^\circ$. However, the interpretation of these maps in terms of physical mechanisms is difficult. Firstly, it is important to realize that, by definition, the anisotropy of the relative intensity maps shows equal contributions of {\it excesses} and {\it deficits}. In other words, an excess in the relative intensity map can be produced by an excess flux in that region or by by deficit flux in all other regions. Secondly, as discussed at length in the previous section, most ground--based observations are insensitive to anisotropies that align with the Earth's rotation axis. This can have a significant effect on the observation of large--scale features (see Fig.~\ref{fig:mock}).

A convenient way to quantify the level of large-- and small--scale anisotropy is by a harmonic analysis. The relative intensity can then be expanded into spherical harmonics $Y_{\ell m}$,
\begin{equation}\label{eq:Yexpansion}
I(\alpha,\delta) = 1 + \sum_{\ell\geq1}\sum_{m=-\ell}^\ell a_{\ell m}Y_{\ell m}(\pi/2-\delta,\alpha)\,,
\end{equation}
where in order for $I$ to be real, the complex coefficients $a_{\ell m}$ obey the relation $a^*_{\ell m} = (-1)^m a_{\ell -m}$. As a rule of thumb, the typical angular size characterized by the expansion terms is related to the multipole number $\ell$ as $\Delta \theta \simeq 180^\circ/\ell$. The large--scale features correspond to the CR dipole ($\ell=1$), quadrupole ($\ell=2$), octupole ($\ell=3$), etc. In the case of a full sky coverage the spherical harmonics are an orthogonal basis and the individual $a_{\ell m}$'s in the expansion (\ref{eq:Yexpansion}) can be extracted as
\begin{equation}\label{eq:alm}
a_{\ell m} = \int {\rm d}\Omega Y^*_{\ell m}(\pi/2-\delta,\alpha)I(\alpha,\delta)\,.
\end{equation}

Note, that the normalization condition (\ref{eq:gauge}) that we introduced for the observation made by ground--based observatories is equivalent to the condition $a_{\ell0} =0$ for $\ell\geq1$ in the harmonic picture. This can be seen by inserting Eq.~(\ref{eq:Yexpansion}) into Eq.~(\ref{eq:gauge}),
\begin{equation}
\int \mathrm{d} \alpha \, \delta I(\hat{n}) =  2\pi\sum_{\ell} a_{\ell 0} \sqrt{\frac{2\ell+1}{4\pi}}P_{\ell} (\sin \delta) = 0\quad (\delta_1<\delta<\delta_2)\,,
\end{equation}
where $\delta_1$ and $\delta_2$ bound the observatory's declination range.

%--------------------------------------------------------------------------------------------------------------------------------
%--------------------------------------------------------------------------------------------------------------------------------

\subsection{Sidereal Dipole}
\label{subsec:SiderealDipole}

The CR dipole anisotropy, i.e., the expansion coefficient with $\ell=1$ are of particular interest for CR phenomenology, since this is expected to be the leading order anisotropy from CR diffusion as we will discuss in the following section. The dipole anisotropy is typically parametrized in terms of a dipole vector $\boldsymbol{\delta}$, defined by
\begin{equation}\label{eq:dipole1}
I(\alpha,\delta) = 1 + \boldsymbol{\delta}\!\cdot\!{\bf n}(\alpha,\delta) + \mathcal{O}\left(\lbrace a_{\ell m}\rbrace_{\ell \geq 2}\right)\,.
\end{equation}
In terms of the harmonic expansion (\ref{eq:Yexpansion}) into spherical harmonics $Y_{1m}$, the dipole vector components in the equatorial coordinate system are related to the expansion coefficients as
\begin{equation}\label{eq:deltageneral}
\boldsymbol{\delta} \equiv \big(\delta_\text{0h},\delta_\text{6h},\delta_\text{N}\big)=\sqrt{\frac{3}{2\pi}}\big(-\Re (a_{11}),\Im (a_{11}),a_{10}\big)\,.
\end{equation}
Note, that we have $a_{1\text{-}1} = -a^*_{11}$ and $a_{10} = a_{10}^*$. Here, we introduced the notation $\delta_\text{0h}$ and $\delta_\text{6h}$ corresponding to the dipole components parallel to the equatorial plane and pointing to the direction of the local hour angle 0h ($\alpha=0^\circ$) and 6h ($\alpha=90^\circ$) of the vernal equinox, respectively. We also introduce $\delta_\text{N}$ as the dipole component pointing north. However, since $a_{10}$ is in general not accessible by ground--based observatories the corresponding dipole component $\delta_\text{N}$ can not be constrained. For that reason, experiments only report the dipole components aligning with the equatorial plane as
\begin{equation}\label{eq:ampphase}
\big(\delta_\text{0h},\delta_\text{6h}\big) = (A_1\cos\alpha_1, A_1\sin\alpha_1)\,,
\end{equation}
with dipole amplitude $A_1$ and phase $\alpha_1$.

\begin{figure}[t]\centering
\includegraphics[width=0.7\linewidth]{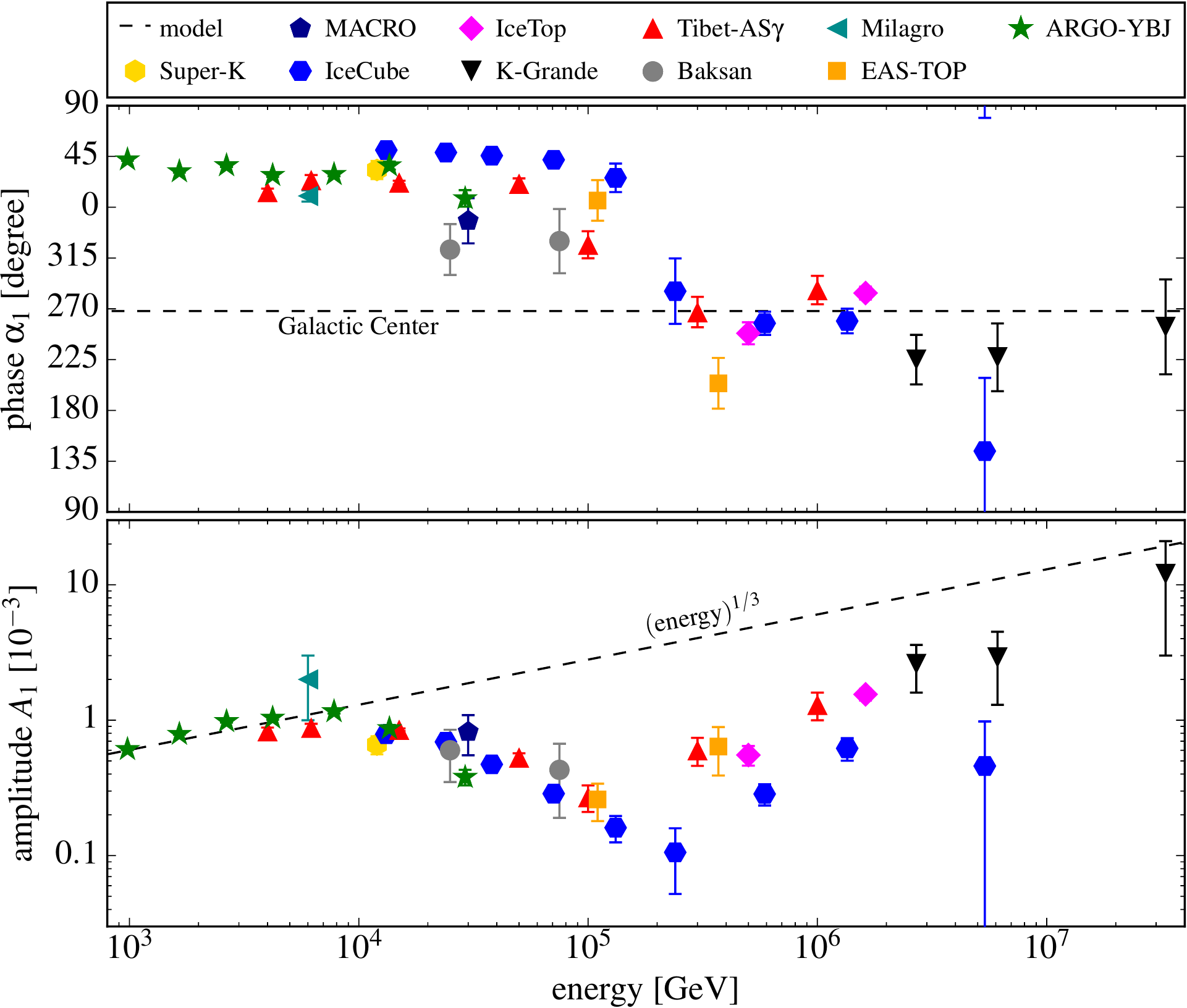}
\caption[]{Inferred phase and amplitude of the (equatorial) dipole anisotropy from recent measurements. See Table~\ref{table:dipoledata} for a description of the data. The dashed line shows a na\"ive model prediction assuming a smooth distribution of CR sources and isotropic diffusion with energy dependence ${\bf K}\propto E^{1/3}$.}\label{fig:dipoledata}
\end{figure}

\begin{table}[p]\centering
\renewcommand{\arraystretch}{0.75}
\begin{tabular}{ccc|cccccc}
\hline\hline
Observatory&Location&$\theta_{\rm max}$&$E_{\rm med}$ [TeV]&$A_1$ [$10^{-4}$]&$\alpha_1$&Method&Reference\\
\hline\hline
K-Grande &$49^\circ$N $8^\circ$E&$40^\circ$&$ 2700 $&$ 26.0 \pm 10.0 $&$ 225^\circ \pm 22^\circ $&EAS, $\langle\delta\rangle$&\cite{KASCADEICRC281}\\
 $\cdots$&$\cdots$&$\cdots$&$ 6100 $&$ 29.0 \pm 16.0 $&$ 227^\circ \pm 30^\circ$&$\cdots$&$\cdots$\\
 $\cdots$&$\cdots$&$\cdots$&$ 33000 $&$ 120.0 \pm 90.0 $&$ 254^\circ \pm 42^\circ $&$\cdots$&$\cdots$\\
Baksan &$43^\circ$N $43^\circ$E&$-$&$ 25 $&$ 6.0 \pm 2.5 $&$ 322^\circ \pm 22^\circ $&EAS, EW&\cite{Alekseenko:2009ew}\\
 &&$-$&$ 75 $&$ 4.3 \pm 2.4 $&$ 330^\circ \pm 28^\circ $&$\cdots$&$\cdots$\\
EAS-TOP &$42^\circ$N $13^\circ$E&$40^\circ$&$ 110 $&$ 2.6 \pm 0.8 $&$ 6^\circ \pm 18^\circ $&EAS, EW&\cite{Aglietta:2009mu}\\
 $\cdots$&$\cdots$&$\cdots$&$ 370 $&$ 6.4 \pm 2.5 $&$ 204^\circ \pm 22^\circ $&$\cdots$&$\cdots$\\ 
MACRO &$42^\circ$N $13^\circ$E&$-$&$ 30 $&$ 8.2 \pm 2.7 $&$ 348^\circ \pm 20^\circ $&muons, $\langle\delta\rangle$&\cite{Ambrosio:2002db}\\
Super-K &$36^\circ$N $137^\circ$E&$-$&$ 12 $&$ 6.6 \pm 1.0 $&$ 33^\circ \pm 8^\circ $&muons, $\langle\delta\rangle$&\cite{Guillian:2005wp}\\
Milagro &$36^\circ$N $107^\circ$W&$50^\circ$&$ 6 $&$ 20.0 \pm 10.0 $&$ 10^\circ \pm 5^\circ $&EAS, FB&\cite{Abdo:2008aw}\\
ARGO-YBJ &$30^\circ$N $91^\circ$E&$45^\circ$&$ 0.98 $&$ 6.1 \pm 0.1 $&$ 42^\circ \pm 1^\circ $&EAS, $\langle\delta\rangle$&\cite{Bartoli:2015ysa}\\
 $\cdots$&$\cdots$&$\cdots$&$ 1.65 $&$ 7.9 \pm 0.1 $&$ 32^\circ \pm 1^\circ $&$\cdots$&$\cdots$\\
 $\cdots$&$\cdots$&$\cdots$&$ 2.65 $&$ 9.8 \pm 0.2 $&$ 37^\circ \pm 1^\circ $&$\cdots$&$\cdots$\\
 $\cdots$&$\cdots$&$\cdots$&$ 4.21 $&$ 10.4 \pm 0.3 $&$ 28^\circ \pm 2^\circ $&$\cdots$&$\cdots$\\
 $\cdots$&$\cdots$&$\cdots$&$ 7.8 $&$ 11.6 \pm 0.4 $&$ 29^\circ \pm 2^\circ $&$\cdots$&$\cdots$\\
 $\cdots$&$\cdots$&$\cdots$&$ 13.6 $&$ 8.7 \pm 0.5 $&$ 37^\circ \pm 4^\circ $&$\cdots$&$\cdots$\\
 $\cdots$&$\cdots$&$\cdots$&$ 29.1 $&$ 3.8 \pm 0.5 $&$ 8^\circ \pm 7^\circ $&$\cdots$&$\cdots$\\
  Tibet-AS$\gamma$ &$30^\circ$N $91^\circ$E&$45^\circ$&$ 4 $&$ 8.3 \pm 0.5 $&$ 14^\circ \pm 3^\circ $&EAS, $\langle\delta\rangle$&\cite{Amenomori:2005dy}\\
 $\cdots$&$\cdots$&$\cdots$&$ 6.2 $&$ 8.8 \pm 0.6 $&$ 24^\circ \pm 4^\circ $&$\cdots$&$\cdots$\\
$\cdots$&$\cdots$&$60^\circ$&$ 15 $&$ 8.5 \pm 0.2 $&$ 22^\circ \pm 2^\circ $&EAS, $\langle\delta\rangle$&\cite{TibetICRC355}\\
 $\cdots$&$\cdots$&$\cdots$&$ 50 $&$ 5.3 \pm 0.4 $&$ 21^\circ \pm 5^\circ $&$\cdots$&$\cdots$\\
 $\cdots$&$\cdots$&$\cdots$&$ 100 $&$ 2.7 \pm 0.6 $&$ 327^\circ \pm 12^\circ $&$\cdots$&$\cdots$\\
 $\cdots$&$\cdots$&$\cdots$&$ 300 $&$ 6.0 \pm 1.4 $&$ 267^\circ \pm 14^\circ $&$\cdots$&$\cdots$\\
 $\cdots$&$\cdots$&$\cdots$&$ 1000 $&$ 13.0 \pm 3.0 $&$ 287^\circ \pm 13^\circ $&$\cdots$&$\cdots$\\
 IceTop &Southpole&$55^\circ$&$ 500 $&$ 5.5 \pm 0.9 $&$ 248^\circ \pm 10^\circ $&EAS, $\langle\delta\rangle$&\cite{Abbasi:2011zka}\\
$\cdots$&$\cdots$&$\cdots$&$ 1600 $&$ 15.5 \pm 1.5 $&$ 284^\circ \pm 6^\circ $&$\cdots$&\cite{Aartsen:2016ivj}\\
IceCube &Southpole&$65^\circ$&$ 13 $&$ 7.9 \pm 0.1 $&$ 51^\circ \pm 1^\circ $&muons, $\langle\delta\rangle$&\cite{Aartsen:2016ivj}\\
 $\cdots$&$\cdots$&$\cdots$&$ 24 $&$ 6.9 \pm 0.2 $&$ 49^\circ \pm 1^\circ $&$\cdots$&$\cdots$\\
 $\cdots$&$\cdots$&$\cdots$&$ 38 $&$ 4.7 \pm 0.2 $&$ 46^\circ \pm 3^\circ $&$\cdots$&$\cdots$\\
 $\cdots$&$\cdots$&$\cdots$&$ 71 $&$ 2.9 \pm 0.3 $&$ 42^\circ \pm 6^\circ $&$\cdots$&$\cdots$\\
 $\cdots$&$\cdots$&$\cdots$&$ 130 $&$ 1.6 \pm 0.4 $&$ 26^\circ \pm 13^\circ $&$\cdots$&$\cdots$\\
 $\cdots$&$\cdots$&$\cdots$&$ 240 $&$ 1.1 \pm 0.5 $&$ 286^\circ \pm 29^\circ $&$\cdots$&$\cdots$\\
 $\cdots$&$\cdots$&$\cdots$&$ 580 $&$ 2.9 \pm 0.5 $&$ 257^\circ \pm 10^\circ $&$\cdots$&$\cdots$\\
 $\cdots$&$\cdots$&$\cdots$&$ 1400 $&$ 6.2 \pm 1.2 $&$ 259^\circ \pm 11^\circ $&$\cdots$&$\cdots$\\
 $\cdots$&$\cdots$&$\cdots$&$ 5400 $&$ 4.6 \pm 5.2 $&$ 144^\circ \pm 65^\circ $&$\cdots$&$\cdots$\\
 \hline\hline
\end{tabular}
\caption[]{Summary of recent dipole data shown in Fig.~\ref{fig:dipoledata}. We indicate the maximal zenith cut $\theta_{\rm max}$ (if available), the observation via extended air showers (EAS) or atmospheric muons and the analysis method: East--West (EW), Forward--Backward (FB), or declination average ($\langle\delta\rangle$).}\label{table:dipoledata}
\end{table}

Figure \ref{fig:dipoledata} shows the dipole amplitude and phase reported by various CR experiments, see Tbl.~\ref{table:dipoledata} for a summary of these data. The dashed lines indicate a simple parametric model of the dipole anisotropy for a CR diffusion model that we are going to discuss in the next sections. The experimental data show a large scatter from experiment to experiment. This has various reasons.
Firstly, ground--based observatories only observe CR indirectly via their showers produced in the atmosphere. This results in a large uncertainty of the reconstructed CR energy from shower--to--shower variations and the assumed chemical composition of CRs. For clarity, these uncertainties are not shown in Fig.~\ref{fig:dipoledata}. Secondly, the amplitude and phase reported by observatories depends on the analysis method that sometimes introduce a rescaling of the amplitude compared to Eq.~(\ref{eq:ampphase}). And lastly, the limited FOV of observatories can introduce a cross talk between large-- and medium--scale anisotropies which does not allow for a precise dipole reconstruction as in Eq.~(\ref{eq:alm}). We will briefly summarize in the following the various dipole reconstruction methods used by experiments.

%--------------------------------------------------------------------------------------------------------------------------------
%--------------------------------------------------------------------------------------------------------------------------------

\paragraph{East--West \& Forward--Backward Method}
\label{subsubsec:EastWest}

One of the standard dipole anisotropy reconstruction methods is the {\it East--West} method~\citep{Bonino:2011nx}.  Here, the dipole is derived from the {\it derivative} of the relative intensity with respect to right ascension. For simplicity, we assume for a moment that the relative intensity is given by a dipole, $I(\alpha,\delta) = 1 + \boldsymbol{\delta}\!\cdot\!{\bf n}(\alpha,\delta)$, where ${\bf n}$ is a unit vector in the equatorial coordinate system. As before, it can be related to the corresponding unit vector ${\bf n}'(\varphi,\theta)$ in the local coordinate system via a time-dependent rotation matrix, ${\bf n}'={\bf R}(t)\cdot{\bf n}$.

A crucial assumption of the {\it East--West} method is that the relative acceptance $\mathcal{A}$ is Forward--Backward symmetric, i.e., $\mathcal{A}(\varphi,\theta) = \mathcal{A}(-\varphi,\theta)$. For simplicity, we will assume for our example that $\mathcal{A}$ depends only on zenith angle $\theta$, i.e., $\mathcal{A}(\varphi,\theta) = \mathcal{A}(\theta)$. At each sidereal time $t$ the CR data is then divided into two bins, covering the east ($0<\varphi<\pi$) and west ($-\pi<\varphi<0$) sectors in the local coordinate system, that are divided by the meridian (cf.\ top panel of Fig.~\ref{fig:mock}). The event numbers observed during a short sidereal time interval $\Delta t$ are then related to the relative intensity $I$ and total accumulated detector exposure $\mathcal{E}$ as
\begin{align}
N_{\rm E}(t) &\simeq \phi^{\rm iso}\Delta t E(t)\int_{0}^{\pi}{\rm d}\varphi\!\!\!\int_0^{\theta_{\rm max}}\!\!{\rm d}\theta\sin\theta\,\mathcal{A}(\theta)I(t,\varphi,\theta)\,,\\
N_{\rm W}(t) &\simeq \phi^{\rm iso}\Delta t E(t)\!\!\int_{-\pi}^{0}\!\!{\rm d}\varphi\!\!\!\int_0^{\theta_{\rm max}}\!{\rm d}\theta\sin\theta\,\mathcal{A}(\theta)I(t,\varphi,\theta)\,.\end{align}
Now, the relative difference between the East and West data is independent of the absolute exposure $E(t)$, and we arrive at
\begin{equation}\label{eq:EW}
\frac{N_{\rm E}(t)-N_{\rm W}(t)}{N_{\rm E}(t)+N_{\rm W}(t)} \simeq \Delta\alpha\frac{\partial}{\partial\alpha}\delta I(\alpha,0)\,,
\end{equation}
where $\Delta\alpha$ is an effective right ascension step size. For the dipole anisotropy, it can be calculated as
\begin{equation}\label{eq:dalpha}
\Delta\alpha = \int_{0}^{\pi}{\rm d}\varphi\!\!\!\int_0^{\theta_{\rm max}}\!\!{\rm d}\theta\sin\theta\,\mathcal{A}(\theta)(\sin\theta\sin\varphi) \bigg/ \int_{0}^{\pi}{\rm d}\varphi\!\!\!\int_0^{\theta_{\rm max}}\!\!{\rm d}\theta\sin\theta\,\mathcal{A}(\theta)\,.
\end{equation}
Note, that the East--West method only allows to study the components of the dipole anisotropy in the equatorial plane, as can be seen in Eq.~(\ref{eq:EW}).

A variation of the East--West method is the {\it Forward--Backward} method used by Milagro~\citep{Abdo:2008aw}.
This method analyzes the relative right-ascension derivative of event rates for individual declination bands and allows to study the two-dimensional anisotropy. However, both methods are only sensitive to the right-ascension derivative and are hence not able to reconstruct the overall normalization of the anisotropy in each declination band. Hence, the full two-dimensional anisotropy can not be recovered. 

%--------------------------------------------------------------------------------------------------------------------------------
%--------------------------------------------------------------------------------------------------------------------------------

\paragraph{Right-Ascension Projection}

Many modern observatories have the sufficient angular resolution and event statistics to analyze the dipole anisotropy in two-dimensional relative intensity maps as shown in Sec.~\ref{sec:Observation}. However, mostly for historical reasons, some observations are reporting the dipole amplitude and phase from a right-ascension projection, i.e.\ declination average of the relative CR intensity $I$ in the form, 
\begin{equation}\label{eq:Aprojected}
\widetilde{A}_1e^{i\widetilde{\alpha}_1} = \frac{1}{\pi}\int_0^{2\pi}{\rm d}\alpha e^{i\alpha}\left[\frac{1}{s_2-s_1}\int_{s_1}^{s_2}{\rm d}s I(\alpha,\delta)\right]\,,
\end{equation}
where $s=\sin\delta$ and the boundaries $s_{1/2}=\sin\delta_{1/2}$ correspond to the declination interval $[\delta_1,\delta_2]$ of the observatory's time--integrated FOV. Again, assuming a pure dipole anisotropy, $I(\alpha,\delta) = 1 + \boldsymbol{\delta}\!\cdot\!{\bf n}(\alpha,\delta)$, in Eq.~(\ref{eq:Aprojected}) provides the correct phase, $\widetilde\alpha_1 = \alpha_1$, but a rescaled amplitude,
\begin{equation}\label{eq:Arelation}
\widetilde{A}_1 \simeq \frac{\delta_1-\delta_2+c_1s_1-c_2s_2}{2(s_1-s_2)}A_1\,.
\end{equation}
Note that in most cases, experiments do not correct the amplitude following Eq.~(\ref{eq:Arelation}) and only report $\widetilde{A}_1$. Unfortunately, this typically underestimates the true dipole amplitude by an $\mathcal{O}(1)$ factor and can be one reason for the scatter of dipole amplitudes among experiments as seen in Fig.~\ref{fig:dipoledata}. Another systematic uncertainty comes from the possible cross talk of higher multipole moments in Eq.~(\ref{eq:Aprojected}).

%--------------------------------------------------------------------------------------------------------------------------------
%--------------------------------------------------------------------------------------------------------------------------------

\paragraph{Reconstruction from Relative Intensity Maps}
\label{subsubsec:2DMaps}

In the case that a full relative intensity map is available, a precise dipole reconstruction is possible via Eq.~(\ref{eq:alm}). However, in reality, individual observatories have a limited integrated FOV. In this case we can not carry out the integral over the full sky that is needed to use the orthogonality of the spherical harmonics in Eq.~(\ref{eq:alm}). Instead, we can introduce a weight function $w$ that parametrizes the limited FOV and reconstruct the {\it pseudo} multipole moments of the combined product ${I}({\bf n}) = w({\bf n})\widehat{I}({\bf n})$

In general, the pseudo multipole moments ${a}_{\ell m}$ are related to the true multipole moments $\widehat{a}_{\ell m}$ via a linear transformation (see, {e.g.}, \cite{2002ApJ...567....2H} or the review by \cite{Efstathiou:2003dj}) 
\begin{equation}\label{eq:Kmatrix}
{a}_{\ell m} = \sum_{\ell'm'}K_{\ell m\ell'm'}\widehat{a}_{\ell'm'}\,,
\end{equation}
where the coupling matrix ${\bf K}$ depends on the multipole spectrum $b_{\ell m}$ of the weight function of the FOV (see~\ref{sec:appendix}). For the particular case of ground--based observatories, the weight function is expected to be azimuthally symmetric, $w(\alpha,\delta) = w(\delta)$. In that case the true multipole moments $\widehat{a}_{\ell 0}$ are a linear superposition of pseudo multipole moments ${a}_{\ell' 0}$ in Eq.~(\ref{eq:Kmatrix}). We can hence use the normalization condition ${a}_{\ell 0} = 0$ for all $\ell$ to ensure $\widehat{a}_{\ell 0} = 0$ for all $\ell$. For all other coefficients with $m\neq0$, the strength of the mixing between moments ${a}_{\ell m}$ and $\widehat{a}_{\ell' m}$ is determined by the moments $b_{k0}$ with $|\ell-\ell'|\leq k\leq \ell+\ell'$ and there is no mixing in $m$. This mixing can be small for individual moments as pointed out by \cite{Denton:2014hfa}. 

However, even in the absence of other experimental limitations of the anisotropy reconstruction, the full multipole spectrum cannot be unambiguously reconstructed from a partial sky coverage. Whereas the full (infinite dimensional) transition matrix ${\bf K}$ cannot be inverted to solve Eq.~(\ref{eq:Kmatrix}), we can attempt an approximate reconstruction of the low-$\ell$ anisotropy via a truncation of the multipole expansion after a maximum $\ell_{\rm max}$. The corresponding truncated matrix ${\bf K}'$ with entries $\ell\leq\ell_{\rm max}$ and $m\neq0$ can then be invertible. For instance, assuming a pure dipole anisotropy, $\ell\leq1$, and a uniform sky coverage between declination $\delta_1$ and $\delta_2$ gives the dipole transition elements
\begin{equation}\label{eq:specialK}
K'_{1111} = K'_{1\text{-}11\text{-}1} =\frac{1}{4} \bigl(3 (s_2 -s_1) 
   +s_1^3 - s_2^3 \bigr) \, ,
\end{equation}
where again $s_{1,2} = \sin \delta_{1,2}$ and $K'_{111-1}=K'_{1-111}=0$. We can reconstruct in this case the dipole components in the equatorial plane via Eq.~(\ref{eq:ampphase}) and 
\begin{equation}
\big(\delta_\text{0h},\delta_\text{6h}\big) = \sqrt{\frac{3}{2\pi}}\left(-\frac{\Re({a}_{11})}{K'_{1111}},\frac{\Im({a}_{11})}{K'_{1111}}\right)\,.
\end{equation}
However, we emphasize that this treatment is only correct under the assumption that the true anisotropy is dominated by a dipole. Here, again, the presence of higher multipoles can significantly effect the dipole reconstruction.

%--------------------------------------------------------------------------------------------------------------------------------
%--------------------------------------------------------------------------------------------------------------------------------

\subsection{Angular Power Spectrum}
\label{subsec:AngularPowerSpectrum}

We have already discussed the presence of small--scale anisotropies in the arrival direction of CRs. These can be emphasized in the equatorial maps by subtracting fits of the large--scale moments, dipole, quadrupole, etc., from the anisotropy (see Fig.~\ref{fig:HAWCmap}).
Another approach to study small--scale structure is via the angular power spectrum defined as
\begin{equation}\label{eq:Cl}
C_\ell \equiv \frac{1}{2\ell+1}\sum_{m=-\ell}^\ell|a_{\ell m}|^2\,.
\end{equation}
The power spectrum quantifies the absolute amplitude of the multipole components independent of their phases. 
This method can uncover the presence of weak small--scale fluctuations randomly distributed across the equatorial anisotropy maps that are {\it individually} not significant. However, as mentioned in the previous section, the true multipole spectrum cannot be recovered unambiguously from a partial sky coverage, and the same is true for the power spectrum. 

To overcome this difficulty, one can make additional assumptions about the {\it ensemble}--averaged expectation values of the multipole components. In the following, we will {\it assume} that the harmonic coefficients are Gaussian random variables, which in the ensemble--average follow $\langle\widehat{a}_{\ell m}\widehat{a}^*_{op}\rangle = \delta_{mp}\delta_{\ell o}\langle\widehat{C}_\ell\rangle$; this is equivalent to demanding that $\delta I(\hat{n})$ is a statistically isotropic, Gaussian random field. In this particular case, we can recover the ensemble--averaged power spectrum $\langle\widehat{C}_\ell\rangle$ via the relation 
\begin{equation}\label{eq:Mmatrix}
\langle C_\ell\rangle =\sum_{\ell'}M_{\ell\ell'}\langle \widehat{C}_{\ell'}\rangle +\mathcal{N}_\ell\,.
\end{equation}
The transfer matrix ${\bf M}$ is known from the study of temperature anisotropies in the cosmic microwave background~\citep{Efstathiou:2003dj}. 

\begin{figure}[t]\centering
\includegraphics[width=0.7\linewidth]{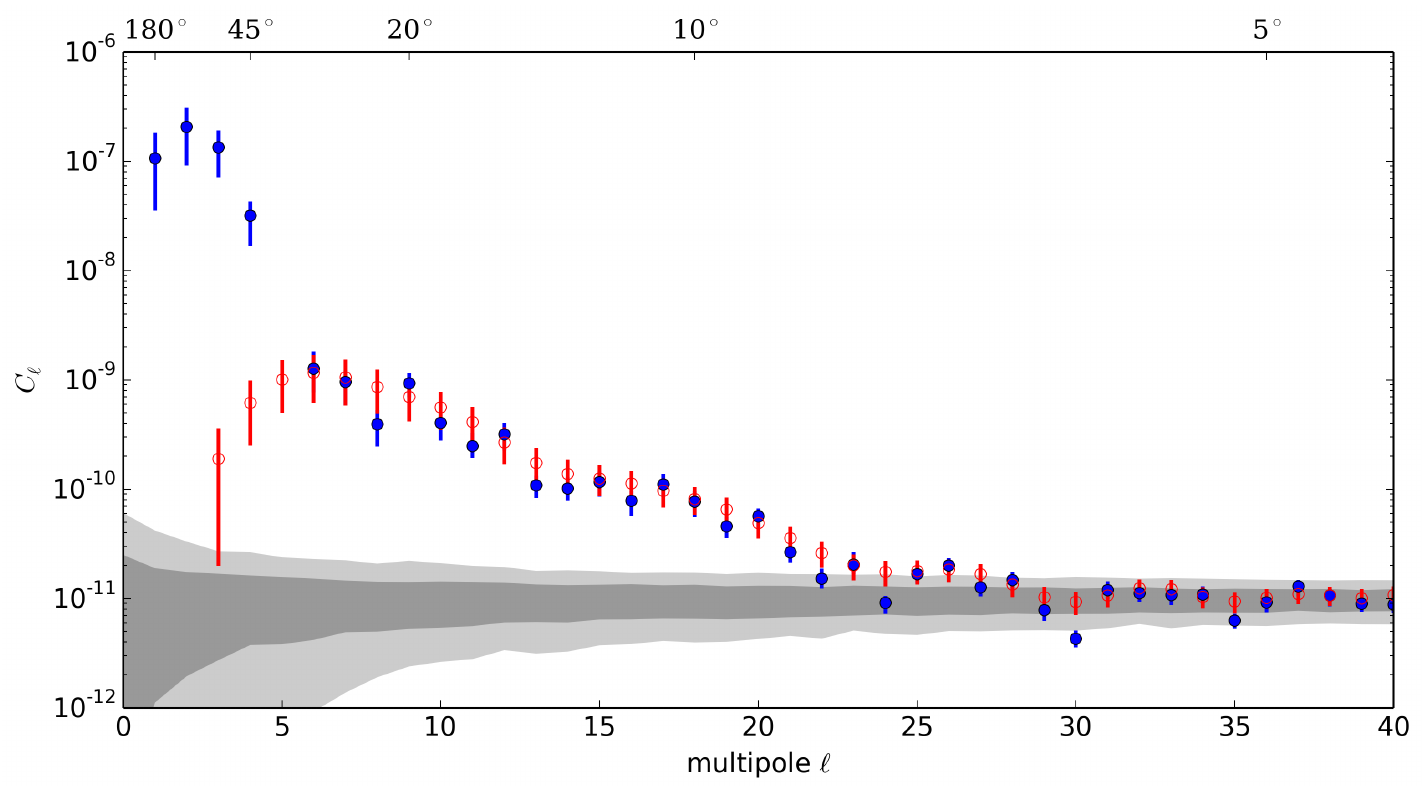}\\[0.5cm]
\includegraphics[width=0.7\linewidth]{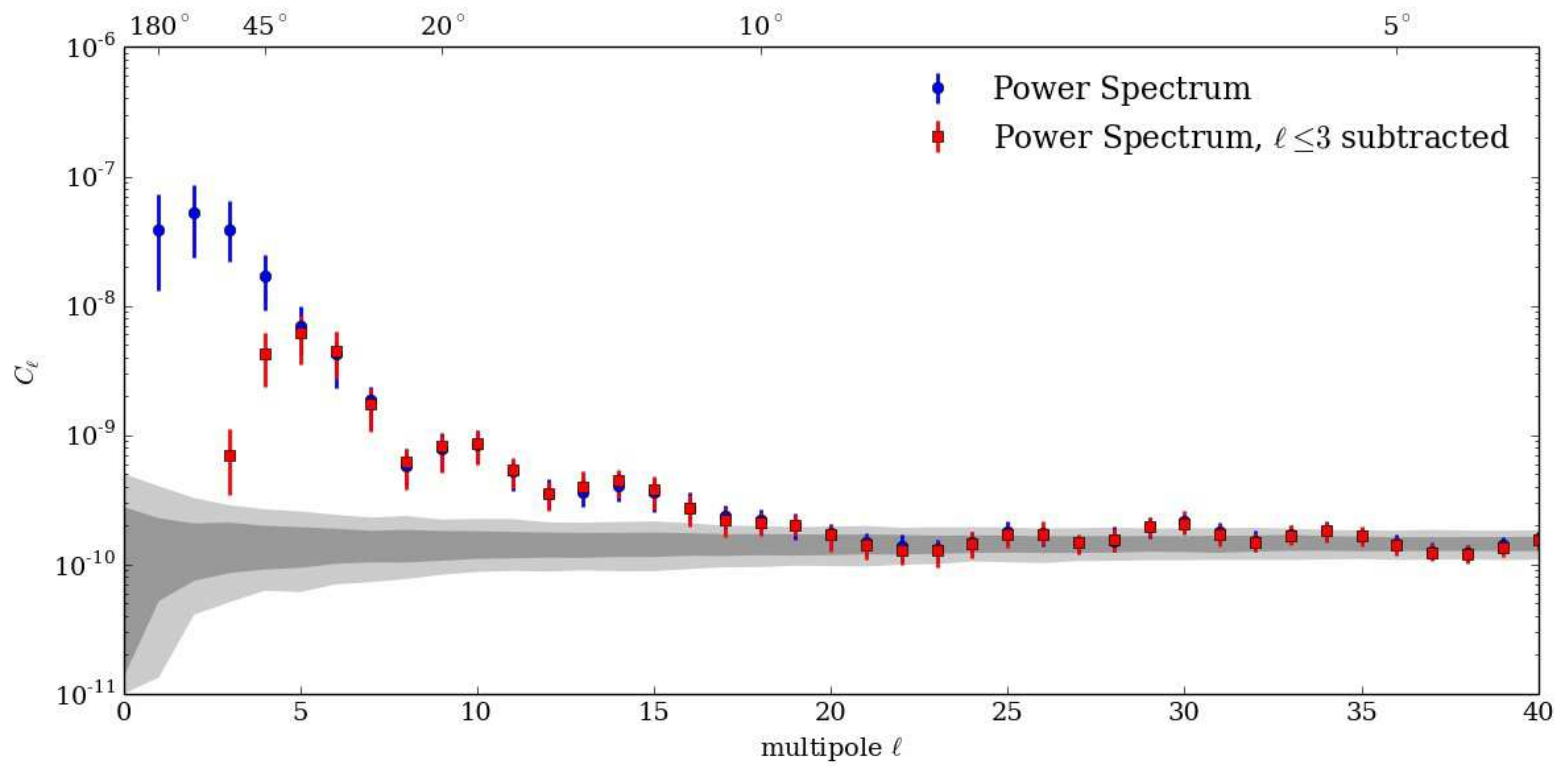}
\caption[]{Angular power spectrum of the anisotropy inferred by IceCube (top) at medium energies of about 20~TeV~\cite{Aartsen:2016ivj} and HAWC (bottom) at median energy of 2~TeV~\cite{Abeysekara:2014sna}. In both plots, the data are shown before (blue) and after (red) fitting and subtracting the dipole, quadrupole and octupole. The gray bands show the 68\% and 95\% intervals for isotropic maps generated from scrambled data.}\label{fig:power}
\end{figure}

As pointed out in Ref.~\cite{Ahlers:2016njl}, for the situation of cosmic-ray anisotropies we have again to account for the fact that the $m=0$ moments are filtered out by the reconstruction. This leads to a modified expression for ${\bf M}$ that we discuss and provide in~\ref{sec:appendix}. However, we would like to caution the reader that the reconstruction of the power spectrum from a limited FOV relies on the assumption that the anisotropy is a statistically isotropic, Gaussian random field. This can only be a possible working assumption, if the source of the anisotropy is random itself. We will later discuss the scenario of small--scale structures induced by local turbulent magnetic field, that could possibly be of this type.

The last term in Eq.~(\ref{eq:Mmatrix}) accounts for the noise power spectrum $\mathcal{N}_\ell$. In the following, we will approximate this quantity by the pixel-by-pixel Poisson noise, which gives a flat spectrum $\mathcal{N}_\ell=\mathcal{N}$ with
\begin{equation}\label{eq:noisediscrete}
\mathcal{N} \simeq \frac{1}{4\pi}\sum_i\frac{w_i^2\Delta\Omega^2}{n_i}\,.
\end{equation}
Here, the sum runs over all pixels $i$ in the FOV with weight $w_i$, angular size $\Delta\Omega$ and total event number $n_i$. For an order of magnitude estimate of the noise level, we can assume a top-hat weight $w_i=1$ with uniform exposure in the FOV with partial sky coverage $f_{\rm sky}$. In this case Eq.~(\ref{eq:noisediscrete}) yields 
\begin{equation}\label{eq:noiseesimtate}
\mathcal{N} \simeq \frac{4\pi f_{\rm sky}^2}{n_{\rm tot}}\,,
\end{equation}
for the total number $n_{\rm tot}$ of CR events.

The top panel of Fig.~\ref{fig:power} shows the angular power spectrum inferred from IceCube data with a median energy of 20~TeV~\cite{Aartsen:2016ivj}. Note, that the IceCube collaboration uses the two-point angular correlation function~\cite{Szapudi2001} rather than the pseudo power spectrum to estimate the power spectrum from the partial sky. With $n_{\rm tot}\simeq3.2\times10^{11}$ and $f_{\rm sky}\simeq0.29$ we can use Eq.~(\ref{eq:noiseesimtate}) to estimate the noise power as $\mathcal{N}\simeq3\times10^{-12}$. This is slightly lower then the noise level shown in Fig.~\ref{fig:power} estimated from a sample of scrambled maps.

Recently, the HAWC collaboration published the power spectrum of the relative intensity map, shown in the lower panel of Fig.~\ref{fig:power}. With $n_{\rm tot}\simeq 4.9\times10^{10}$ and $f_{\rm sky}\simeq0.67$ the noise level can be estimated as $\mathcal{N}\simeq10^{-10}$, which is presently much larger than for the IceCube observation. Nevertheless, the power of medium-$\ell$ moments shows a trend very similar to that observed at IceCube. However, the HAWC analysis is not compensating for the loss of power at $\ell\leq2$ due to the limited instantaneous FOV. As discussed in Ref.~\cite{Ahlers:2016njl}, the true power at $\ell\leq2$ is expected to be significantly larger and, therefore, closer to the power observed with IceCube.

%--------------------------------------------------------------------------------------------------------------------------------
%--------------------------------------------------------------------------------------------------------------------------------
%--------------------------------------------------------------------------------------------------------------------------------

\section{The Standard Diffusive Picture}
\label{sec:StandardPicture}

As a preparation for our later discussion of small--scale anisotropies, we will first review the microscopic theory~\cite{1966ApJ...146..480J,1966PhFl....9.2377K,1967PhFl...10.2620H,1970ApJ...162.1049H} of CR transport in the presence of turbulent magnetic fields. For the discussion of CR anisotropy it is convenient to introduce the phase--space density, $f(t,{\bf r},{\bf p})$, as the expected number of CRs at time $t$ with position within $[{\bf r}, {\bf r}+{\rm d}{\bf r}]$ and momentum within $[{\bf p}, {\bf p}+{\rm d}{\bf p}]$. In the absence of energy losses and sources, the phase--space density follows Liouville's theorem, $\dot{f}=0$. (Here and in the following we denote the temporal derivative $(\mathrm{d}/\mathrm{d} t)$ with a dot.) In other words, the phase--space density along a CR trajectory \ $({\bf r}(t), {\bf p}(t))$ is constant, $f(t_1,{\bf r}(t_1),{\bf p}(t_1)) = f(t_2,{\bf r}(t_2),{\bf p}(t_2))$. Liouville's theorem can be written in the form
\begin{equation}\label{eq:Boltzmann}
\dot{f} = \partial_t f + \dot{\bf r}\!\cdot\!\nabla_{\bf r} f + \dot {\bf p}\!\cdot\!\nabla_{\bf p}f  = 0\,.
\end{equation}
In a static magnetic field configuration and neglecting static electric fields the equation of motion of a charged particle with energy $p_0$ is given as $\dot {\bf p} = {\bf p}\times e{\bf B}/p_0$ and $\dot p_0=0$. (For simplicity, we also assume relativistic CRs, $p_0\simeq p = |{\bf p}|$, working in natural units, $c=1$.)

As usual, we split the magnetic field into a regular and turbulent component, ${\bf B} = {\bf B}_0 +{\delta {\bf B}}$. We also split the phase--space density into the average and fluctuating parts, $f = \langle f\rangle +\delta{f}$. The ensemble--averaged, collisionless Boltzmann equation can be written as~\citep{Jokipii1972,JonesApJ1990}
\begin{equation}\label{eq:genBoltzmann}
\partial_t\langle f\rangle  + \langle\mathcal{L}({\bf r},{\bf p})\rangle\langle f\rangle  = -\langle\delta\mathcal{L}({\bf r},{\bf p})\delta f\rangle\,,
\end{equation}
with the Liouville operator
\begin{equation}\label{eq:LOperator}
\mathcal{L}({\bf r},{\bf p}) \equiv \dot{\bf r}\!\cdot\!\nabla_{\bf r} + \dot {\bf p}\!\cdot\!\nabla_{\bf p} = \widehat{\bf p} \!\cdot\!\nabla_{\bf r} -i[\boldsymbol{\Omega}+\boldsymbol{\omega}({\bf r})]\!\cdot\!{\bf L}\,,
\end{equation}
and its ensemble-fluctuation
\begin{equation}
\delta\mathcal{L}({\bf r},{\bf p}) \equiv -i\boldsymbol{\omega}({\bf r})\!\cdot\!{\bf L}\,.
\end{equation} 
Here we have introduced the angular momentum operator ${L}_a = -i\epsilon_{abc}{p}_b\partial/\partial p_c$ and the rotation vectors ${\boldsymbol \Omega} = e{\bf B}_0/p_0$ and ${\boldsymbol \omega} = e{\delta {\bf B}}/p_0$.
Subtracting this expression from the collisionless Boltzmann equation gives an evolution equation of the ensemble-fluctuation $\delta f$,
\begin{equation}\label{eq:evoldeltaf}
\partial_t\delta f +  \mathcal{L} \delta f   =  - \delta\mathcal{L}\langle f\rangle + \langle \delta\mathcal{L} \delta f\rangle \, .
\end{equation}
Assuming that $ \delta\mathcal{L}\langle f\rangle \gg \langle \delta\mathcal{L}\delta f\rangle$ this expression can be solved to leading order as 
\begin{equation}
\delta f(t,{\bf r},{\bf p}) \simeq \delta f(t-T,{\bf r}(t-T),{\bf p}(t-T))-\int_{t-T}^t{\rm d}t'\Big[\delta\mathcal{L}\langle f\rangle\Big]_{P(t')}\,, \label{eqn:deltaf}
\end{equation}
where $[\cdot]_{P(t')}$ indicates that the quantity is evaluated along the particle trajectory with position ${\bf r}(t')$ and momentum ${\bf p}(t')$ subject to the boundary condition ${\bf r}(t)={\bf r}$ and ${\bf p}(t)={\bf p}$. The last term in Eq.~(\ref{eq:evoldeltaf}) can hence be approximated as
\begin{equation}\label{eq:scattering}
\langle\delta\mathcal{L}\delta f\rangle \simeq -\left\langle\delta\mathcal{L}\int_{-\infty}^t{\rm d}t'\Big[\delta\mathcal{L}\langle f\rangle\Big]_{P(t')}\right\rangle \, .
\end{equation}
Here, we have assumed finite correlation times such that the lower integration boundary can be extended to ($- \infty$). There are different approximations for evaluating Eq.~(\ref{eq:scattering}).

%--------------------------------------------------------------------------------------------------------------------------------
%--------------------------------------------------------------------------------------------------------------------------------

\paragraph{Pitch--Angle Diffusion}

A typical ansatz of quasi--linear theory (QLT)~\cite{1966ApJ...146..480J,1966PhFl....9.2377K,1967PhFl...10.2620H,1970ApJ...162.1049H,Jokipii1972} is that the phase--space density only depends on the  position and momentum along the (strong) regular magnetic field. We therefore assume that $\langle f\rangle=f(t,z,\mu)$, where we choose ${\bf B}_0 = B_0{\bf e}_z$ and parametrize the parallel momentum via the cosine of the pitch--angle, $p_z = p\mu$. The trajectory $P(t')$ in Eq.~(\ref{eq:scattering}) can then be approximated by the unperturbed path, i.e.\ the CR trajectory computed under the influence of only the regular field ${\bf B}_0$. The scattering term in QLT can then be transformed into a diffusion term for $\mu$
\begin{equation}\label{eq:pitch}
\langle\delta\mathcal{L}\delta f\rangle \simeq - \frac{\partial}{\partial\mu} D_{\mu\mu}\frac{\partial}{\partial \mu}\langle f\rangle\,,
\end{equation}
where $D_{\mu\mu}$ is the pitch--angle diffusion coefficient. Uniform pitch--angle diffusion corresponds to the case $D_{\mu\mu}/(1-\mu^2) = D_0 = {\rm const}$. Together with Eq.~(\ref{eq:genBoltzmann}) and after averaging over gyrophase, this gives rise to the Fokker--Planck equation,
\begin{equation}
\partial_t\langle f\rangle + v \mu \frac{\partial}{\partial z} \langle f\rangle  = \frac{\partial}{\partial\mu} D_{\mu\mu}\frac{\partial}{\partial \mu}\langle f\rangle \,.
\end{equation}

%--------------------------------------------------------------------------------------------------------------------------------
%--------------------------------------------------------------------------------------------------------------------------------

\paragraph{BGK Approximation}

The influence of the turbulence can also be approximated as a friction term, suggested by Bhatnagar, Gross \& Krook (BGK)~\cite{Bhatnagar:1954zz}. In its original version this term drives the ensemble--averaged distribution $\langle f\rangle$ to its isotropic mean $\phi/4\pi$ with an effective relaxation rate $\nu$, {\it i.e.}
\begin{equation}\label{eq:BGK}
\langle\delta\mathcal{L}\delta f\rangle \simeq \nu\left(\!\langle f\rangle-\frac{\phi}{4\pi}\right)\,.
\end{equation} 
This corresponds to the case where the ensemble average in Eq.~(\ref{eq:scattering}) can be approximated as
\begin{equation}
\int_{-\infty}^t \mathrm{d} t' \, \langle\omega_i({\bf r})\omega_j({\bf r}')\rangle \simeq (\nu/2)\delta_{ij} \, ,
\end{equation} 
and where we also assumed $L_i(t') \langle f({\rm r}(t'), {\rm p}(t')) \rangle \simeq L_i(t) \langle f({\rm r}(t), {\rm p}(t)) \rangle$. The scattering term is then approximately 
\begin{equation}\label{eq:BGK2}
\langle\delta\mathcal{L}\delta f\rangle \simeq \nu\frac{{\bf L}^2}{2}\langle f\rangle\,,
\end{equation} 
which is identical to the BGK ansatz, if we expand $\langle f\rangle$ into monopole (${\bf L}^2 = 0$) and dipole (${\bf L}^2 = 2$) contributions.

%--------------------------------------------------------------------------------------------------------------------------------
%--------------------------------------------------------------------------------------------------------------------------------

\subsection{Diffuse Dipole Anisotropy}
\label{subsec:DiffuseDipole}

We will first focus on the first two contributions, monopole and dipole, in the multipole expansion of the phase--space density (see, e.g.\ Ref.~\cite{JonesApJ1990}). It is convenient to introduce the angular-integrated quantities
\begin{align}
\phi(t,{\bf r},p) &= \int {\rm d}\widehat{\bf p}f(t,{\bf r},{\bf p})\,,\\
\boldsymbol{\Phi}(t,{\bf r},p) &= \int {\rm d}\widehat{\bf p}\widehat{\bf p}f(t,{\bf r},{\bf p})\,.
\end{align}
The first term is related to the local CR spectral density [${\rm GeV}^{-1} {\rm cm}^{-3}$] at ${\bf r} = {\bf r}_\oplus$ as
\begin{equation}\label{eq:nCR}
n_{\rm CR}(p) = p^2 \phi(t_\oplus,{\bf r}_\oplus,p)\,.
\end{equation}
Note, that due to the extra $p^2$ term the monopole follows $\phi\propto p^{-2-\Gamma}$ compared to the CR spectrum $n_{\rm CR}\propto p^{-\Gamma}$. The relative intensity can then be expressed in terms of the phase--space density as
\begin{equation}\label{eq:dipole2}
I = 4\pi \frac{ {f(t_\oplus,{\bf r}_\oplus,-{\bf p})}}{ {\phi(t_\oplus,{\bf r}_\oplus,p)}} = 1 - 3\widehat{\bf p}\!\cdot\!\frac{\boldsymbol{\Phi}(t_\oplus,{\bf r}_\oplus,p)}{\phi(t_\oplus,{\bf r}_\oplus,p)} + \mathcal{O}\left(\lbrace a_{\ell m}\rbrace_{\ell \geq 2}\right)
\end{equation}
Keep in mind that we {\it visualize} the relative intensity by the apparent origin of CRs, i.e., $- {\bf p}$, which explains the reversed sign in the previous equation. Comparing Eqs.~(\ref{eq:dipole1}) and (\ref{eq:dipole2}) we see that the dipole vector can be identified as 
\begin{equation}\label{eq:delta1}
\boldsymbol{\delta} = - 3\frac{\boldsymbol{\Phi}(t,{\bf r}_\oplus,p)}{\phi(t,{\bf r}_\oplus,p)} \, .
\end{equation}
Expanding the average phase--space density in terms of the monopole and dipole components and using the simple BGK ansatz of Eq.~(\ref{eq:BGK}), we arrive at a coupled set of differential equations,
\begin{gather}\label{eq:diff1}
\partial_t\langle\phi\rangle  + \nabla_{\bf r}\langle\boldsymbol{\Phi}\rangle \simeq 0\,, \\
\partial_t\langle\boldsymbol{\Phi}\rangle+ (1/3)\nabla_{\bf r} \langle\phi\rangle +\boldsymbol{\Omega}\times \langle\boldsymbol{\Phi}\rangle\simeq-\nu\langle\boldsymbol{\Phi}\rangle\,.\label{eq:diff2}
\end{gather}

In the diffusion approximation it is typically assumed that the dipole component is only slowly varying over the relaxation time $1/\nu$ and hence $\partial_t \langle\boldsymbol{\Phi}\rangle \simeq 0$. In this case we can combine Eqs.~(\ref{eq:diff1}) and (\ref{eq:diff2}) to the familiar diffusion equation 
\begin{equation}
\partial_t\langle\phi\rangle \simeq \nabla_{\bf r}\!\cdot\!({\bf K}\!\cdot\!\nabla_{\bf r}\langle\phi\rangle)\,,
\end{equation}
with diffusion tensor
\begin{equation}\label{eq:K}
{K}_{ij} = \frac{\widehat{B}_i\widehat{B}_j}{3\nu_\parallel}+\frac{\delta_{ij}-\widehat{B}_i\widehat{B}_j}{3\nu_\perp}+\frac{\epsilon_{ijk}\widehat{B}_k}{3\nu_A}\,.
\end{equation}
Here, $\widehat{\bf B}$ is a unit vector pointing in the direction of the regular magnetic field, $\nu_\parallel$ and $\nu_\perp$ denote the effective scattering rates along and perpendicular to the magnetic field, respectively, and $\nu_A$ is the axial scattering rate. In the BGK ansatz the effective diffusion rates are related as $\nu_\perp = \nu_\parallel+\Omega^2/\nu_\parallel$ and $\nu_A = \Omega+\nu_\parallel^2/\Omega$ where $\Omega= c/r_L$ is the gyrofrequency in the regular magnetic field. The solution to the diffusion equation with initial condition $\langle \phi(0,{\bf r})\rangle= \delta^{(3)}({\bf r})$ is simply
\begin{equation}\label{eq:diffsol}
\langle\phi(t,{\bf r})\rangle = \frac{1}{(4\pi t)^{3/2}\sqrt{\det{\bf K}^s}}\exp\left(-\frac{{\bf r}^t({\bf K}^s)^{-1}{\bf r}}{4t}\right)\,,
\end{equation}
where ${\bf K}^s$ is the symmetric part of the diffusion equation, i.e., the first two terms of Eq.~(\ref{eq:K}). Finally, we can solve Eq.~(\ref{eq:diff2}) as, 
\begin{equation}\label{eq:Fick}
\langle\boldsymbol{\Phi}\rangle \simeq -{\bf K}\!\cdot\!\nabla_{\bf r}\langle\phi\rangle\,,
\end{equation}
which is also known as {\it Fick's law}. Comparing this to Eqs.~(\ref{eq:nCR}) and (\ref{eq:delta1}) we arrive at the final expression of the dipole anisotropy 
\begin{equation}\label{eq:DA}
{\boldsymbol \delta^\star} = 3{\bf K}\!\cdot\!\nabla \ln n_{\rm CR}\,.
\end{equation}

In general, the parameters $\nu_\parallel$, $\nu_\perp$ and $\nu_A$ depend on details of CR transport in turbulent magnetic fields. In the case of strong regular magnetic fields we have $\nu_\parallel \ll \nu_\perp$ and $\nu_\parallel \ll \nu_A$ and the diffusion tensor (\ref{eq:K}) reduces to the first term corresponding to a projection of the CR gradient onto the magnetic field direction~\cite{JonesApJ1990,Mertsch:2014cua,Schwadron2014,Kumar:2014dma}. Hence, anisotropic diffusion predicts that the dipole anisotropy should align with the local ordered magnetic field. This ordered magnetic field corresponds to the sum of the large--scale regular magnetic field and the contribution of the turbulent component averaged over distance scales set by the gyroradius. We refer to reviews on (Galactic) magnetic fields~\cite{Beck:2008ty} and the local environment~\cite{Salvati:2010pq,Frisch2011}.

%--------------------------------------------------------------------------------------------------------------------------------
%--------------------------------------------------------------------------------------------------------------------------------

\subsection{Compton--Getting Effect}
\label{subsec:GCeffect}

The relative motion of the solar system through the local plasma frame can introduce a dipole anisotropy known as the {\it Compton--Getting} (CG) effect~\cite{CG1935,1968Ap&SS...2..431G}. In the following we will consider a phase--space density $f^\star(p^\star)$ in the plasma rest frame, denoted by starred quantities. Let's assume that the observer moves along a velocity vector $\boldsymbol{\beta} = {\bf v}/c$ relative to the plasma rest frame. To leading order, the momentum vector ${\bf p}^\star$ in the plasma frame is related to ${\bf p}$ measured in the observer's frame as
\begin{equation}
{\bf p}^\star = {\bf p} + p\boldsymbol{\beta} +\mathcal{O}(\beta^2)\,.
\end{equation}
The CG effect can now be easily derived from the Lorentz-invariance of the phase--space density~\citep{Forman1970}, $f({\bf p}) = f^\star({\bf p}^\star)$. A Taylor-expansion of $f^\star({\bf p}^\star)$ around ${\bf p}$ yields
\begin{equation}
f({\bf p})  \simeq f^\star({\bf p}) + ({\bf p}^\star-{\bf p})\!\cdot\!\nabla_{{\bf p}}f^\star({\bf p})+\mathcal{O}(\beta^2)
\end{equation}
To leading-order, this transformation leaves the isotropic contribution invariant, $\phi = \phi^\star$, whereas the second term on the right hand side gives an additional contribution to the dipole,
\begin{equation}
\boldsymbol{\Phi} \equiv\int{\rm d}\widehat{\bf p}\widehat{\bf p} f({\bf p}) \simeq \boldsymbol{\Phi}^\star + \int{\rm d}\widehat{\bf p}\widehat{\bf p} (\boldsymbol{\beta}\!\cdot\!\widehat{\bf p})p\frac{\partial f^\star({\bf p})}{\partial p} \simeq \boldsymbol{\Phi}^\star + \frac{\boldsymbol{\beta}}{3}\frac{\partial \phi}{\partial \ln p}.
\end{equation}
Therefore, the dipole ${\boldsymbol \delta} = -3\boldsymbol{\Phi}/\phi$ in the observer's frame transforms to 
\begin{equation}\label{eq:deltarel}
{\boldsymbol \delta} \simeq {\boldsymbol \delta}^\star + (2+\Gamma){\boldsymbol \beta}\,,
\end{equation}
where we assume that the isotropic part of the phase--space density follows $\phi \propto p^{-2-\Gamma}$, where $\Gamma\simeq2.7$ is the CR spectral index. 

From a conceptual point of view, there is a variety of possible sources of the motion of the solar system with respect to the frame in which the cosmic rays are isotropic. An average star at the position of the Sun has a circular speed around the Galactic center of $v_0 \simeq 220\pm20$ km/s pointing into Galactic longitude and latitude $l\simeq 270^\circ$ and $b\simeq0^\circ$~\cite{Binney2008}. The observable CG effect after projection onto the equatorial plane would be $\delta_{\rm GC}\simeq 2.4\times 10^{-3}$, independent of energy. However, these values are inconsistent with the observed dipole amplitude and phase shown in Fig.~\ref{fig:dipoledata}. 

This indicates that the CR rest frame is mostly co-rotating with the average motion of the stars. Therefore, a more suitable choice as a CR rest frame seems to be the local standard of rest (LSR). In this frame the Sun moves towards $l \simeq 47.9^\circ\pm2.9^\circ$ and $b\simeq23.8^\circ\pm2.0^\circ$, called the solar apex, with velocity $v_{\rm LSR} \simeq 18.0\pm0.9$ km/s~\cite{Schoenrich2010}. Yet another choice would be the relative movement through the local interstellar medium (ISM) with velocity $v_{\rm ISM} \simeq 23.2\pm0.3$ km/s in the direction $l \simeq 5.25^\circ\pm0.24^\circ$ and $b\simeq12.0^\circ\pm0.5^\circ$~\cite{McComas2012}. These latter estimates of the CG effects are at the level of $10^{-4}$ and would contribute to the observed dipole anisotropy as an overall shift in the data.

It has been argued~\cite{Biermann:2012tc} that the coherent, magnetized flow from an old, nearby supernova remnant could be the source of a large--scale, not necessarily dipole anisotropy. Flow speeds of $\beta \sim 3.3 \times 10^{-4} (v / 100 \, \text{km} \, \text{s}^{-1})$ which give the right order of magnitude large--scale anisotropy are in fact inferred for old supernova remnants like Loop~I~\cite{1984ApJS...55..585H}. However, it is not clear how this could possibly lead to a non--dipolar anisotropy as argued by the authors of Ref.~\cite{Biermann:2012tc}: A Compton--Getting like effect leads to first approximation always to a dipole and in any event (that is even at higher orders) always to a symmetric anisotropy, i.e.\ an excess in one direction is mirrored by a deficit in the opposite direction. Furthermore, at variance with what is claimed by the authors, the CR density inside the flow would need to be dominated by CRs from the source of the flow (which is virtually impossible for TeV particles and a necessarily old supernova remnant) or from an even closer source which would then again dominate the flow. Ambient CRs on the other hand would need many mean free paths to isotropize in the flow.

On the other hand, a well--known contribution to the CG effect is induced by the motion of the Earth around the Sun, the {\it solar} dipole anisotropy. The solar CG (SCG) effect in the sidereal frame depends on the time of observation. By definition, mean solar time is given by the hour angle of the mean Sun shifted by 12 hours, $\omega_{\rm solar}t \,\,({\rm mod}\,\, 2\pi) = h_{\odot}+ \pi$. On the other hand, the right ascension of the Sun and the local hour angle are related via  $\omega t\,\, ({\rm mod}\,\, 2\pi) = \alpha_\odot + h_\odot$. Therefore, the position of the Sun can be expressed as $\alpha_\odot = \omega_{\rm orbit}t +\pi$. In the sidereal frame, the orientation of the SCG dipole anisotropy is then proportional to the velocity vector of the Earth. 

For simplicity, we will treat the Earth's orbit in the following as circular with average orbital velocity $\bar{v}_{\rm orb} \simeq 29.8 {\rm km}/{\rm s}$. Given the small eccentricity of the Earth's orbit and the limited sensitivity of present observatories this is a sufficient first order approximation. The SCG effect in the sidereal frame can then be expressed as
\begin{equation}
\boldsymbol\delta_{\rm SCG} = \delta_{\rm SCG}\big(-\sin (\omega_{\rm orb}t),\cos\epsilon\cos(\omega_{\rm orb}t),\sin\epsilon\cos(\omega_{\rm orb}t)\big)\,,
\end{equation}
where $\epsilon\simeq23.4^\circ$ is the tilt of the Earth's rotation axis with respect to the ecliptic and $\delta_{\rm SCG} = (2+\Gamma)\bar{v}_{\rm orb}/c$. However, most observatories collect CR data over a period of multiple years. When time--averaging over full years, the SCG effect in the {\it sidereal} frame vanishes, $\langle\boldsymbol\delta_{\rm SCG}\rangle_{\rm year} \simeq (0,0,0)$. 

It is possible to study the SCG effect in the {\it solar} frame, given by the transformation $\boldsymbol\delta_{\rm SCG}'' = {\bf R}^T((\omega_{\rm sol}/\omega)t)\cdot{\bf R}(t)\cdot\boldsymbol\delta_{\rm SCG}$ or, explicitly,
\begin{equation}
\boldsymbol\delta_{\rm SCG}'' =\delta_{\rm SCG}\big(-\sin^2(\epsilon/2) \sin (2\omega_{\rm orb}t), \cos^2(\epsilon/2)-\sin^2(\epsilon/2)\cos(2\omega_{\rm orb}t), \sin\epsilon\cos (\omega_{\rm orb}t) \big)
\end{equation}
Now, the yearly average is $\langle\boldsymbol\delta_{\rm SCG}''\rangle_{\rm year} = \delta_{\rm SCG}\left(0,\cos^2(\epsilon/2),0\right)$. For $\Gamma\simeq2.7$ the average SCG strength is thus $|\langle\boldsymbol\delta_{\rm SCG}''\rangle| \simeq 4.5\times10^{-4}$ and points into the direction of $t=6$h. The amplitude and orientation of this solar dipole has been studied and confirmed by various experiments (see Ref.~\cite{DiSciascio:2014jwa}). These observations can also be used to infer the spectral index of CRs~\cite{Amenomori:2007ug}.

%--------------------------------------------------------------------------------------------------------------------------------
%--------------------------------------------------------------------------------------------------------------------------------

\subsection{Comparison to the Sidereal Dipole Anisotropy}
\label{subsec:Comparison}

In the case of isotropic diffusion with a smooth distribution of sources, the diffusive dipole anisotropy (\ref{eq:DA}) is expected to simply align with the direction of the Galactic center at $\alpha\simeq266^\circ$ due to the off--center position of the solar system. The dipole amplitude is expected to simply follow the energy scaling of the diffusion tensor (\ref{eq:K}), typically a power-law ${\bf K}\propto E^\beta$ with $\beta\simeq0.3$--$0.6$ (cf.\ e.g.\ Ref.~\cite{Strong:2007nh}). These predictions are shown as dashed lines in the summary plot of Fig.~\ref{fig:dipoledata}, where we use $\beta=1/3$ for the amplitude's energy scaling and normalize to the TeV data. Clearly, this na\"ive extrapolation does not follow the general trend of the amplitude at higher energies shown in this plot. Even worse, the absolute scale of the CR gradient computed from a pulsar--like or supernova remnant like distribution of sources is that large, that with the normalisation of the diffusion coefficient from nuclear secondary--to--primary ratios, like Boron--to--Carbon, the predicted dipole is up to two orders of magnitude larger than measured~\cite{Blasi:2011fm,Pohl:2012xs,Evoli:2012ha,Kumar:2014dma}. This discrepancy is known as the ``anisotropy problem''~\cite{2005JPhG...31R..95H}. In addition, the data indicate that the TeV--PeV dipole anisotropy undergoes a rapid phase flip at an energy of 0.1--0.3~PeV. 

Note, that the measured dipole amplitudes and phases shown in Fig.~\ref{fig:dipoledata} only include the statistical error of the observation. There are many systematic uncertainties that are related to the partial sky coverage, reconstruction methods, and different analysis methods (see Section~\ref{subsec:HarmonicAnalysis}). These factors can account for the relative systematic shift of data from observations in the same energy region. However, it is unlikely that the large deviation from the model prediction (dashed line) can be accounted for by these systematic errors.

A simple way of reducing the model prediction for the dipole amplitude $A_1$ would be to postulate a very small {\it local} diffusion coefficient~\cite{Zirakashvili:2005gz}, much smaller than the one derived, e.g.\ from Boron--to--Carbon. While nuclear secondary--to--primary ratios are indeed testing the propagation over kiloparsec distances, and local structure, like the local bubble (a cavity filled with hot, tenuous gas, extending to $\sim 100 \, \text{pc}$, cf.\ e.g.\ the review~\cite{Frisch2011}) could in principle modify the diffusion coefficient locally, it is not clear whether this would indeed lead to a {\it smaller} diffusion coefficient. Another idea would be to reduce the CR gradient by modifying CR escape in the inner Galaxy, e.g.\ by a much larger rate of diffusion and escape perpendicular to the Galactic plane~\cite{Evoli:2012ha}. This model has the added benefit that a smaller CR gradient is also indicated from observations of  diffuse gamma--rays by the \text{Fermi}--LAT experiment~\cite{collaboration:2009ag,Collaboration:2010cm}.

In general, the distribution of CR sources is not smooth, in particular in our local environment. The presence of local and young CR sources can introduce modulations of phase and amplitude, even if their individual contribution to the total CR flux is only subdominant~\citep{Erlykin:2006ri,Blasi:2011fm,Pohl:2012xs,Sveshnikova:2013ui,Kumar:2014dma}. However, most of these scenarios discuss dipole modulations from local sources under the simplifying assumption of isotropic CR diffusion. This ansatz seems appropriate for the prediction of the local CR density since CR diffusion from distant sources typically averages over local properties of the diffusion medium. However, the properties of our local medium are important in calculating the diffusion dipole (\ref{eq:DA}). Locally, the strength of the turbulent component magnetic field is smaller than that of the regular component~\cite{Beck:2008ty}. This indicates a strong anisotropic diffusion in our local environment. In this case the diffusion tensor reduces to the first term of Eq.~(\ref{eq:K}), corresponding to a projection of the CR gradient onto the magnetic field direction~\cite{JonesApJ1990,Mertsch:2014cua,Schwadron2014,Kumar:2014dma,Battaner:2014cza,Ahlers:2016njd}. 

\begin{figure}[p]\centering
\includegraphics[width=0.6\linewidth]{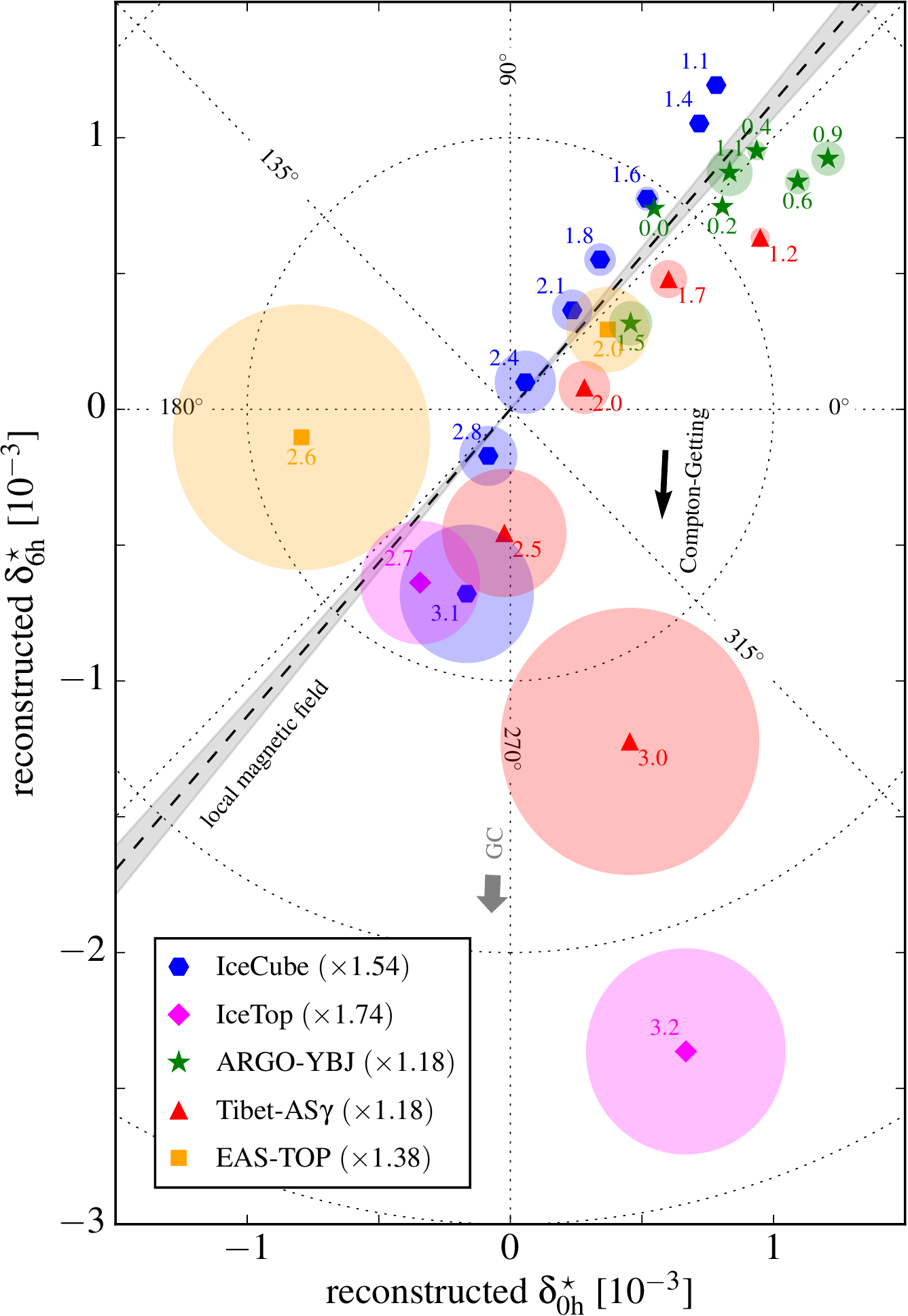}
\caption[]{Summary plot of the reconstructed TeV--PeV dipole components $\delta^\star_{0{\rm h}}$ and $\delta^\star_{6{\rm h}}$ in the equatorial plane (from Ref.~\cite{Ahlers:2016njd}). The black arrow indicates the Compton--Getting effect from the solar motion with respect to the local standard of rest that we subtracted from the data following Eq.~(\ref{eq:deltarel}). The declination-averaged data of ARGO-YBJ~\cite{Bartoli:2015ysa}, Tibet-AS$\gamma$~\cite{TibetICRC355}, and IceCube/IceTop~\cite{Aartsen:2012ma,Aartsen:2016ivj} is rescaled following Eq.~(\ref{eq:Arelation}). The result by EAS-TOP~\cite{Aglietta:2009mu} was derived by the East--West method and uses a rescaling factor to account for the relation (\ref{eq:dalpha}). The numbers attached to the data indicate the median energy of the bins as $\log_{10}(E_{\rm med}/{\rm TeV})$. The colored disks show the $1\sigma$ error range. The dashed line and gray-shaded area indicate the magnetic field direction and its uncertainty (projected onto the equatorial plane) inferred from IBEX observations~\cite{Funsten2013}. We also indicate the direction towards the Galactic center (GC).}\label{fig:reconstructed}
\end{figure}

Indeed, the analysis of Ref.~\cite{Schwadron2014} pointed out that the combined anisotropy maps of IceCube at 20~TeV and Tibet-AS$\gamma$ at 5~TeV show a close alignment of large--scale features with the local magnetic field direction inferred from the emission of energetic neutral atoms (ENA) from the outer heliosphere observed by the {\it Interstellar Boundary Explorer} (IBEX)~\cite{McComas2009}. The emission of ENA is enhanced along a circular region that defines a magnetic field axis along $l \simeq 210.5^\circ$ and $b\simeq-57.1^\circ$ with an uncertainty of $\sim1.5^\circ$~\cite{Funsten2013}. This is consistent with the magnetic field direction inferred from the polarization of local stars within $40$~pc with best--fit field direction $l\simeq216.2^\circ$ and $b\simeq-49.0^\circ$, and statistical uncertainties $\sim16^\circ$~\cite{Frisch2015}. The projection of the CR gradient onto the local magnetic field can also be responsible for the low amplitude of the observed dipole anisotropy in the TeV--PeV range as pointed out in Ref.~\cite{Mertsch:2014cua}. In fact, if the orientation of the CR gradient passes across the magnetic horizon as the CR energy changes, the dipole is expected to experience a phase-flip. 

Equipped with this list of ideas, we can now re-access the ``anisotropy problem'' of Fig.~\ref{fig:dipoledata}. Figure~\ref{fig:reconstructed} shows the reconstructed dipole components ($\delta^\star_{\rm 0h}$ and $\delta^\star_{\rm 6h}$) following Eq.~(\ref{eq:ampphase}) inferred from recent measurements by ARGO-YBJ~\cite{Bartoli:2015ysa}, EAS-TOP~\cite{Aglietta:2009mu}, IceCube/IceTop~\cite{Aartsen:2012ma,Aartsen:2016ivj}, and Tibet-AS$\gamma$~\cite{TibetICRC355} (figure from Ref.~\cite{Ahlers:2016njd}). These observations are derived from declination-averaged data or via the East--West method (see Section~\ref{subsec:HarmonicAnalysis}). To account for systematics errors of these analysis methods, the  amplitudes are rescaled following Eq.~(\ref{eq:Arelation}) or corrected for the effective right ascension step size (\ref{eq:dalpha}), respectively, as indicated in the legend (see Ref.~\cite{Ahlers:2016njd} for details). The data is also corrected for the Compton--Getting effect from the motion of the solar system in the local standard of rest, following Eq.~(\ref{eq:deltarel}).

The data in the TeV--PeV energy range clearly shows a trend to align with the local magnetic field inferred from the IBEX measurement, which is predicted by anisotropic diffusion. Note, that the Compton--Getting shift in the equatorial plane is only a small correction of the dipole as indicated by the black arrow in Fig.~\ref{fig:reconstructed}. The shaded circles indicate the statistical uncertainty quoted by experiments. After rescaling the data and correcting for a CG shift, there is still a large systematic shift between the data sets, in particular at high energy. This can be explained by additional systematic effects, like the cross talk of multipole moments in a harmonic analysis with a partial sky coverage.

As pointed out in Ref.~\cite{Ahlers:2016njd}, the dipole data indicates that the CR gradient below the phase flip at 100~TeV is oriented towards Galactic longitudes $120^\circ\lesssim l \lesssim 300^\circ$. The Vela SNR ($l = 263.9^\circ$, $b=-3.3^\circ$) is one of the closest known SNRs~\cite{Green:2014cea} at a distance of about $0.3$~kpc~\cite{Cha:1999pn} and with an age of about $11$~kyr~\cite{Reichley1970}. The estimated ejecta energy is of the order of $10^{51}$~erg~\cite{Jenkins1995}, the required energy scale if SNRs are the sources of CRs~\cite{Baade1934,TheOriginofCosmicRays1964}. Using the diffusion kernel (\ref{eq:diffsol}), one can show that Vela is expected to be one of the strongest contributors to the CR anisotropy~\cite{Sveshnikova:2013ui,Ahlers:2016njd}. For typical values of the diffusion tensor (assumed to be effectively isotropic over large distance scales) the analysis of Ref.~\cite{Ahlers:2016njd} finds that the Vela SNR naturally dominates the CR gradient below 100~TeV and its relative location to the local magnetic field is consistent with the observed dipole phase and phase-flip.

\begin{figure}[t]
\centering
\includegraphics[width=0.6\linewidth]{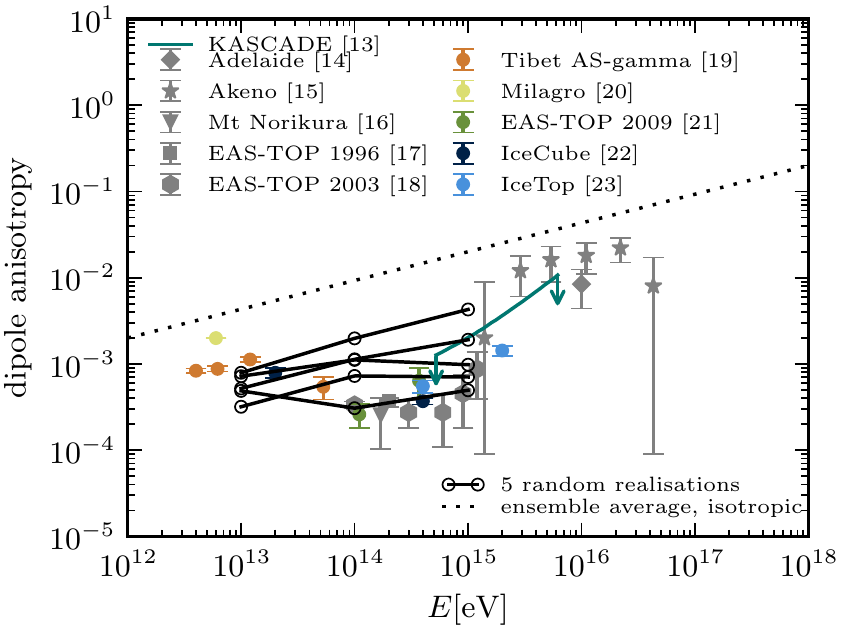}
\caption[]{Measured dipole amplitude as a function of CR energy~\cite{Antoni:2003jm,Gerhardy:1984wq,Kifune:1985vq,Nagashima:1990ze,Aglietta:1996sz,Aglietta:2003uc,Amenomori:2005dy,Abdo:2008aw,Aglietta:2009mu,Abbasi:2011zka,Aartsen:2012ma}. (Compare with Fig.~\ref{fig:dipoledata}.) The dotted line is the average prediction from an ensemble of source~\cite{Blasi:2011fm}. The open circles, connected by solid lines show the dipole amplitude computed in Ref.~\cite{Mertsch:2014cua} for five different realisations of the turbulent magnetic field if gradient and ${\bf B}_0$ directions are at close to $90^{\circ}$.}
\label{fig:MertschFunk2014}
\end{figure}

Another important effect is the intermittency of the turbulent magnetic field, in the sense that a single realisation of the phase--space density $f$ (for a particular realisation of the turbulent field ${\delta {\bf B}}$) will deviate from its ensemble average $\langle f \rangle$ (in the notation of Sec.~\ref{sec:StandardPicture}). This influences macroscopic observables like the dipole direction and amplitude~\cite{Mertsch:2014cua} that get generated only over distances of a few scattering lengths even if the CR gradient extends over many coherence lengths of the turbulent field. In Fig.~\ref{fig:MertschFunk2014} we show the dipole amplitudes for five different realisations of the turbulent magnetic field and assuming a CR gradient almost perpendicular to ${\bf B}_0$. Interestingly, there are non--trivial rigidity dependencies. These are due to the fact that the anisotropy for particles of different rigidities is formed over different distances. Those particles are hence experiencing different effective regular fields (the sum of the background field plus the average of the large--scale turbulent modes). This changes the effective projection of the gradient and thus the overall amplitude. Even for purely turbulent magnetic fields, the dipole direction in a particular realisation of the field is in general not along the CR gradient and can be off by as much as $90^{\circ}$. In addition, the amplitude can be deviating from its ensemble average rather strongly. We will revisit intermittency effects as a possible source of the small--scale anisotropies in Sec.~\ref{subsec:MagneticTurbulence}.

%--------------------------------------------------------------------------------------------------------------------------------
%--------------------------------------------------------------------------------------------------------------------------------
%--------------------------------------------------------------------------------------------------------------------------------

\section{Interpretations of Small--Scale Sidereal Anisotropies}
\label{sec:InterpretationSmall}

In the previous section we have discussed the expected anisotropy from the diffusion of CRs from distant Galactic sources. In the (idealized) case of isotropic scattering of CRs in magnetic turbulence, the leading order effect is a dipole with amplitude and orientation determined by the source distribution, the diffusion medium, and the relative motion of the observer. However, recent observations indicate the presence of significant small--scale anisotropies as shown in Sec.~\ref{sec:Observation}. Diffusion is expected to wash out any memory of the initial direction of CR trajectories over distances larger then the effective scattering length. The appearance of small--scale anisotropies in the arrival direction of CRs is therefore an indication of a local effect, that is either (though unlikely) related to a local source or local electromagnetic field configurations. In this section we will discuss various scenarios that have been proposed as the origin of small--scale features.

A recurring problem in studying the influence of local electromagnetic field configuration on the arrival direction of CRs is the simulation of CRs propagation in the magnetic turbulence over large Galactic distances.
Ideally, one would want to compute the trajectories of particles from their source(s) to the observer. The problem here is that given the small size of a typical detector relative to the system, only a vanishingly small fractions of simulated particles will reach the detection volume in a finite amount of time. A viable alternative is to give up on simulating the transport from (an) individual source(s) and instead consider the phase--space density intermediate between source(s) and the observer as a starting point at earlier time. (Ideally, the results of this approach would not depend too sensitively on the precise details of the assumed phase--space density, but see below.) 

Instead of forward--tracking particles towards the observer, particles are {\it back--tracked} from the observer, that is the equation of motions are solved for an oppositely charged particle leaving the observer in the direction from which a particle would be observed. This is equivalent to reversing the arrow of time for the original particle, hence the name ``back--tracking''. This is done for a number of particles at least as large as the number of pixels (if an anisotropy map is to be computed), and their directions at the observer are usually uniformly distributed. Making use of Liouville's theorem, that is the conservation of phase--space density along trajectories, we can connect the phase--space density at observation (the anisotropy map) to the phase--space density assumed at an earlier time. This method is used frequently in the scenarios discussed below.

%--------------------------------------------------------------------------------------------------------------------------------
%--------------------------------------------------------------------------------------------------------------------------------

\subsection{Heliosphere}

The possibility that the Galactic CR anisotropy could be influenced by the heliosphere was first suggested in Ref.~\cite{Nagashima1998}. The authors studied the global anisotropy of CR arrival directions in the 10~GeV to 10~TeV energy range by a combination of several declination-averaged anisotropy measurements. It was argued that the combined large--scale anisotropy could be described by a broad Galactic deficit region centered at $\alpha\simeq 180^\circ$ and $\delta\simeq 20^\circ$ (``loss--cone'' region) and a broad excess region close to the direction of the heliotail (``tail--in'' region) at $\alpha\simeq 90^\circ$ and $\delta\simeq-24^\circ$. Although, the authors of Ref.~\cite{Nagashima1998} did not provide a particular mechanism that explains the large--scale anisotropy from first principles, the name ``tail--in'' and ``loss-cone'' have become synonymous for the large--scale excess and deficit regions in the anisotropy maps.

Later observations by Tibet-AS$\gamma$~\cite{Amenomori:2006bx}, Milagro~\cite{Abdo:2008kr}, and ARGO-YBJ~\cite{Vernetto:2009xm} identified narrow excess regions (regions A and B in Fig.~\ref{fig:HAWCmap}) that seem to bracket the direction of the heliotail. The Tibet-AS$\gamma$ collaboration showed~\cite{Amenomori:2009ha} that the large--scale anisotropy could be modeled by the superposition of a dipole (``uni--directional flow'') and quadrupole (``bi--directional flow'') component. The remaining features were described as two excess regions oriented along the hydrogen deflection plane (HDP), the great circle containing the heliotail and the interstellar magnetic field direction. This was then interpreted as a lensing effect from coherent magnetic fields in the heliotail with opposite polarity and orthogonal to the HDP~\cite{Amenomori:2009ha}. However, this model is again {\it ad hoc} and does not provide an explanation of the heliospheric magnetic environment and its orientation with respect to the HDP from first principles.

\begin{figure}
\centering
\setlength\fboxsep{0pt}
\fbox{\includegraphics[width=0.7\textwidth]{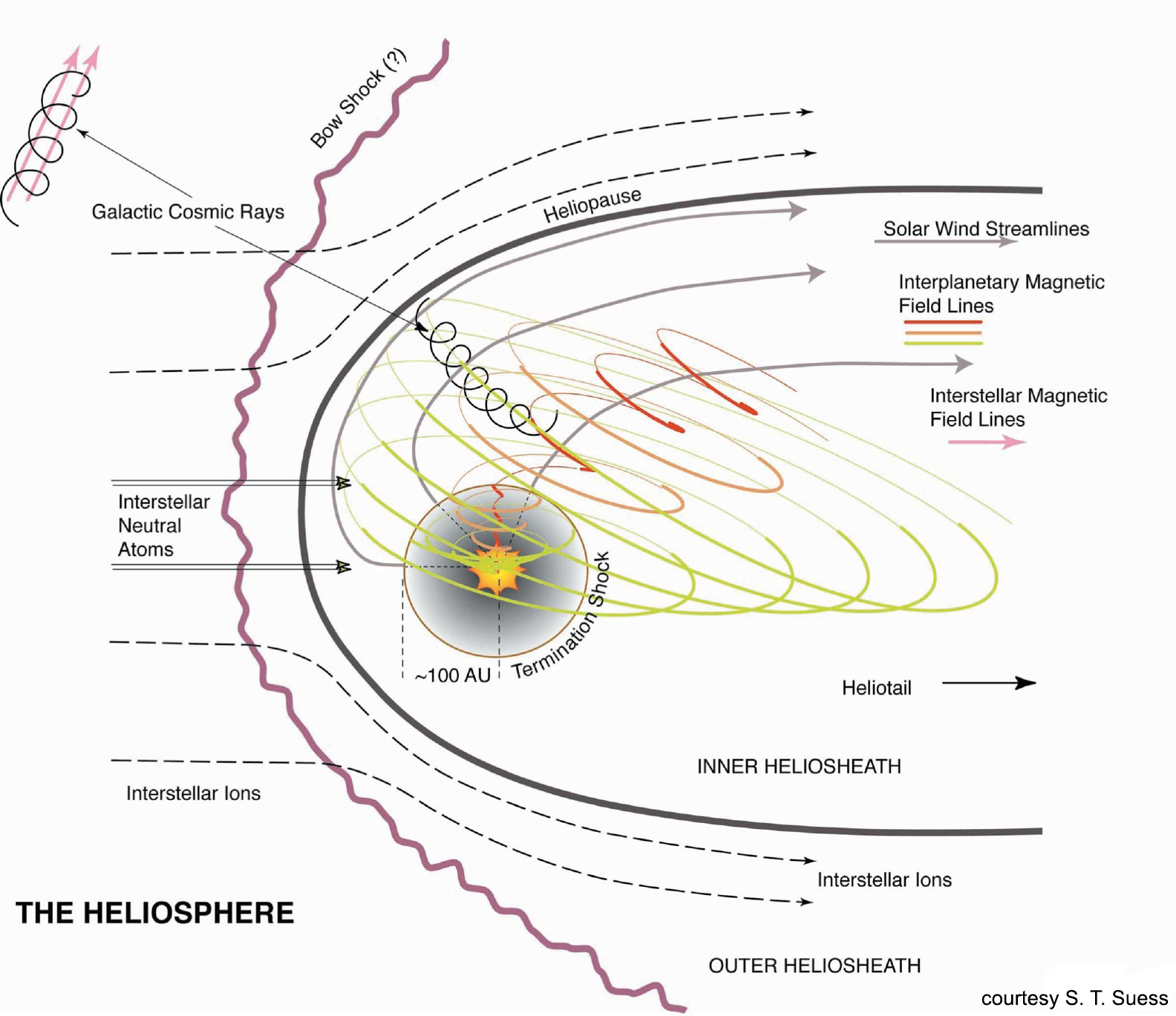}}
\caption{A schematic view of the heliosphere (from Ref.~\cite{USSSB2004}).}
\label{fig:heliosphere}
\end{figure}

Before we continue the discussion let us briefly review the mechanisms that shape the heliosphere. For a review on the structure of the heliosphere, see Ref.~\cite{1990RvGeo..28...97S}. Figure~\ref{fig:heliosphere} shows a schematic view of the interaction of the interstellar medium with the solar environment (from Ref.~\cite{USSSB2004}). The plasma flow and with it the frozen--in magnetic field around the Earth is governed by two effects: the interstellar wind, caused by the motion of the solar system through the interstellar medium, and the solar wind, the hot coronal plasma continuously launched by the sun at supersonic speeds (up to $800 \, \text{km} \, \text{s}^{-1}$ closer to the poles and in quiet solar conditions) since there is no pressure equilibrium in the solar corona. The solar wind is initially radial and discontinuously slows down to subsonic speeds at a distance of $\sim 80-90 \, \text{AU}$ ($1~{\rm AU} \simeq1.5\times 10^{13}$~cm), the so--called ``termination shock''. At even larger distances, the heliopause marks the surface inside of which (``inner heliosheath'') the solar flow and magnetic fields dominate and outside of which (``outer heliosheath''), the interstellar flow and magnetic fields dominate. Past the termination shock, in the inner heliosheath, the solar wind curves from its radial form, turns over and flows towards the interstellar downstream direction, the heliotail. 

The structure of the magnetic field inside the heliosphere is influenced by the magnetic field structure launched with the solar wind and by rotation: The sun is rotating with a period of $\sim 25 \, \text{days}$ and the rotation axis is misaligned with the magnetic axis. As the magnetic field is frozen into the plasma, together with the solar wind this leads to a twist of magnetic field lines into the well--known Parker spiral~\cite{Parker:1958zz} (see Fig.~\ref{fig:heliosphere}). In one hemisphere, the magnetic field is directed inward and in the other hemisphere it is directed outward. The transition between both polarity regions is taking place in a surface which due to the misalignment of rotation and magnetic axis by the so--called tilt--angle takes a wavy shape: the ``wavy heliospheric current sheet''.

At first, it might seem surprising that the heliosphere could imprint itself on TeV CRs. It is well--known that the solar wind modulates interstellar CR {\it fluxes} only below rigidities of a few GV. This can be accounted to the fact that nuclear spectra computed from CR propagation models fit the observed (solar modulated) spectra reasonably well when the effect of modulation is modeled as an electro--static potential (``force--field approximation'',~\cite{Gleeson:1968zza}) with potentials of a few hundred MeV. At rigidities of a few GV, the effect is then a few percent at most. This argument does however not preclude the possibility that the heliosphere could influence the CR {\it arrival directions} even if the flux remains largely the same. In fact, the gyro--radius of a TV particle in a micro--Gauss magnetic field, $r_g \simeq 200 \, (R / \text{TV}) (B / \mu\text{G})^{-1} \, \text{AU}$ is smaller than the heliosphere.

It has been pointed out~\cite{Drury:2013uka} that the small--scale features in TeV--PeV anisotropy maps, like region A, could be caused by the typical solar modulation potentials of a few hundred GV. The presence of such localised electric fields ${\bf E}$ could be due to the relative velocity ${\bf v}$ between the interstellar plasma frame (in which electric fields are shorted out due to the high conductivity of the ISM) and the solar wind, ${\bf E} = - {\bf v} \times {\bf B}$, where ${\bf B}$ would be the heliospheric magnetic field. In a magnetic field of $10 \, \mu\text{G}$ and with a speed of $10 \, \text{km} \, \text{s}^{-1}$ this would generate an electric field of $1.5 \, \text{MV}/\text{AU}$. Most CRs experience electric fields of different polarities on their way through the heliosphere and thus no net acceleration or deceleration. Along special direction, however, CRs could experience the same electric field along a trajectory of order $100 \, \text{AU}$ and thus encounter a potential of, say $150 \, \text{MeV}$ as needed for an anisotropy level of $10^{-4}$~\cite{Drury:2013uka}. Turbulence on scales smaller than $100 \, \text{AU}$ would also prevent CRs to experience the same electric field over distances of $100 \, \text{AU}$ and therefore wash out the effect. This would be the case at energies lower than $\sim \text{TeV}$ where the gyroradius becomes smaller than the size of the heliosphere and particle trajectories quickly diverge.

\begin{figure}
\centering
\includegraphics[width=0.6\textwidth]{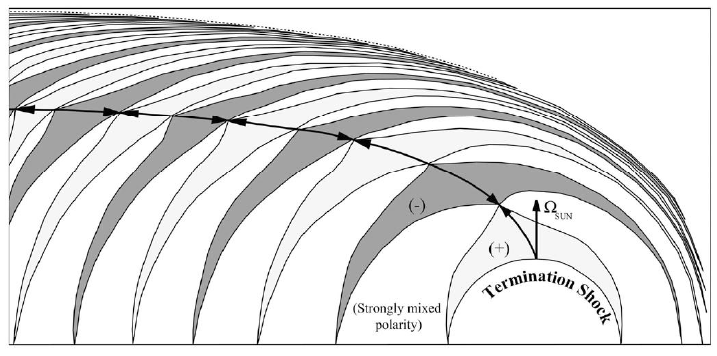}
\caption{Unipolar regions generated over many periods of the $22 \, \text{years}$ solar cycle and advected with the solar wind into the heliotail direction to the left (from Ref.~\cite{Nerney1995}).}
\label{fig:Nerney1995}
\end{figure}

The Sun's polarity is also reversing every $11 \, \text{years}$ at the maxima of solar activity, giving rise to a $22 \, \text{year}$ period, and with the solar wind the different polarity regions are advected into the heliotail. Coronal regions close to the poles (that is beyond the tilt angle of the current sheet) emit unipolar regions, especially around solar minimum, whereas around the heliographical equator strongly mixed domains dominate~\cite{Nerney1995}. In the diverging solar wind past the termination shock, these regions can grow to maximum sizes of $\sim$ 200--300 $\text{AU}$; the particular latitudinal structure of these domains is in part due to the misalignment of the sun's magnetic and rotational axis and is shown in Fig.~\ref{fig:Nerney1995}. With the solar wind decelerating down the heliotail, the unipolar domains are compressed and regions of opposite polarity can reconnect. This reconnection has been suggested~\cite{Lazarian:2010sq} as the cause of the unusually hard spectrum of CRs from the directions of the small scale features. This would naturally be able to accommodate the fact that at least one of these small--scale features is coincident with the heliotail direction.

It has been observed in MHD simulations of the heliosphere~\cite{2013ApJ...772....2P} that various plasma instabilities generate turbulence in the heliotail. The turbulence extends over many hundreds, maybe thousand of AU downstream into the heliotail before being damped. If the turbulence injection scale is (a few) hundred AU and the turbulence velocity at the injection scale is of the order of the ISM flow speed $\sim 25 \text{km} \, \text{s}^{-1}$, then turbulence at $\mathcal{O}(10) \, \text{AU}$ (resonant with TeV particles in a few $\mu\text{G}$ fields) is super--Alfv\'enic and strong, that is $\delta B/B_0 \gtrsim 1$, where $\delta B$ and $B_0$ are the regular and turbulent magnetic fields strengths. It can be argued~\cite{Desiati:2011xg} that the mean--free path (or scattering length) of TeV particles is of the order of the gyroradius, that is much smaller than typically in the ISM. An anisotropic contribution coming in from the nose direction could thus be ``back--scattered'' and contribute to an apparent excess from the heliotail direction. A non--dipolar anisotropy could be due to the fact that the cosmic ray gradient rapidly changes across the plane perpendicular to the interstellar magnetic field (IMF)---ultimately a consequence of the inclination of the IMF direction to the heliotail direction and the transport of turbulence in the heliosphere. Finally, the authors of Ref.~\cite{Desiati:2011xg} also entertain the possibility that small--scale features can be due to reconstruction errors in the presence of a large angular gradient in CR flux.

A much more detailed analysis of how the heliosphere is shaping interstellar CR anisotropies is presented in Ref.~\cite{Zhang:2014dsu}. Given a state--of--the--art MHD model of the heliosphere~\cite{2013ApJ...772....2P}, particles can be back--tracked through the heliosphere by solving the Newton equation in the presence of the Lorentz force term. Liouville's theorem can then be applied to this set of trajectories, resulting in the CR anisotropy map at Earth for an assumed distribution of CR distribution far enough outside the heliosphere. (See Sec.~\ref{subsec:MagneticTurbulence} below for more details on back--tracking, in particular through the ISM.) The authors discuss three effects which can affect the particle trajectories and therefore influence the observed anisotropy: (1) acceleration in electric fields, (2) non-uniform pitch--angle scattering along the regular magnetic field, and (3) drift diffusion perpendicular to the field (``B--cross--gradient'' forces).

For the external anisotropy in the interstellar medium the authors of Ref.~\cite{Zhang:2014dsu} chose 
the ansatz (in the notation of Sec.~\ref{subsec:DiffuseDipole}) $\delta I \simeq 3\widehat{\bf p}\cdot{\bf K}\cdot\nabla \ln n+A_{2\parallel}P_2(\widehat{\bf p}\cdot\widehat{\bf B}) + ({\bf r}-{\bf r}_\oplus)\cdot\nabla_\perp \ln n$ with $1/\nu_\perp \to 0$ in the diffusion tensor (\ref{eq:K}). The first term corresponds to the familiar dipole term and the second term accounts for a quadrupole along the magnetic field from non-uniform pitch-angle diffusion (see next section). The last term corresponds to a Taylor expansion in a perpendicular density gradient. The authors argue that the parallel density gradient is negligible for anisotropic diffusion from a point source. However, one should keep in mind that the density gradient receives contributions from many sources. In that case the gradient rather reflects the distribution of sources than the tilted gradient from a local source. Also, it is unlikely that the local magnetic field is homogeneous up to the position of sources. See Ref.~\cite{Ginzburg:1990sk} for the question how on scales larger than the coherence length diffusion can be isotropic even if locally it is anisotropic.

Fitting to Tibet AS$\gamma$ data, the free parameters of this model, that is the strength of the interstellar unipolar and dipolar anisotropies as well as the CR gradients perpendicular to the interstellar field can be determined. Temporal variations due to solar cycle variations and effects of the solar corona's and inner heliosphere's magnetic fields are found to be subdominant. The authors argue that reduced drift diffusion (``B--cross--gradient'' forces) in the low magnetic field of the heliotail can lead to narrow features that align with the HDP and could be responsible for region A and B seen in anisotropy maps. There are also additional ring-like features in the plane orthogonal to the interstellar magnetic field. However, it remains to be seen if the inclusion of a parallel density gradient could have an influence on these pronounced features.

A complementary approach could be to use forward--tracking of particles from large distance from the Earth through simulated magnetic fields. (See again Sec.~\ref{subsec:MagneticTurbulence} for a discussion of the validity of back--tracking.) In Ref.~\cite{Lopez-Barquero:2016wnt} it is shown that the heliosphere can affect the arrival directions at Earth although it is not discussed in as much detail as in Ref.~\cite{Zhang:2014dsu} which physical process is the origin. Particles of gyroradii $r_{\text{g}} \leq L$ (where $L$ is the largest scale of turbulence) are affected differently than those with gyroradii $r_{\text{g}} > L$. Due to the computational expense of forward--tracking compared to back--tracking there could be systematic uncertainties introduced by the finite size of the detector sphere (which ideally would be $\ll r_{\text{g}}$).

In conclusion, the interesting possibility that the heliosphere could influence the arrival directions of Galactic CRs at TeV energies has been investigated in a variety of studies, ranging from back--of--the--envelope arguments to Monte Carlo studies using state--of--the-art MHD models of the heliosphere. Contrary to the usual lore that CR are affected by the solar wind only below energies of a few GeV, simple arguments prove that the heliosphere can still influence CR arrival directions at the observed level of $10^{-4}$. The most intuitive argument for the effect of the heliosphere is the apparent alignment of the large--scale excess (and a particular small--scale feature) with the heliotail direction in CR anisotropy maps below 100~TeV. However, this alignment is not that clearly reproduced in the numerical models. We therefore turn to other possible explanations for the small--scale anisotropies, in particular considering effects that are located well outside the heliosphere.

%--------------------------------------------------------------------------------------------------------------------------------
%--------------------------------------------------------------------------------------------------------------------------------

\subsection{Non-Uniform Pitch--Angle Diffusion}

It is possible to compute the pitch--angle distribution from quasi--linear theory (see Sec.~\ref{sec:StandardPicture}), given a specific turbulence model. For example, it was shown~\cite{Malkov:2010yq} that an anisotropy at the source can be sustained in the presence of a non--uniform pitch--angle diffusion coefficient $D_{\mu\mu}$, i.e.\ $D_{\mu\mu} / (1- \mu^2) \neq \text{const.}$ (cf.\ Eq.~(\ref{eq:pitch})), that strongly peaks close to $\mu=1$ (also around $\mu=-1$ and $0$). Such a peak results, for instance, for resonant interaction of CRs with Alfv\'enic Goldreich--Sridhar~\cite{Goldreich:1994zz} turbulence. The evolution of the pitch--angle distribution is then described by the usual Fokker--Planck equation in $\mu$ and assuming the narrow feature to be relatively weak, it can be solved with a perturbative approach. The result is a pitch--angle distribution that is in fact exhibiting a very narrow feature close to $\mu = 1$, in width and amplitude in agreement with region A (see Sec.~\ref{sec:Observation} above). The angular size of the small--scale feature and its relative amplitude (with respect to the large--scale anisotropy) are set by the same parameter $l$, the maximum wavelength, that the CRs interact with. Requiring stability of the narrow feature in pitch--angle with respect to self--excited waves allows inferring a maximum energy of the feature of $10 \, \text{TeV}$ (for a median energy of $1\, \text{TeV}$) that is consistent with the cut--off energy observed by Milagro in region A~\cite{Abdo:2008kr}. 

Connecting all three observables (angular size, relative amplitude and cut--off energy) with one physical scale obviously makes this model very attractive. However, generically the narrow feature would be expected to coincide with the minimum of the large--scale distribution which is {\it not} observed. An additional source of a large--scale CR gradient would not help because in anisotropic diffusion the maximum/minimum of the large scale distribution are always aligned with the local magnetic field direction (at least in the ensemble average, see Sec.~\ref{subsec:MagneticTurbulence} below). Furthermore, with observational evidence that the small--scale anisotropies are not composed of one or a few isolated narrow features, but that there is structure on all angular scales, anisotropic turbulence is probably not the last word on the origin of the small--scale anisotropies.

Non--uniform pitch--angle diffusion can also modify the large--scale anisotropy from a pure dipole, which can be seen by considering the higher--order terms of the Legendre series in $\mu$ for the anisotropy. A detailed study~\cite{Giacinti:2016tld} has been carried out, computing in quasi--linear theory~\cite{1966ApJ...146..480J,1966PhFl....9.2377K,1967PhFl...10.2620H,1970ApJ...162.1049H} the anisotropy (necessarily axisymmetric around the background field) resulting from pitch--angle scattering along a magnetic flux tube containing the observer. Various turbulence models were considered: anisotropic Goldreich--Sridhar~\cite{Goldreich:1994zz} with different parameterizations of the anisotropy and isotropic fast--mode turbulence, in both cases with two different resonance functions. For a large set of Alfv\'en speeds and ratios of particle gyroadius to outer scale of turbulence (a proxy for the particle rigidity) the large--scale anisotropy is {\it not} dipolar but has much broader maxima and minima (along the magnetic field direction) and wider regions of zero relative CR intensity (approximately perpendicular to the magnetic field direction). This is primarily due to a peak of the scattering rate $D_{\mu\mu} / (1- \mu^2)$ around $\mu \approx 0$. It was further found that Goldreich--Sridhar turbulence with a particular parameterization of the power spectrum~\cite{Yan:2002qm} and with a resonance function that takes into account focussing and defocussing of the local magnetic field results in the best fit to IceTop data~\cite{Aartsen:2012ma}, able to reproduce the narrowing of the observed large--scale deficit from $2 \, \text{PeV}$ to $400 \, \text{TeV}$.

%--------------------------------------------------------------------------------------------------------------------------------
%--------------------------------------------------------------------------------------------------------------------------------

\subsection{Non-Diffusive Galactic Cosmic Ray Transport}
\label{subsec:NonDiffusive}

Non-diffusive transport of CRs over distances much larger than the size of the heliosphere was first considered by Ref.~\cite{Salvati:2008dx} as an explanation of the excess region A. The authors entertained the idea that a small--scale excess of CRs could originate from a nearby supernova remnant, possibly the one associated with the Geminga pulsar, that would stream along magnetic flux tubes. The close proximity of the observed local regular magnetic field with the direction of region A would support this idea. However, scatter--free particle streaming on distances of the order of 100~pc is unfeasible, as was pointed out in Ref.~\cite{Drury:2008ns}. Rather, the observed anisotropy should be that of a large--scale dipole oriented along the magnetic field direction as discussed in Sec.~\ref{subsec:Comparison}.

If pitch--angle scattering were not the dominant transport process, CR would stream along magnetic flux tubes and the pitch--angle would change while conserving the first adiabatic invariant $\propto p_\perp^2/B$ if changes in the field were small over the size of a gyroradius. If a point source was situated behind a magnetic mirror, the decrease in the field experienced by CRs passing through the mirror and streaming away would lead to a focussing of CRs into a cone around the magnetic field direction~\cite{Drury:2008ns}. If this was the source of region A, its alignment with the heliotail would be purely coincidental. Over long enough timescales, such a phase--space configuration would generate turbulence by the streaming instability~\cite{Kulsrud:1969zz}, so for it to survive, the source would need to be relatively close to the observer (unless there is strong Landau or ion--neutral damping). 

Other magnetic field configurations can also distort the image of near-by sources. The authors of Ref.~\cite{Battaner:2010bd} discuss the effect of magnetic lenses that consist of axial symmetric magnetic field loops. In principle, these configurations can make anisotropies appear larger or smaller, depending on the orientation of the magnetic field. Analogous to magnetic mirrors, these configurations can only have an effect if they appear within the scattering length from turbulent magnetic fields. In fact, turbulence itself is expected to generate magnetic mirrors and lenses close to the observer and can distort the image of the large--anisotropy. We will return to this point in the next section.

A setup~\cite{Harding:2015pna} that combines elongated regions of purely coherent and regions of purely turbulent magnetic fields can be considered a variation on this theme. (The observational evidence for the existence of such structures though is missing.) While in the purely turbulent region, transport is considered diffusive, particles can leak into the purely coherent field regions and then propagate essentially ballistically. As CRs are not scattering back out of the purely coherent field regions, this leads to enhanced CR flux with respect to the purely turbulent field regions. The authors of Ref.~\cite{Harding:2015pna} argue that a two--dimensional slice through their three-dimensional simulation volume can be mapped to the sky observed from Earth, but this seems far from obvious, given that the size of a putative detector will most likely be much smaller than the cross--sectional area of any one coherent field region. Also, the stability of such a purely coherent field region with respect to the CR streaming instability needs exploring.

Another (trivial) form of non--diffusive CR transport is the case of neutral messengers. The authors of Ref.~\cite{Drury:2008ns} consider the possibility that the small--scale features with relative excess of order $10^{-4}$ could be due to secondary neutrons produced in nearby clumps of matter. However, while the energy--dependence $\propto E^{-1}$ of the decay rate could explain the relatively hard spectrum observed for features like region A or B, the needed intensity enhancement compared to the ambient medium is of the order of $10^6$. Such a high--density clump could not be a caustic produced by gravitational focussing of the Sun and would have also likely been observed in gas surveys of the local interstellar medium. On the other hand, residual $\gamma$-rays of strong Galactic sources, e.g.\ the Crab pulsar wind nebula or the Cygnus region are clearly visible in many TeV anisotropy maps from extended air-shower observatories and have to be treated with some care.

%--------------------------------------------------------------------------------------------------------------------------------
%--------------------------------------------------------------------------------------------------------------------------------

\subsection{Small--Scale Anisotropy induced by Magnetic Turbulence}
\label{subsec:MagneticTurbulence}

In the previous section we have already encountered the idea that special electromagnetic field configuration (mirrors and lenses) at a distance closer than the effective scattering length of magnetic turbulence could lead to distortions of large scale anisotropies. The heliosphere was considered to be the source of these distortions. Some alignments of small--scale structure with the orientation of the heliotail seem to support this idea. Turbulent magnetic fields, which are the source of CR diffusion in the first place, have also been considered as a possible source of small--scale anisotropies, as first proposed in the seminal paper by Giacinti \& Sigl~\cite{Giacinti:2011mz}.

In general, the small--scale magnetic field that is determining the phase--space density is unknown. Therefore, the phase--space density can only be predicted as an average over a statistical ensemble of random small--scale magnetic fields, characterized for instance by its power spectrum. It is usually assumed that CRs that have diffused over distances larger than the coherence length of the turbulent magnetic field have been subject to many different field configurations, such that by ergodicity arguments their phase--space density is close to the average over the ensemble of turbulent magnetic fields, denoted in the following by $\langle \mathellipsis \rangle$. This is however not quite correct in that ergodicity only guarantees that the {\it spatially averaged} phase--space density is close to the ensemble average. Therefore, at any one point the phase--space density $f = \langle f \rangle + \delta f$ can deviate from the ensemble average, $\langle f \rangle$. The angular variations of $\delta f$ also do not necessarily follow those of $\langle f \rangle$. This is the central idea for explaining small--scale anisotropies by magnetic turbulence. As we will discuss below, the anisotropy pattern observed at one particular position is a reflection of the particular realisation of the {\it local} turbulent magnetic field and has been generated over the last few scattering times before observation.

According to Liouville's theorem we can relate the local (i.e.\ ${\bf r}={\bf r}_\oplus$) phase--space density $f(t,{\bf r}_\oplus,{\bf p})$ to the contribution back--tracked along CR trajectories to an arbitrary time $t-T$, 
\begin{equation}
f(t,{\bf r}_\oplus,{\bf p}) = f(t-T,{\bf r}(t-T),{\bf p}(t-T))\,.
\end{equation}
This technique was already applied in the studies presented in the previous section discussing the effect of the heliosphere. In that case, the natural choice of the back--tracking time $T$ was provided by the time-scales required to leave the influence of the heliosphere and the {\it external} anisotropy was considered a large scale anisotropy of the form
\begin{equation}
4\pi f(t-T,{\bf r}(t-T),{\bf p}(t-T)) \simeq \phi + ({\bf r}_i(t-T)-{\bf r}_\oplus)\!\cdot\!\nabla \phi-3\widehat{\bf p}_i(t-T)\!\cdot\!{\bf K}\!\cdot\!\nabla \phi\,.
\end{equation} 
In principle, the same technique can be applied to the case of small--scale anisotropies from local turbulence. The appropriate choice of the back--tracking time $T$ is here the effective scattering time-scale $1/\nu$ of magnetic turbulence. 

\begin{figure}[tb]
\centering
\includegraphics[width=0.3\linewidth]{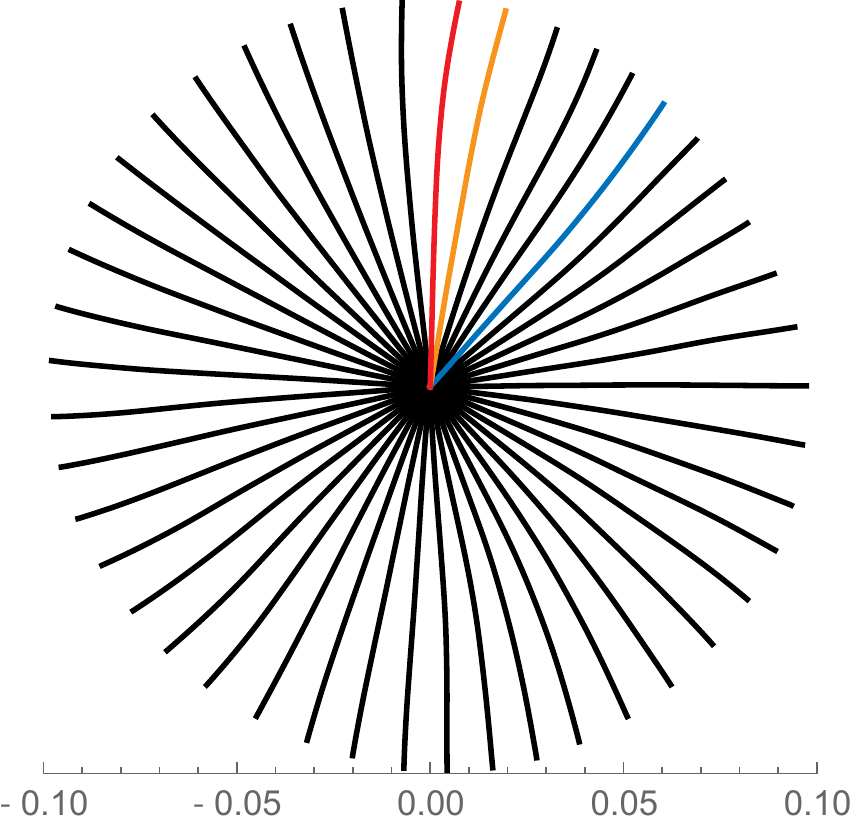}\hspace{0.04\linewidth}\includegraphics[width=0.3\linewidth]{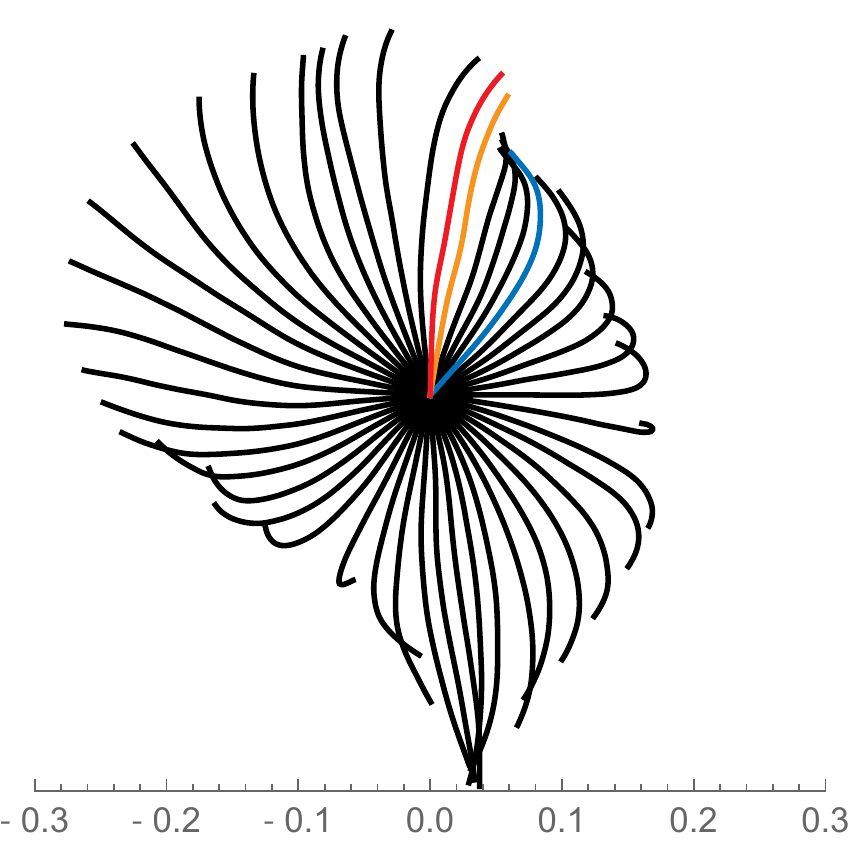}\hspace{0.04\linewidth}\includegraphics[width=0.3\linewidth]{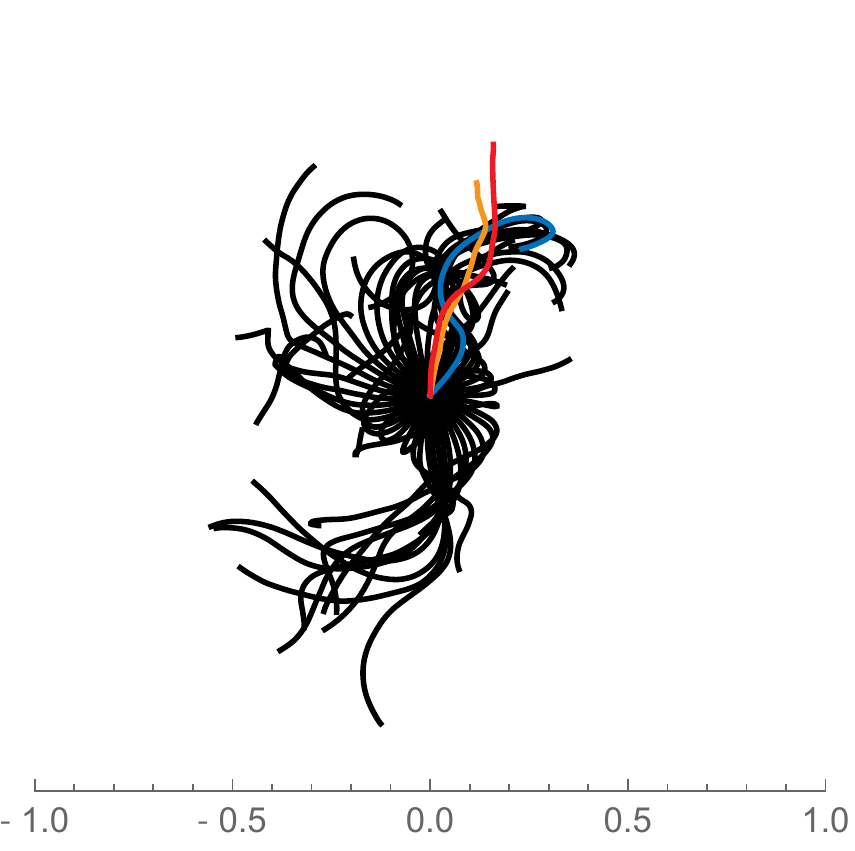}
\caption[]{Sketch of particle back--tracking in a turbulent magnetic field. For simplicity, we do not consider the presence of a regular magnetic field. In that case, the CR back--tracking ``flow'' starts ballistic $\nu T\ll 1$ (left panel), remains laminar for $\nu T\simeq 1$ (middle panel), and starts to become turbulent for $\nu T\gg 1$ (right panel).}
\label{fig:backtracking}
\end{figure}

The formation of small--scale anisotropies can be understood in the following thought experiment: Assume a homogeneous, but anisotropic dipolar state~\cite{Ahlers:2013ima}. This means the phase--space density is the same at every point in space, but its angular dependence is $\propto (\widehat{\bf p} \cdot {\bf \Phi})$. We also assume, for simplicity, that the magnetic field is dominated by turbulence. We now back--track particles from the observer for a fixed amount of time $T$ (see Fig.~\ref{fig:backtracking}) and exploit Liouville's theorem to compute the anisotropy map from the set of trajectories and the assumed distribution. This is equivalent of preparing the system into the initial state of the assumed distribution and then observing the anisotropy a time $T$ later at the position of the observer. 

At early times, $T\nu \ll 1$, (cf.\ middle panel of Fig.~\ref{fig:backtracking}) the back--tracked particles will have travelled away from the observer only ballistically and the observed sky map will be the same as the assumed dipole. However, as $T\nu$ becomes larger (cf.\ middle panel of Fig.~\ref{fig:backtracking}; the details depend on the scales and strength of the turbulent field), the anisotropy map will show the first small--scale structures: Particles will have travelled sufficiently far, that particles sent out back--tracking into very different directions will have experienced different magnetic fields and their momenta will lose correlation. (Compare the red with the blue trajectory in the middle panel of Fig.~\ref{fig:backtracking}.) However, neighboring CRs (cf.\ the red and orange trajectory) will have experienced similar magnetic configurations and their moment correlate over much larger times. On average, the strength of correlation is expected to decrease with the opening angle of initial momenta. This argument indicates the the power of small--scale anisotropy induced by turbulence should experience a hierarchical structure~\cite{Ahlers:2013ima}. 

%--------------------------------------------------------------------------------------------------------------------------------
%--------------------------------------------------------------------------------------------------------------------------------

\paragraph{Simulation}

\begin{figure}[t]\centering
\includegraphics[width=\linewidth]{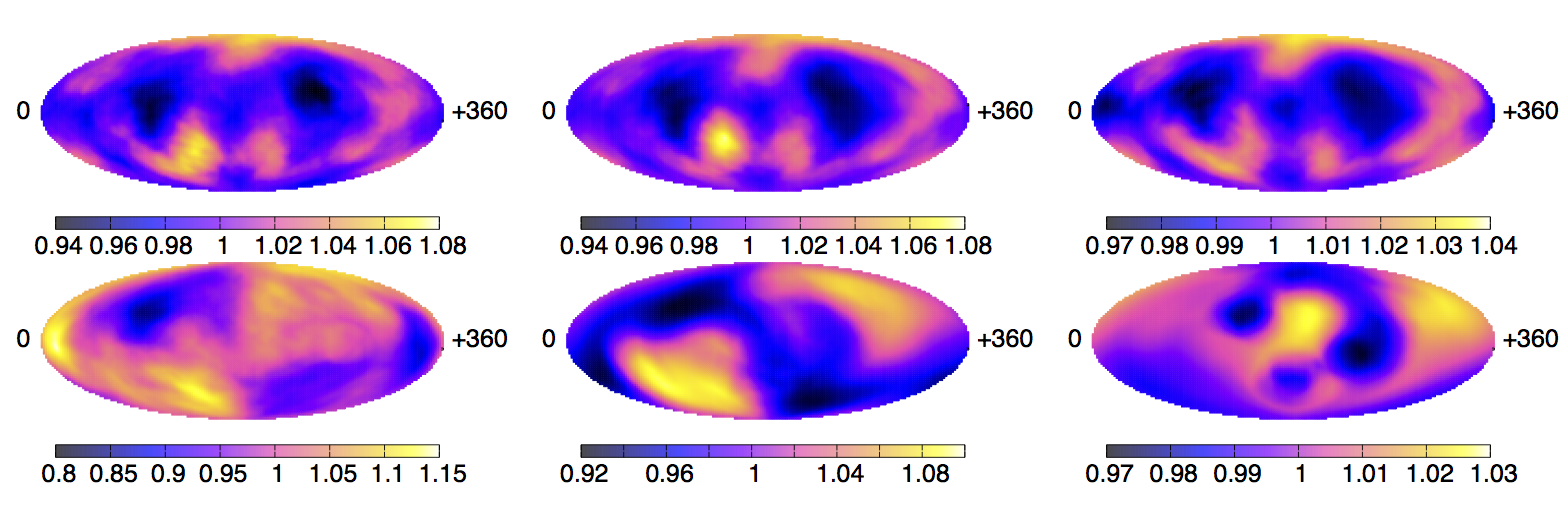}
\caption[]{Simulated relative intensity maps after subtraction of the dipole. In the left, middle and right column, the initial gradient condition is enforced on spheres of radii of $100$, $50$ and $25 \, \text{pc}$, respectively; the top and bottom row are for particles of rigidities of $10$ and $50 \, \text{PV}$, respectively. Note that the assumed gradient is rather large and that rescaling also to TeV energies would bring the amplitude down to $\sim 10^{-3}$. From Ref.~\cite{Giacinti:2011mz}.}\label{fig:GiacintiSigl}
\end{figure}

The authors of Ref.~\cite{Giacinti:2011mz} supported their idea with numerical simulations of CR maps computed by backtracking CRs through a particular realization of a turbulent magnetic field, with an outer scale of $150 \, \text{pc}$ and $\sqrt{\langle B^2 \rangle} = 4\, \mu\text{G}$. In Fig.~\ref{fig:GiacintiSigl} we show their simulated relative intensity maps for different rigidities and different initial conditions. It can be clearly seen that the transport of CRs through a particular realization of the turbulent magnetic field generates power on small scales even though the distribution was isotropic initially, but with a CR gradient.  Other numerical studies~\cite{Giacinti:2011mz,Ahlers:2015dwa,Pohl:2015fdp,Lopez-Barquero:2015qpa} find similar sky maps although it can be difficult to compare them because of the widely varying assumptions.

While the basic idea of simulating the small--scale anisotropy from turbulence via back--tracking is rather straight--forward, a variety of prescriptions have been adopted for the back--tracking time $T$ and corresponding initial condition: The initial phase--space density can be assumed at a fixed distance from the observer, and the back--tracking of particles is stopped as soon as they cross this sphere the first time~\cite{Giacinti:2011mz,Lopez-Barquero:2015qpa}. Note that in general the different particles cross this sphere at different times. Alternatively, the backtracking can be stopped after a fixed propagation time~\cite{Ahlers:2015dwa}. After large propagation times, deep in the diffusive regime, the particles will be approximately drawn from a multivariate Gaussian centered on the observer. We here take the view that, conceptually, the latter option is more transparent, as this corresponds to the forward propagation of particles for a fixed amount of time from a prepared initial distribution. Also, the form of the initial distribution function can be different, e.g.\ inhomogeneous, isotropic~\cite{Giacinti:2011mz,Ahlers:2015dwa} or homogeneous, anisotropic~\cite{Lopez-Barquero:2015qpa} (cf.\ also the thought experiment in Ref.~\cite{Ahlers:2013ima}).

A convenient way to quantify the statistical properties of the sky maps is via the angular power spectrum of the CR phase--space density $f(t,{\bf r},{\bf p})$ defined as
 \begin{equation}\label{eq:Celldef}
C_\ell = \frac{1}{4\pi}\int {\rm d}\widehat{\bf p}_1 \int {\rm d}\widehat{\bf p}_2P_\ell(\widehat{\bf p}_1 \!\cdot\!\widehat{\bf p}_2) f_1f_2\,,
\end{equation}
where $\widehat{\bf p}_{i}$ denotes a unit vector of ${\bf p}_{i}$ and $P_\ell$ are Legendre polynomials of degree $\ell$. This definition is equivalent to the one in Eq.~(\ref{eq:Cl}). Here and in the following we use the abbreviation $f_i = f(t,{\bf r}_\oplus,{\bf p}_i)$ for the local phase--space density. Note that in the standard diffusion theory (cf.\ Sec.~\ref{sec:StandardPicture}), only $\langle f \rangle$ can be predicted and therefore only $\langle f_1 \rangle \langle f_2 \rangle$ contributes to the angular power spectrum. Instead we here study the average angular power spectrum computed from $\langle f_1 f_2 \rangle$ which is larger than $\langle f_1 \rangle \langle f_2 \rangle$ if $f_1$ and $f_2$ are correlated. Scattering by local turbulence is providing just this correlation.

According to Liouville's theorem we can relate the local (i.e.\ ${\bf r}={\bf r}_\oplus$) phase--space density $f_i = f(t,{\bf r}_\oplus,{\bf p}_i)$ to the contribution backtracked along CR trajectories to an arbitrary time, 
\begin{equation}\label{eq:backtrack}
4\pi f_i \simeq 4\pi\delta{f}(t-T,{\bf r}_i(t-T),{\bf p}_i(t-T))  + \phi 
+ ({\bf r}_i(t-T)-{\bf r}_\oplus)\!\cdot\!\nabla \phi-3\widehat{\bf p}_i(t-T)\!\cdot\!{\bf K}\!\cdot\!\nabla \phi\,,
\end{equation}
where $\phi$ and $\nabla \phi$ denotes the local CR density and gradient and ${\bf r}_i(t-T)$ and ${\bf p}_i(t-T)$ are the position and momentum of a CR (that is at position ${\bf r}_i(t) = {\bf r}_\oplus$ and $\widehat{\bf p}_i(t)=\widehat{\bf p}_i$ at time $t$) after backtracking for a time $T$. Now, in the limit of large $T$ the right--hand side of Eq.~(\ref{eq:backtrack}) is dominated by the third term scaling with the position of the particle. Also, for two momenta ${\bf p}_1\neq{\bf p}_2$ we can assume that the ensemble--average of fluctuations are uncorrelated, $\langle\delta{f}_1(t-T)\delta{f}_2(t-T)\rangle \simeq 0$, for sufficiently large backtracking times when the CR trajectories eventually separate. 
In the degenerate case ${\bf p}_1={\bf p}_2$ the two backtracked CR trajectories stay correlated over arbitrarily long backtracking times. It is sufficient for the later discussion to assume that $\langle (\delta f(t-T))^2\rangle$ remains finite. We can then express the multipole spectrum of the ensemble--averaged relative intensity as the limit
\begin{equation}\label{eq:Cell}
\frac{1}{4\pi}{\langle C_\ell\rangle}\simeq \int \frac{{\rm d}\widehat{\bf p}_1}{4\pi} \int \frac{{\rm d}\widehat{\bf p}_2}{4\pi}P_\ell(\widehat{\bf p}_1\!\cdot\!\widehat{\bf p}_2)\lim_{T\to\infty}\langle {d}_{1 i}(t-T){d}_{2 j}(t-T)\rangle \frac{\partial_i n\partial_j n}{n^2}\,,
\end{equation}
$\partial_i$ being shorthand for $\partial/\partial x_i$ and ${\bf d}_{1} = {\bf r}_1 - {\bf r}_\oplus$, etc.

A few comments are in order. Firstly, the ensemble average of the two back--tracked trajectories can be expressed as
\begin{equation}\label{eq:split}
\langle {d}_{1 i}(t-T){d}_{2 j}(t-T)\rangle  = \langle {d}_{1 i}(t-T)\rangle\langle{d}_{2 j}(-T)\rangle + \langle \delta{d}_{1 i}(t-T)\delta{d}_{2 j}(t-T)\rangle\,,
\end{equation}
with $\delta{d}_{1i}(t-T) = {d}_{1i}(t-T) - \langle{d}_{1i}(t-T) \rangle$, etc. The first term in Eq.~(\ref{eq:split}) corresponds to the contribution from standard diffusion theory. In general, it is given in the form $C_\ell = \sum_{m=-\ell}^\ell|\langle a_{\ell m}\rangle|^2/(2\ell+1)$ and expected to be dominated by the diffusive dipole. The additional contribution is expected from the variance $\langle \delta{d}_{1 i}(t-T)\delta{d}_{2 j}(t-T)\rangle$ of the particle trajectories.

Secondly, the $\ell\geq1$ multipole spectrum is generated through {\it relative} diffusion. As discussed in Ref.~\cite{Ahlers:2015dwa}, the sum of multipoles $\ell\geq1$ is given by
\begin{equation}\label{eq:id1}
\frac{1}{4\pi}\sum_{\ell\geq1}(2\ell+1){\langle C_\ell\rangle}(T) \simeq 2T\widetilde{K}^{\rm s}_{ij}\frac{\partial_i n\partial_j n}{n^2}\,,
\end{equation}
where $\widetilde{\bf K}^{\rm s}$ is the symmetric part of the relative diffusion tensor, i.e.\ diffusion of particle separations $\Delta{\bf r}_{12} \equiv {\bf r}_1-{\bf r}_2$,
\begin{equation}\label{eq:Krel}
\widetilde{K}^{\rm s}_{ij} =\int\frac{{\rm d}\widehat{\bf p}_1}{4\pi}\int\frac{{\rm d}\widehat{\bf p}_2}{4\pi} \lim_{T\to\infty} \frac{1}{4T}\big\langle\Delta{r}_{12i}(t-T)\Delta{r}_{12j}(t-T)\big\rangle\,.
\end{equation}
For uncorrelated particle trajectories this expression reduces to the normal diffusion tensor. However, particle trajectories with small relative opening angle will follow similar trajectories and the relative contribution (\ref{eq:Krel}) remains small over long time scales. Note that the multipoles in Eq.~(\ref{eq:Cell}) are expected to be finite in the limit of large backtracking times since particle trajectories with arbitrarily small opening angles will eventually become uncorrelated, $\langle {r}_{1i}(t-T){r}_{2j}(t-T)\rangle \to 0$. 

Lastly, only for a statistically isotropic, Gaussian random field does the angular power spectrum contain all statistically meaningful information. Given that the anisotropy maps are likely neither, we would therefore miss information (contained in the individual $a_{\ell m}$'s) when comparing the power spectrum of simulated maps with measured ones. In addition, comparing simulated $C_\ell$ to measured ones with limited sky coverage introduces additional problems, e.g.\ because for cut skies the $C_\ell$'s are correlated and might not be faithfully reconstructed under the assumption of statistically isotropic, Gaussian maps.

In the following we illustrate the general trends in the computed angular power spectrum Eq.~(\ref{eq:Cell}) based on the results of Ref.~\cite{Ahlers:2015dwa} (cf. also Figs.~5 and 6 of Ref.~\cite{Lopez-Barquero:2015qpa}). The magnetic field is the superposition of a regular field ${\rm B_0}$ and a turbulent field ${\delta {\bf B}}$ with $\sigma^2 \equiv \langle {\delta {\bf B}}^2 \rangle / B_0^2 = 1$. The adopted power spectrum is $\propto  (1+(k L_c)^\gamma)^{-1}$~\cite{Giacalone:1999} with $2 \pi/k \in [\lambda_{\text{min}}, \lambda_{\text{max}}]$ and a coherence length $L_c$. With $L_c \simeq 100 \, \text{pc}$~\cite{Haverkorn:2008tb}, the gyroradius of $0.1 \, L_c$ corresponds to rigidities of $\sim 30 \, \text{PV}$ in a $3 \, \mu\text{G}$ field~\cite{Beck:2008ty}.

Figure~\ref{fig:power100} shows the power spectrum of Eq.~(\ref{eq:Cell}) at different backtracking times determined via a multipole expansion of simulated relative intensity sky maps~\cite{Ahlers:2015dwa}. The results are shown for the case of a CR gradient parallel (solid lines) and perpendicular (dotted lines) to ${\bf B}_0$. We first focus on the dipole result with $\ell=1$. The magenta lines show the contribution of the standard dipole anisotropy, which can be reproduced via Eq.~(\ref{eq:Cell}) after replacing $\langle{d}_{1i}{d}_{2j}\rangle \to \langle{d}_{1i}\rangle\langle{d}_{2j}\rangle$. On the other hand, the evolution of the ensemble--averaged dipole of Eq.~(\ref{eq:Cell}) is shown via black lines. It can be seen that the standard dipole estimate is in general smaller, as expected from our earlier arguments. In the case of perpendicular diffusion (dotted lines), the relative difference is about one order of magnitude for the particular set of model parameters used in Ref.~\cite{Ahlers:2015dwa}.

\begin{figure}[t]\centering
\includegraphics[width=0.5\linewidth]{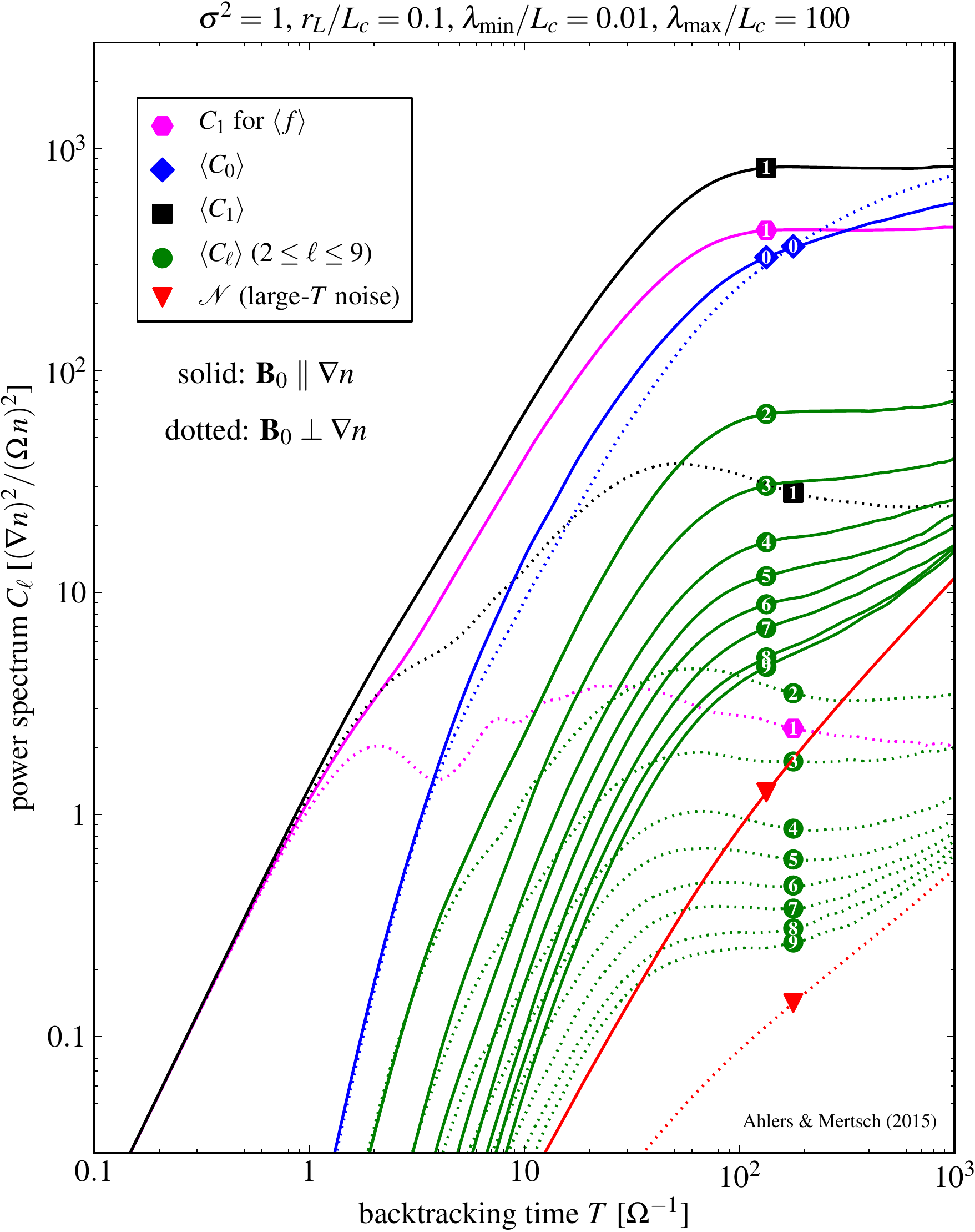}
\caption[]{The evolution of the ensemble--averaged power spectrum (\ref{eq:Cell}) for a CR gradient parallel (solid lines) and perpendicular (dotted lines) to the regular magnetic field for a turbulence model discussed in the main text (from Ref.~\cite{Ahlers:2015dwa}). The lines show the dipole $\langle C_1\rangle$ (black), monopole $\langle C_0\rangle$ (blue), medium-$\ell$ multipoles (green), asymptotic noise level (\ref{eq:noise}) (red), and the dipole prediction of standard diffusion (magenta).}
\label{fig:evolution}
\end{figure}
 
The green lines in Fig.~\ref{fig:power100} show the multipole power for $2\leq\ell\leq9$. Note that at large times the high-$\ell$ components in the multipole expansion are dominated by simulation noise, due to the finite number of trajectories sampling the sky maps. The noise level can be estimated for large backtracking times $T$ as pixel shot noise~\cite{Ahlers:2015dwa}
\begin{equation}\label{eq:noise}
\mathcal{N}
 \simeq \frac{4\pi}{N_{\rm pix}}2TK^{\rm s}_{ij}\frac{\partial_in\partial_jn}{n^2}\,,
\end{equation}
and is indicated as red lines in Fig.~\ref{fig:evolution}. The noise level clearly influences the level of high-$\ell$ multipoles (green lines) in the map. The true power spectrum can then be estimated via $\widehat{C}_\ell = \langle C_\ell\rangle-\mathcal{N}$ and the variance (excluding cosmic variance) is approximated by ${\rm var}(\widehat{C}_\ell) \simeq 2\mathcal{N}^2/(2\ell+1)$, see {\it e.g.}~\cite{Campbell:2014mpa}. The noise--corrected power spectrum is shown in Fig.~\ref{fig:power100} at a backtracking time $\Omega T=100$, corresponding to the time scale of the transition into the diffusion regime (cf.~Fig.~\ref{fig:evolution}). We normalize the estimators $\widehat{C}_\ell$ to the dipole $\widehat{C}_1$.

\begin{figure}[t]\centering
\includegraphics[width=0.6\linewidth]{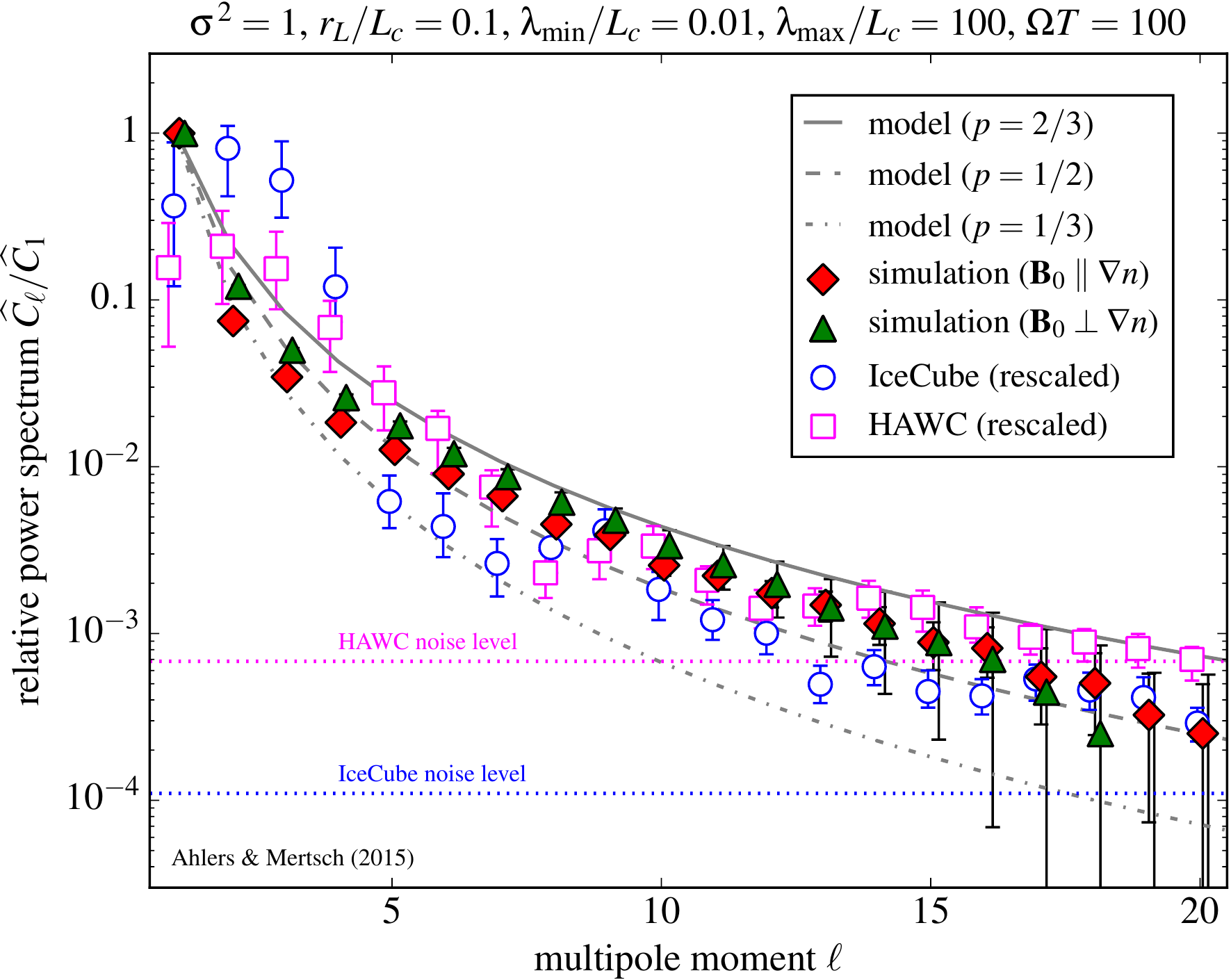}
\caption[]{The estimated power spectrum, $\widehat{C}_\ell = \langle C_\ell\rangle -\mathcal{N}$ (normalized to $\widehat{C}_1$), for parallel (filled red diamonds) and perpendicular (filled green triangles) CR gradient for the simulation shown in Fig.~\ref{fig:evolution} at a backtracking time $\Omega T =100$ (from Ref.~\cite{Ahlers:2015dwa}). The power spectra derived by IceCube~\citep{Aartsen:2013lla} (blue open circles) and  HAWC~\citep{Abeysekara:2014sna} (magenta open squares) are rescaled for a better comparison of the high-$\ell$ spectrum. The different noise level of the data is estimated via Eq.~(\ref{eq:noiseesimtate}) and indicated as dotted lines. The three gray lines correspond to the prediction of a relative scattering term $\nu_r(x) \propto (1-x)^p$ in Eq.~(\ref{eq:finalCl}) for three different values of $p$.}
\label{fig:power100}
\end{figure}

Figure~\ref{fig:power100} also shows the re-normalized power spectrum of IceCube and HAWC, that we already showed in Fig.~\ref{fig:power}. As discussed in Sec.~(\ref{subsec:HarmonicAnalysis}), the sensitivity of experimental data to high-$\ell$ multipoles is also limited by shot noise, estimated by Eq.~(\ref{eq:noiseesimtate}). This is indicated as dotted lines in the plot. After rescaling, the simulation agrees well with the high-$\ell$ IceCube and HAWC data. Note that, as pointed out earlier, the limited FOV of the observatories introduces additional systematic errors, in particular for low-$\ell$ multipoles, which are not included in the error bars of Fig.~\ref{fig:power100}.

The general trend is that of a falling angular power spectrum with a marked convex shape, see Fig.~\ref{fig:power100}. Mathematically, the formation of the angular power spectrum (cf.\ the thought experiment outlined above) can be described as a system of differential equations for the $\widehat{C}_\ell$ with mixing between the different $\widehat{C}_\ell$ induced by the different rotations of particle momenta at different positions. It can be shown under simplifying assumptions~\cite{Ahlers:2013ima,Ahlers:2015dwa} that the angular power spectrum reaches an asymptotic shape with the type of convex shape reproduced in the numerical studies.

It has been claimed~\cite{Lopez-Barquero:2015qpa} that the angular power spectrum is observed to be energy--dependent, with the spectrum being markedly flatter at lower energies. Given that the gyroradius is at the inner turbulence scale in two of the three setups of Ref.~\cite{Lopez-Barquero:2015qpa} this could however also be an effect of the limited dynamical range of the magnetic field used.

%--------------------------------------------------------------------------------------------------------------------------------
%--------------------------------------------------------------------------------------------------------------------------------

\paragraph{Generalized BGK Formalism}
\label{sec:BGK}

\begin{figure}[t]\centering
\includegraphics[width=0.4\linewidth]{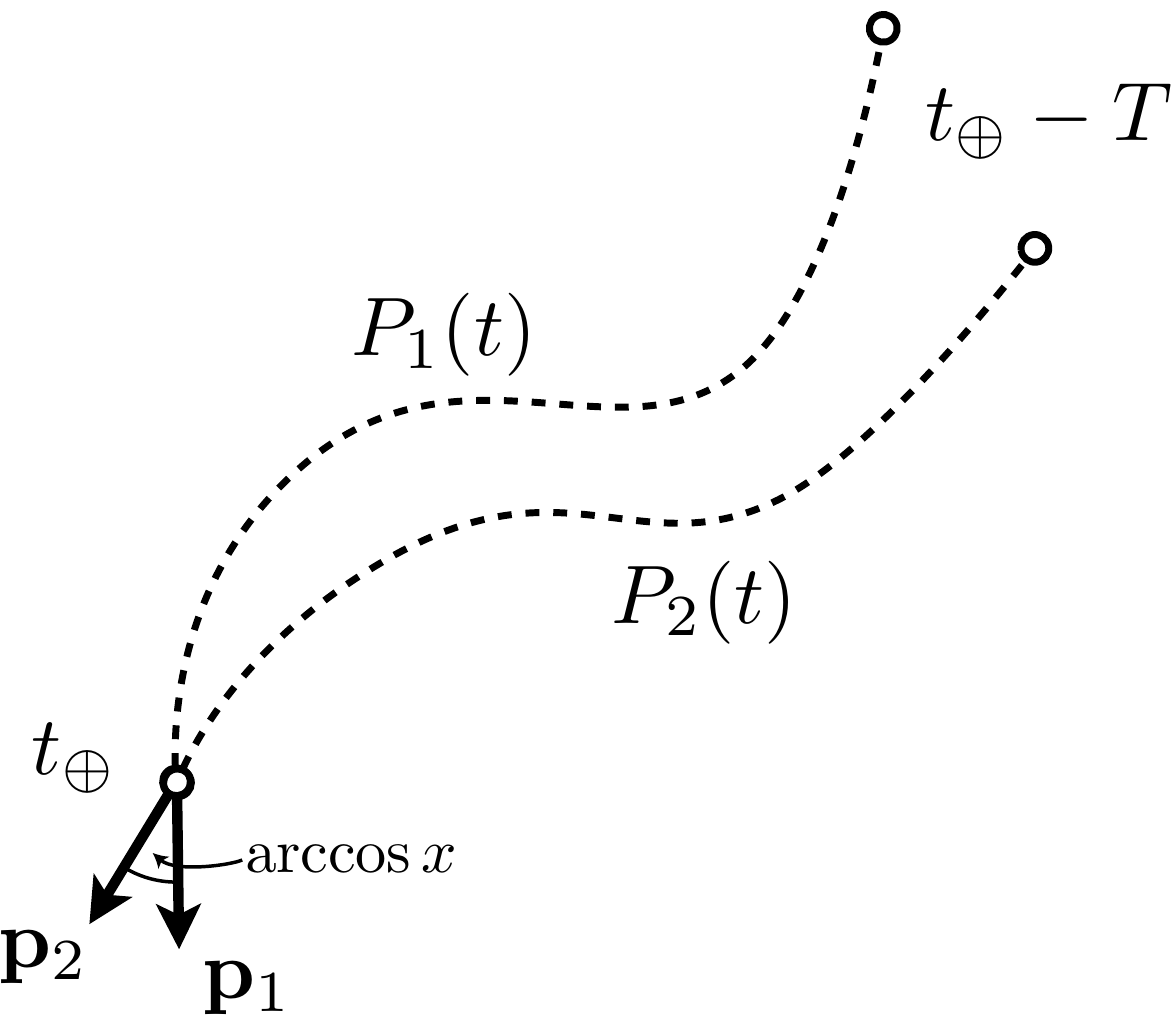}
\caption[]{Back--tracked CR trajectories for different CR momenta at $t=t_\oplus$. For similar arrival directions and small back--tracking times $T$, the CRs experience correlated magnetic environments that result in similar trajectories. A strong correlation is expected to prevail over longer back--tracking times, the closer the observed momenta.}\label{fig:paths}
\end{figure}

In Section (\ref{sec:StandardPicture}) we outlined a derivation of the effective scattering term of the ensemble--averaged collisionless Boltzmann equation. We can apply an analogous formalism to the evolution of the ensemble--averaged product $\langle f_1f_2\rangle$ following
\begin{equation}\label{eq:f12average}
\partial_t\langle f_1f_2\rangle = -\langle \mathcal{L}_1+\mathcal{L}_2\rangle \langle f_1f_2\rangle -\langle(\delta\mathcal{L}_1+\delta\mathcal{L}_2)\mathcal{F}_{12}\rangle\,.
\end{equation}
Here, we have defined the quantity $\mathcal{F}_{12} = f_1f_2 -\langle f_1 f_2\rangle$, which vanishes in the ensemble average, and introduced the Liouville operators $\mathcal{L}_{1}$ and $\mathcal{L}_2$ that are identical to the definition of Eq.~(\ref{eq:LOperator}), but only act on the phase--space densities $f_1$ or $f_2$, respectively.
From (\ref{eq:f12average}) and $\partial_t(f_1f_2) = -(\mathcal{L}_1+\mathcal{L}_2)f_1f_2$ we can derive the evolution of the quantity $\mathcal{F}_{12}$ as
\begin{equation}
\partial_t\mathcal{F}_{12} = -(\mathcal{L}_1+\mathcal{L}_2)\mathcal{F}_{12}-(\delta\mathcal{L}_1+\delta\mathcal{L}_2)\langle f_1 f_2\rangle + \langle(\delta\mathcal{L}_1+\delta\mathcal{L}_2)\mathcal{F}_{12}\rangle\,.
\end{equation}
Following the same line of arguments as in Section~(\ref{sec:StandardPicture}) we can then derive a leading-order solution for the scattering term in (\ref{eq:f12average}) as
\begin{equation}\label{eq:scattering12}
\langle(\delta\mathcal{L}_1+\delta\mathcal{L}_2)\mathcal{F}_{12}\rangle\simeq -\left\langle (\delta\mathcal{L}_1+\delta\mathcal{L}_2)\int_{-\infty}^t{\rm d}t'[(\delta\mathcal{L}_1+\delta\mathcal{L}_2)\langle f_1 f_2\rangle]_{P_1(t') \& P_2(t')}\right\rangle\,,
\end{equation}
where $[\cdot]_{P_1(t') \& P_2(t')}$ indicates that the quantity is evaluated along the two particle trajectories, as indicated in Fig.~\ref{fig:paths}.

Following Ref.~\cite{Ahlers:2015dwa}, we now look for the power spectrum of a stationary solution of Eq.~(\ref{eq:f12average}). The first term on the right hand side of Eq.~(\ref{eq:f12average}) consists of a global rotation (which does not contribute to the power spectrum) and a gradient term that is dominated by the $\langle f_1\rangle\langle f_2\rangle$ contribution,
\begin{equation}\label{eq:C1source}
(\widehat{\bf p}_1\!\cdot\!\nabla_1+\widehat{\bf p}_2\!\cdot\!\nabla_2)\langle f_1 f_2\rangle  \simeq-\frac{3}{(4\pi)^2}\widehat{\bf p}_1\!\cdot\!\nabla \phi\,\widehat{\bf p}_2\!\cdot\!{\bf K}\!\cdot\!\nabla \phi+ (1\leftrightarrow2)\,.
\end{equation}
The scattering term~(\ref{eq:scattering12}) is approximated by a BGK-type ansatz~\citep{Bhatnagar:1954zz} as in Eq.~(\ref{eq:BGK2}),
\begin{equation}\label{eq:BGK3}
\langle(\delta\mathcal{L}_1+\delta\mathcal{L}_2)\mathcal{F}_{12}\rangle \simeq  -\left[\nu_{\rm r}(x)\frac{{\bf L}_1^2+{\bf L}_2^2}{2} +\nu_{\rm c}(x){\bf J}^2\right]\langle f_1f_2\rangle\,,
\end{equation}
with $x=\widehat{\bf p}_1  \!\cdot\!\widehat{\bf p}_2$ and total angular momentum operator ${\bf J}={\bf L}_1+{\bf L}_2$. The BGK-type scattering rates, $\nu_{\rm r}$ and $\nu_{\rm c}$, correspond to the relative and correlated rotation of particle momenta, respectively, along the trajectories and are expected to have different dependencies on the initial opening angle $\arccos(x)$ (cf.~Fig.~\ref{fig:paths}). The correlation proportional to $\nu_{\rm r}(x)$ does not contribute to the stationary power spectrum as discussed in Ref.~\cite{Ahlers:2015dwa}. It can then be shown~\cite{Ahlers:2015dwa} that with ansatz (\ref{eq:BGK3}), the stationary average $C_\ell$ spectrum is given by
\begin{equation}\label{eq:finalCl}
C_\ell = \frac{3}{2}\frac{\mathcal{Q}_1}{\ell(\ell+1)}\int\limits_{-1}^1{\rm d}x\frac{x\,P_\ell(x)}{\nu_{\rm r}(x)}\,,
\end{equation}
where $\mathcal{Q}_1= {K}^{\rm s}_{ij}{\partial_i \phi\partial_j \phi}/(6\pi)$ corresponds to the effective dipole source term.

Cosmic rays observed from similar directions will separate relative late in back--tracking, as indicated in Fig.~\ref{fig:paths}. For identical momenta, ${\bf p}_1={\bf p}_2$, the relative scattering rate should vanish completely. This can be seen from the thought experiment that we discussed earlier~\citep{Ahlers:2013ima}, following the time evolution of a homogeneous and anisotropic initial phase--space density $f$ in a turbulent magnetic field. It can be shown~\citep{Ahlers:2013ima} that the phase--space density in this special setup obeys $\partial_t \int {\rm d}\widehat{\bf p} \langle f^2(t,{\bf r},{\bf p})\rangle = 0$ and this implies $\int {\rm d}\widehat{\bf p}\nu_{\rm r}(1){\bf L}^2\langle f^2(t,{\bf r},{\bf p})\rangle = 0$ and, therefore, $\nu_{\rm r}(1)=0$.

The $x$-dependence of the relative scattering rate $\nu_r(x)$ can be estimate via a simple geometrical argument. Two CRs back--tracked under a small opening angle will only slowly separate with a reduced relative velocity $\Delta v\sim \sqrt{2(1-x)}$ as long as both particles are experiencing a strongly correlated magnetic field (cf.~Fig.~\ref{fig:paths}). Locally, this can therefore be treated via the ansatz $\nu_{\rm r}(x) \propto \sqrt{1-x}$. However, to allow for a weaker or stronger $x$-dependence, Ref.~\citep{Ahlers:2013ima} considers the general form $\nu_{\rm r}(x) \propto (1-x)^p$, with $0<p<1$. Figure~\ref{fig:power100} shows examples of the power spectrum (\ref{eq:finalCl}) for the three cases $p=1/3$, $1/2$ and $2/3$.

%--------------------------------------------------------------------------------------------------------------------------------
%--------------------------------------------------------------------------------------------------------------------------------

\paragraph{Validity of Liouville's Theorem}

Recently, the validity of Liouville's theorem has been questioned in the context of the validity of back--tracking particles through turbulent magnetic fields~\cite{Lopez-Barquero:2015qpa,Lopez-Barquero:2016wnt}. Liouville's theorem states that under certain conditions the convective derivative of phase--space density vanishes, meaning that the phase--space density along any trajectory remains constant. (Together with the conservation of particle number, conservation of {\it density} is therefore also equivalent to conservation of phase space {\it volume} along any trajectory.) Liouville's theorem follows from the divergence theorem, which assumes conservation of particles, so absence of sources or sinks, and Hamilton's equations, meaning that the $p$--divergence of the force term vanishes, $\nabla_{{\bf p}}{\bf F} = 0$ which is equivalent to demanding conservative, differentiable forces. The latter in particular implies the absence of collisions. Specifically the Lorentz--force fulfills Liouville's theorem: While the force depends on the velocity, it is not $p$--divergent. Even small--scale turbulence does not change this: If there is a smallest turbulent scale, the magnetic field is always differentiable and so is the Lorentz--force.

In Ref.~\cite{Lopez-Barquero:2015qpa,Lopez-Barquero:2016wnt}, the authors numerically check whether the computed trajectories preserve the magnetic moment (also known as the first adiabatic invariant in plasma physics) and conclude that depending on conservation or non--conservation Liouville's theorem is satisfied or violated, respectively. However, the validity of Liouville's theorem is not equivalent to the conservation of the first adiabatic invariant. The pitch--angle scattering does not conserve the magnetic moment, as a particle must experience a force coherently over (many) gyrations before it ``scatters''. Cosmic rays in the situations in question (ISM, interplanetary medium) are a collisionless plasma, i.e.\ the collision length is much larger than the size of the system (the Galaxy, the solar system). The presence of a magnetic field does not change that. Cosmic rays resonantly interacting with a turbulent B-field, will ``scatter'', but this only refers to a continuous change of their momenta.

Trajectories of charged particles in a turbulent magnetic field exhibit a certain level of chaos: The position of a particle in phase space becomes very quickly very dependent on its initial position. This chaotic behavior does however not violate the determinism of Hamiltonian dynamics: Liouville's theorem also applies to chaotic systems. A ``collision'' term, such as the BGK ansatz (\ref{eq:BGK}), is only an effective term that approximates the propagation of charged particles in turbulent magnetic fields. As we illustrated in Sec.~\ref{sec:StandardPicture}, the term that is replaced by the BGK ansatz is {\it derived} from Liouville's theorem in the first place. We conclude that Liouville's theorem is valid and can be used to compute the anisotropy maps unless non--conservative forces (e.g.\ energy losses), sources, or sinks are considered.

In above disucssion, we have ignored the presence of sources and considered the generation of small--scale anisotropies {\it locally}, that is on scales of a few scattering lengths and therefore smaller than the scales on which the gradient develops. This can be justified as the formation of the gradient is taking place over many scattering lengths and possibly also coherence lengths of the magnetic field. Nevertheless, one might wonder if sources or sinks (which are not present in the approach based on Liouville's theorem) could alter the picture. This question has been investigated~\cite{Pohl:2015fdp} in a specific setup where particles are injected on a sphere of radius $R_{\text{inj}}$ and either reach the detector sphere of radius $R_{\text{det}}$ or escape at $R_{\text{esc}}$ with $R_{\text{det}} < R_{\text{inj}} < R_{\text{esc}}$. This setup produces large-- and small--scale anisotropies in the arrival directions at $R_{\text{det}}$ even for a random, but uniformly distributed injection at $R_{\text{inj}}$. The anisotropies are therefore a consequence of the {\it non--uniform} initial distribution of particles that reach $R_{\text{det}}$ or again of the ensemble fluctuations of the local magnetic field. For $R_{\text{esc}} \simeq 2 R_{\text{inj}} \sim 0.1 - 100 L$ (where $L$ is the outer scale of turbulence) the anisotropy level is $\mathcal{O}(1)$, much higher than observed at PeV energies. It is not obvious how this would change when scaling the geometry to a more realistic, Galactic setup.

%--------------------------------------------------------------------------------------------------------------------------------
%--------------------------------------------------------------------------------------------------------------------------------

\subsection{Exotic Scenarios}
\label{subsec:ExoticScenarios}

So far, we have only considered scenarios that describe the small--scale anisotropies as a result of CR source distributions and electromagnetic field configurations. Some authors have speculated that the small--scale anisotropy, in particular region A first observed by Milagro~\cite{Abdo:2008kr}, could originate via exotic scenarios. One of these scenarios~\cite{Kotera:2013mpa} considers the emission of {\it strangelets} via the collapse of a neutron stars to a more compact astrophysical object consisting of strange quark matter (SQM). 

Using energy and symmetry arguments, it has been speculated~\cite{Bodmer:1971we,Witten:1984rs,Farhi:1984qu} that SQM, a hypothetical form of matter with roughly equal numbers of up ($u$), down ($d$), and strange ($s$) quarks, could be the true ground state of quantum chromodynamics (QCD).  For a plasma of quarks in thermodynamical equilibrium it might be energetically preferable to condense into a phase containing strange quarks instead of ordinary matter with neutrons ($udd$) and protons ($uud$). Lumbs of SQM, so-called {\it stranglets}, can have a large atomic mass number $A$ and charge $Z$. Classical strangelets have a quark charge $Z\sim 0.1A$ for low mass numbers ($A\ll700$). For total quark charges exceeding $Z\sim\alpha^{-1}\sim 137$ strong field QED corrections lead to screening and $Z\sim 8A^{1/3}$ ($A\gg700$)~\cite{Farhi:1984qu}. It has also been speculated that color and flavor symmetries at high baryon densities might be broken simultaneously by the condensation of quark Cooper pairs~\cite{Madsen:2001fu}. In this scenario, the ``color-flavor-locked'' strangelets have charges of $Z\sim0.3A^{2/3}$~\cite{Madsen:2000kb}.

Now, the model of Ref.~\cite{Kotera:2013mpa} considers the acceleration of intermediate-mass stranglets ($A\simeq10^3$) in the rapid transition of a neutron star to a strange star. If the ejecta reach bulk Lorentz factors of the order of $\Gamma\simeq10$, the strangelets will reach energies of the order of $10$~TeV. These strangelets could be misidentified as high energy CRs in extend air shower observatories. Similar to ordinary CRs, the high charge of the strangelets would scramble the arrival direction and only contribute to large--scale anisotropy features. However, if the neutron star is close or embedded in the dense molecular cloud, the strangelets passing the cloud can become neutral by electron capture or spallation. From that moment the neutral strangelets propagates undeflected in Galactic magnetic fields and could be visible as small--scale features, corresponding to the projected size of the molecular cloud onto the sky. The authors of Ref.~\cite{Kotera:2013mpa} argue that the hotspot region A is located close to the Taurus molecular cloud, a star-forming region at a distance of 140~pc, which could be responsible for the feature.

Another study~\cite{Harding:2013qra} argues that the annihilation in close dark matter (DM) sub-halos could be responsible for small--scale features. Dark matter annihilation can proceed via various channels, that lead to the production of cosmic rays (nuclei and leptons), gamma rays, and neutrinos. In general, the DM halo is expected to peak close to the Galactic Center. The emitted hadronic cosmic rays would diffuse through the Galactic environment and would only contribute to the large--scale anisotropy. The authors of Ref.~\cite{Harding:2013qra} consider the case that a DM sub-halo with mass of the order of $10^6M_\odot$ exists within 100~pc of the solar system and close to a magnetic stream that maps the CR arrival directions onto region A. This model is very similar to that considered in Refs.~\cite{Salvati:2008dx}, including its problems mentioned in Section~\ref{subsec:NonDiffusive}. Here, however, the source of CRs is not an ordinary CR accelerator, but the DM annihilation in the sub-halo. 

In general, the CR spectrum of DM annihilation will show a sharp cutoff close to the DM mass $m_{\rm DM}$ and can be harder than  typical astrophysical spectra. This is consistent with the energy spectrum of region A reported by the Milagro~\cite{Abdo:2008kr} and HAWC~\cite{Abeysekara:2014sna}. Ref.~\cite{Harding:2013qra} shows that the spectrum can be reproduced by DM in the mass range of $30~{\rm TeV}\lesssim m_{\rm DM}\lesssim200~{\rm TeV}$ with annihilation channels into electroweak bosons ($W^+W^-$ or $Z^0Z^0$) and bottom quarks ($bb$). These multi-TeV DM models have a natural Sommerfeld-enhanced annihilation cross section due to electroweak boson exchange. Besides the general problem of CR transport described by Ref.~\cite{Salvati:2008dx}, the DM sub-halo scenario will also contribute to other channels. The authors of Ref.~\cite{Harding:2013qra} have argued that anti-proton bounds from PAMELA and ARGO-YBJ and gamma-ray bounds from Fermi and HESS are consistent with the emission. Future observatories like HAWC or CTA would have the ability to test the gamma-ray emission of the sub-halo.

%--------------------------------------------------------------------------------------------------------------------------------
%--------------------------------------------------------------------------------------------------------------------------------
%--------------------------------------------------------------------------------------------------------------------------------

\section{Summary and Outlook}
\label{sec:SummaryOutlook}

We have reviewed the observation and interpretation of small--scale anisotropies in the arrival directions of CRs. These subtle features have only become accessible in recent years due to the increased event statistics and improved angular resolution of ground--based CR observatories. Typically, these results are quantified in terms of a harmonic analysis of two-dimensional relative intensity maps. We have highlighted that the various methods used by observatories do not always yield the same result and need to be rescaled to connect to predictions. Moreover, due to the subtleties of the reconstruction of relative intensity maps, observatories are not sensitive to anisotropies aligned with Earth's rotation axis.

Standard diffusion theory predicts the emergence of large--scale anisotropies that are related to the distribution of sources, the properties of the diffusion medium, or the relative motion of the observer. For completeness, we have summarized the well--known formalism that predicts a CR dipole anisotropy based on these quantities. Experimental observations correspond to the projection of the CR dipole anisotropy onto the equatorial plane that can be parametrized by the projected amplitude and a phase. We argue that the strong energy--dependence of the dipole amplitude and phase observed by various recent measurements can be understood in the context of standard diffusion theory.

On the other hand, the appearance of strong small--scale anisotropies in relative intensity maps is somewhat surprising. Classical scenarios can account for small--scale features via electromagnetic mechanisms in the local environment, i.e.\ within the typical scattering length of diffusion. One possibility consists of magneto-hydrodynamical effects in the heliosphere, that create electric potentials or magnetic reconnections. Particle acceleration in these localized environments can create intensity hot-spots that align with features of the heliosphere, e.g.\ the heliotail. Static configurations of magnetic fields can introduce magnetic mirrors and lenses that distort the large--scale arrival direction of CRs. Possible mechanism that create these configurations could be, again, the heliosphere or local magnetic turbulence, that is responsible for diffusion in the first place. Non--uniform pitch--angle diffusion in the local ordered magnetic field can itself create axisymmetric small-- and medium--scale anisotropies.

The possibility that the intermittent turbulent magnetic field is creating the small--scale anisotropies is certainly very attractive. Not only is turbulence a necessary ingredient in any model of diffusive CR transport, it also provides a theoretical framework that allows making {\it quantitative} predictions as the anisotropy mirrors the local structure of the turbulent magnetic field. This could be similar to how diffusion coefficient can be computed from the magnetic correlation functions in quasi--linear theory. It might be interesting to ask whether observations of anisotropies will help us to determine the structure of the local turbulent magnetic field. Solving the inverse problem of determining the field from the distribution of arrival directions might seem hopeless. In a statistical sense, it should however be possible to learn something about the statistical properties of magnetic turbulence from the observations of CR anisotropies. An apparent difficulty in this might be that the models presented in Sec.~\ref{subsec:MagneticTurbulence} are already giving a good description of the observations, see e.g.\ Fig.~\ref{fig:power100}. This is quite surprising, given that in (most of) these models turbulence is assumed to be isotropic whereas there are strong indications that it must be anisotropic~\cite{Goldreich:1994zz}. It is by no means clear that the remaining discrepancies between the predictions and the data can be used to find evidence for more realistic models of turbulence, especially in the presence of the various observational systematics and conceptual difficulties as discussed.

We have already stressed the fact that the angular power spectrum contains all the statistical information only for a Gaussian, statistically isotropic random field. However, it is not clear at all that the anisotropy maps should in fact be of this kind. On the contrary, the presence of a regular magnetic field should manifest itself to a certain degree in a breaking of the isotropy. Extracting the additional information from sky maps and confronting it with models could be a promising future line of attack. A difficulty here will be ``cosmic variance'', i.e.\ the fact that we can only observe one anisotropy map, at least at any one energy. Cross--correlating maps at different energies can play another important role for scrutinising the structure of magnetic turbulence in our Galactic neighbourhood.

An apparent difficulty in comparing maps at different energies and from different experiments is the fact that each map is composed of a distribution of energies or a distribution of rigidities. Many of the proposed explanations for the formation of the small--scale anisotropies are in fact rigidity--dependent, resulting in different maps (different random phases) at different rigidities. The fact that observations can only be made with a finite energy resolution must to a certain degree wash out the structure predicted for individual energies. An additional difficulty can be that the chemical compositions assumed to convert the distribution of observed energies into a distribution of rigidities can vary between different observatories.

A few exotic scenarios of small--scale anisotropies have also been considered. It was considered that the production of cosmic rays via dark matter annihilation in sub-halos can introduce anisotropyies. However, this scenario will most likely be observable as an additional contribution to the large--scale anisotropy. On the other, an exotic neutral messenger produced in extended Galactic sources, e.g.~strangelets, could be observable as a local excess in anisotropy maps. This type of scenario could be testable, e.g.\ by correlating the small--scale features in anisotropy maps with known sources in our local environment. However, this has to be done with some care, as $\gamma$-rays from the same sources are a potential background in CR anisotropy measurements.

Finally, the interpretation of large-- and small--scale anisotropies is affected by systematic uncertainties related to the methods used for the reconstruction of anisotropy maps and their harmonic analysis. We hope that this review can be a useful guideline for, both, observers and analyzers regarding the reporting of results and their interpretation.

\paragraph{Acknowledgments}

This work is supported in part by the National Science Foundation (Grants No.~PHY-1306958 and No.~PLR-1600823) and Danmarks Grundforskningsfond (Grant No.~1041811001).

\appendix

\section{Multipole Coupling Matrix}\label{sec:appendix}

For the special case of ground--based observatories the weight function $w$ of the field of view is expected to be azimuthally symmetric~(see Ref.~\cite{Ahlers:2016njl}). In this case, the matrix elements of the matrix ${\bf K}$ in Eq.~(\ref{eq:Kmatrix}) are block-diagonal, $K_{\ell m\ell'm'} = \delta_{mm'}T^m_{\ell\ell'}(w)$, with block elements defined via a sum over Wigner-$3j$ coefficients,
\begin{equation}
T^m_{\ell\ell'}( b) = (-1)^{m}\sum_{k=|\ell-\ell'|}^{\ell+\ell'} b_{k0}\sqrt{\frac{(2\ell+1)(2\ell'+1)(2k+1)}{4\pi}}\begin{pmatrix}\ell&\ell'&k\\0&0&0\end{pmatrix}\begin{pmatrix}\ell&\ell'&k\\m&-m&0\end{pmatrix}\,.
\end{equation}
Under the additional assumption that the anisotropy is an isotropic, Gaussian random field we can evaluate the transfer matrix  ${\bf M}$ in Eq.~(\ref{eq:Mmatrix}) to
\begin{equation}
M_{\ell\ell'} = \frac{2\ell'+1}{4\pi}\sum_{k}(2k+1)W_{k}\begin{pmatrix}\ell&\ell'&k\\0&0&0\end{pmatrix}^2 - \frac{[T^0_{\ell\ell'}( w)]^2}{2\ell+1}\,.
\end{equation}
Note that the unfamiliar last term in the previous equation accounts for the projection of the pseudo angular momentum onto $m\neq0$ terms.

\section*{References}

\bibliography{smallscale}

\end{document}